\pdfoutput=1

\documentclass[pdftex,twocolumn,epjc3]{svjour3}          

\usepackage[margin=0.9cm,bmargin=1.1cm]{geometry} 

\RequirePackage[T1]{fontenc}
\smartqed  
\RequirePackage{mathptmx}      
\RequirePackage{caption}
\RequirePackage[numbers,sort&compress]{natbib}
\RequirePackage[colorlinks,citecolor=blue,urlcolor=blue,linkcolor=blue]{hyperref}

\journalname{Eur. Phys. J. C}
\usepackage{booktabs}
\usepackage{xspace}
\usepackage{color}\definecolor{darkblue}{rgb}{0,0,0.5}
\usepackage{hyperref}
\hypersetup{
    bookmarks=true,         
    unicode=false,          
    pdftoolbar=true,        
    pdfmenubar=true,        
    pdffitwindow=true,      
    pdftitle={My title},    
    pdfauthor={Author},     
    pdfsubject={Subject},   
    pdfnewwindow=true,      
    pdfkeywords={keywords}, 
    colorlinks=true,       
    linkcolor=darkblue,          
    citecolor=darkblue,        
    filecolor=magenta,      
    urlcolor=darkblue,         
}

\usepackage[latin2]{inputenc}
\usepackage{indentfirst}
\usepackage{enumerate}

\usepackage{amsmath}
\usepackage{bm}
\usepackage{amssymb}
\usepackage[english]{babel}
\usepackage{url}
\usepackage{natbib}

\usepackage{multirow}
\usepackage[abs]{overpic}

\usepackage{subcaption} 

\usepackage{graphicx}

\usepackage{epstopdf}

\usepackage{stmaryrd}

\usepackage[switch]{lineno}

\journalname{Eur. Phys. J. C}

\newcommand{\eV}{\ensuremath{\mbox{e\kern-0.1em V}}\xspace}
\newcommand{\GeV}{\ensuremath{\mbox{Ge\kern-0.1em V}}\xspace}
\newcommand{\MeV}{\ensuremath{\mbox{Me\kern-0.1em V}}\xspace}
\newcommand{\GeVc}{\ensuremath{\mbox{Ge\kern-0.1em V}\!/\!c}\xspace}
\newcommand{\GeVcc}{\ensuremath{\mbox{Ge\kern-0.1em V}\!/\!c^2}\xspace}
\newcommand{\AGeV}{\ensuremath{A\,\mbox{Ge\kern-0.1em V}}\xspace}
\newcommand{\AGeVc}{\ensuremath{A\,\mbox{Ge\kern-0.1em V}\!/\!c}\xspace}
\newcommand{\MeVc}{\ensuremath{\mbox{Me\kern-0.1em V}\!/\!c}\xspace}

\newcommand{\mfkp}{\ensuremath{\mbox{(Ge\kern-0.1em V}\!/\!c\mbox{)kG}^{-1}\mbox{cm}^{-1}}\xspace}

\newcommand{\cm}{\ensuremath{\mbox{cm}}\xspace}
\newcommand{\mm}{\ensuremath{\mbox{mm}}\xspace}

\newcommand{\micron}{\ensuremath{\mu\mbox{m}}\xspace}

\newcommand{\mrad}{\ensuremath{\mbox{mrad}}\xspace}
\newcommand{\mus}{\ensuremath{\mu\mbox{s}}\xspace}
\newcommand{\ns}{\ensuremath{\mbox{ns}}\xspace}
\newcommand{\ps}{\ensuremath{\mbox{ps}}\xspace}

\newcommand{\kHz}{\ensuremath{\mbox{kHz}}\xspace}

\newcommand{\dd}{\ensuremath{{\textrm d}}\xspace}
\newcommand{\dedx}{\ensuremath{\dd E\!/\!\dd x}\xspace}
\newcommand{\tof}{\ensuremath{tof}\xspace}
\newcommand{\mtof}{\ensuremath{\mbox{m}^2_{tof}}\xspace}

\newcommand{\prot}{\ensuremath{\textup{p}}\xspace}

\newcommand{\pim}{\ensuremath{\pi^-}\xspace}
\newcommand{\pip}{\ensuremath{\pi^+}\xspace}

\newcommand{\kz}{\ensuremath{K^0_{s}}\xspace}

\newcommand{\ep}{\ensuremath{\textup{e}^+}\xspace}
\newcommand{\lm}{\ensuremath{\Lambda}\xspace}

\newcommand{\numu}{\ensuremath{\nu_{\mu}}\xspace}
\newcommand{\anumu}{\ensuremath{\bar{\nu}_{\mu}}\xspace}

\newcommand{\km}{\ensuremath{K^-}\xspace}
\newcommand{\kp}{\ensuremath{K^+}\xspace}

\newcommand{\FlukaEleven}{{\scshape Fluka2011}\xspace}

\newcommand{\GeantThree}{{\scshape Geant3}\xspace}

\newcommand{\GeantFT}{{\scshape Geant4.10}\xspace}
\newcommand{\QGSP}{{\scshape QGSP\_BERT}\xspace}
\newcommand{\NuBeam}{{\scshape NuBeam}\xspace}

\newcommand{\GCALOR}{{\scshape GCALOR}\xspace}

\newcommand{\CernVM}{\textsc{Cern\-\kern-0.05emVM}\xspace}

\newcommand{\sk}{\textit{Super-Kamiokande}\xspace}      

\newcommand{\NASixtyOne}{NA61/\-SHINE\xspace}
\begin{document}
\newcommand{\ra}[1]{\renewcommand{\arraystretch}{#1}}
\DeclareGraphicsExtensions{.pdf,.png,.eps,.jpg,.ps}
\cleardoublepage


\title{\large \bf
  Measurements of $\pi^\pm$, $K^\pm$ and proton double differential yields from the surface of the T2K replica target for incoming 31~GeV/$c$ protons with the NA61/SHINE spectrometer at the CERN SPS
}
	
\clearpage

\institute{
{National Nuclear Research Center, Baku, Azerbaijan}\label{inst0}
\and{Faculty of Physics, University of Sofia, Sofia, Bulgaria}\label{inst1}
\and{Ru{\dj}er Bo\v{s}kovi\'c Institute, Zagreb, Croatia}\label{inst2}
\and{LPNHE, University of Paris VI and VII, Paris, France}\label{inst3}
\and{Karlsruhe Institute of Technology, Karlsruhe, Germany}\label{inst4}
\and{Fachhochschule Frankfurt, Frankfurt, Germany}\label{inst5}
\and{University of Frankfurt, Frankfurt, Germany}\label{inst6}
\and{Wigner Research Centre for Physics of the Hungarian Academy of Sciences, Budapest, Hungary}\label{inst7}
\and{University of Bergen, Bergen, Norway}\label{inst8}
\and{Jan Kochanowski University in Kielce, Poland}\label{inst9}
\and{H. Niewodnicza\'nski Institute of Nuclear Physics of the
      Polish Academy of Sciences, Krak\'ow, Poland}\label{inst10}
\and{National Centre for Nuclear Research, Warsaw, Poland}\label{inst11}
\and{Jagiellonian University, Cracow, Poland}\label{inst12}
\and{AGH - University of Science and Technology, Cracow, Poland}\label{inst13}
\and{University of Silesia, Katowice, Poland}\label{inst14}
\and{University of Warsaw, Warsaw, Poland}\label{inst15}
\and{University of Wroc{\l}aw,  Wroc{\l}aw, Poland}\label{inst16}
\and{Warsaw University of Technology, Warsaw, Poland}\label{inst17}
\and{Institute for Nuclear Research, Moscow, Russia}\label{inst18}
\and{Joint Institute for Nuclear Research, Dubna, Russia}\label{inst19}
\and{National Research Nuclear University (Moscow Engineering Physics Institute), Moscow, Russia}\label{inst20}
\and{St. Petersburg State University, St. Petersburg, Russia}\label{inst21}
\and{University of Belgrade, Belgrade, Serbia}\label{inst22}
\and{University of Geneva, Geneva, Switzerland}\label{inst23}
\and{Fermilab, Batavia, USA}\label{inst24}
\and{University of Colorado, Boulder, USA}\label{inst25}
\and{University of Pittsburgh, Pittsburgh, USA}\label{inst26}
\and{Los Alamos National Laboratory, Los Alamos, USA}\label{inst27}
\and{Institute for Particle and Nuclear Studies, KEK, Tsukuba, Japan}\label{inst28}
\and{Kavli Institute for the Physics and Mathematics of the Universe (WPI), The University of Tokyo Institutes for Advanced Study, University of Tokyo, Kashiwa, Japan}\label{inst30}
\and{TRIUMF, Vancouver, BC, Canada}\label{inst31}
\and{Department of Physics and Astronomy, York University, Toronto, ON, Canada}\label{inst34}
\and{Tokyo Institute of Technology, Department of Physics, Tokyo, Japan}\label{inst35}
\and{Oxford University, Department of Physics, Oxford, United Kingdom}\label{inst36}
\and{University of Bern, Bern, Switzerland}\label{inst37}
}

\author{
{N.~Abgrall}\thanksref{inst23}
\and{A.~Aduszkiewicz}\thanksref{inst15}
\and{E.V.~Andronov}\thanksref{inst21}
\and{T.~Anti\'ci\'c}\thanksref{inst2}
\and{B.~Baatar}\thanksref{inst19}
\and{M.~Baszczyk}\thanksref{inst13}
\and{S.~Bhosale}\thanksref{inst10}
\and{A.~Blondel}\thanksref{inst23}
\and{M.~Bogomilov}\thanksref{inst1}
\and{A.~Brandin}\thanksref{inst20}
% ...
\\(\NASixtyOne Collaboration) 
}

 \date{\today}
\advance\hoffset by 0.5cm
\advance\voffset by -0.7cm
 \maketitle
\advance\voffset by 0.7cm
\advance\hoffset by -0.5cm

 \begin{abstract}
	Measurements of the $\pi^\pm$, $K^\pm$, and proton double differential yields emitted from the surface of the $90$-\cm-long carbon target (T2K replica) were performed for the incoming $31\:$\GeVc protons with the \NASixtyOne spectrometer at the CERN SPS using data collected during 2010 run. The double differential $\pi^\pm$ yields were measured with increased precision compared to the previously published \NASixtyOne results, while the $K^\pm$ and proton yields were obtained for the first time. A strategy for dealing with the dependence of the results on the incoming proton beam profile is proposed. The purpose of these measurements is to reduce significantly the (anti)neutrino flux uncertainty in the T2K long-baseline neutrino experiment by constraining the production of (anti)neutrino ancestors coming from the T2K target.

\end{abstract}


 \tableofcontents


 \pagenumbering{arabic}


\section{Introduction} \label{sec:Introduction}

\NASixtyOne (SPS Heavy Ion and Neutrino Experiment)~\cite{NA61detector_paper} is a large hadron spectrometer at the CERN Super Proton Synchrotron (SPS). The \NASixtyOne collaboration is pursuing several physics goals including hadron production measurements for T2K (Tokai to Kamioka)~\cite{T2K} - an accelerator-based long-baseline neutrino experiment in Japan. The \NASixtyOne measurements are used to reduce the systematic uncertainties associated to the prediction of the (anti)neutrino fluxes in T2K. 
New measurements of $\pi^{\pm}$, $K^{\pm}$ and \prot yields coming from the surface of the T2K replica target are presented here. These results aim at reducing the T2K (anti)neutrino flux uncertainties down to a 3--4\% level. 
They can further be used to tune hadron interaction and transport models.

The paper is structured as follows: 
a motivation for these hadron measurements is first presented. 
In Section~\ref{sec:ExpSetup}, a brief overview of the \NASixtyOne setup is shown followed by the description of the analysis in Section~\ref{sec:Analysis}. Section~\ref{sec:Syst} gives a concise description of all systematic uncertainties. Results and comparison with Monte Carlo (MC) models and with the previous \NASixtyOne measurements are presented in Section~\ref{sec:Results}. Finally, in Section~\ref{sec:ReWeightFactors}, proper usage of these results in the T2K neutrino beam simulation is discussed, followed by a short conclusion.

\subsection{The T2K neutrino beam}
The T2K neutrino beam is produced at the Japan Proton Accelerator Research Complex (J-PARC)~\cite{jparc:2003} by directing a $30\:$\GeV (kinetic energy) proton beam towards a $90$-\cm-long graphite target. Produced mesons, mostly pions and kaons~\cite{T2Kflux}, are focused by a set of three magnetic horns~\cite{T2K} and decay to neutrinos in the decay volume. Focusing horns are aluminum conductors that create a toroidal magnetic field with respect to the beam direction. The polarity of the current in the horns can be changed, so positively or negatively charged pions (kaons) can be focused. In this way, T2K can produce either a neutrino-enhanced (\numu) or an antineutrino-enhanced  (\anumu) beam. Since direct hadron production measurements in situ are not possible, Monte Carlo models are used to predict the (anti)neutrino fluxes at the T2K near and far detectors. However, significant differences between hadron production models induce considerable systematic uncertainties on the (anti)neutrino fluxes. Without any experimental data to constrain the hadron production models, these systematic uncertainties would be larger than $25$\%. For this reason, T2K uses available hadron production measurements, but mostly relies on dedicated \NASixtyOne hadron production data~\cite{T2Kflux}. 

\subsection{The \NASixtyOne measurements for T2K}
The \NASixtyOne
experiment took data
for T2K in 2007, 2009 and 2010. In 2007 and 2009, measurements were performed with a 30.92~\GeVc proton beam and a $2$-\cm-thick carbon target (4\% of a nuclear interaction length, $\lambda_i$) to measure hadron multiplicities ($\pi^{\pm}$, $K^{\pm}$, \prot, \kz, \lm) and the production cross section~\cite{V0_2007, pion_paper, kaon_paper, thin2009paper}. Production events represent the fraction of inelastic events in which at least one new hadron in the final state is produced. By using these measurements, the uncertainty on the neutrino flux in T2K was reduced to about $10\%$~\cite{T2Kflux}. However, the hadron production component of the uncertainty still dominated the total uncertainty due to the insufficient precision of the production cross section measurements
and re-interactions inside the target and aluminum horns, which cannot be directly constrained by the measurements of the primary proton-carbon interactions. To further reduce the hadron production uncertainty of the neutrino flux, hadron production measurements with a replica of the T2K target were needed. Measurements with the T2K replica target can directly constrain up to $90$\% of the neutrino flux because hadron yields at the surface of the target are measured not just for primary interactions, but also for re-interactions inside the target~\cite{LTpaper}. These measurements were also done in 2007, 2009 and 2010. Measurements from the 2007 run~\cite{LTpaper} were used as a proof-of-principle, while measurements from the 2009 run~\cite{LTpaper2009, Hasler:2039148} are being incorporated in the T2K neutrino flux simulation. The expected (anti)neutrino flux uncertainty is around $5$\%~\cite{Zambelli:2017pvq}, representing a significant improvement with respect to the previously published uncertainty of $10\%$. Measurements from the 2010 run are the main topic of this paper. With respect to the previous results, these measurements of $\pi^{\pm}$ yields have smaller statistical and systematic uncertainties. Furthermore, $K^{\pm}$ and \prot yields are measured for the first time. This is expected to further reduce the uncertainties on the (anti)neutrino fluxes in T2K from about 10\% down to the level of $3-4\%$. 
Moreover, since (anti)neutrino fluxes in T2K in the energy region above $\sim$2~GeV are produced mainly by kaon decays, the corresponding uncertainties will be greatly reduced.

\subsection{The T2K requirements}
T2K imposes strict requirements on hadron production measurements with the T2K replica target~\cite{LTpaper2009}. Hadron yields must be measured as a function of the momentum, polar angle and longitudinal position of the hadron exit point on the target surface. In other words, vertices inside the target are not reconstructed. Tracks measured in a detector are only extrapolated to the target surface. Essentially, the target is treated as a black box and measured hadron yields are the sum of yields coming from primary interactions and re-interactions. 
In addition, T2K requires that the longitudinal position along the target surface is binned in at least six bins: five longitudinal bins $18\:$\cm in size and the downstream target face as a sixth bin~\cite{PavinThesis}. Since the T2K target is inserted in the first focusing horn, hadrons coming from different parts of the target will have different paths through the magnetic field and they will be focused differently by the horns. The previous \NASixtyOne measurements with 2007 and 2009 data have already proven that the \NASixtyOne setup is suitable for measuring hadron yields on the T2K replica target surface, although the target is placed outside of the tracking system.

\section{NA61/SHINE experimental setup} \label{sec:ExpSetup}

\begin{figure*}[ht]
    \begin{center}
                \begin{subfigure}[t]{0.98\textwidth}
                        \includegraphics[trim={0cm, 5cm, 0cm, 0cm},clip,width=1\textwidth]{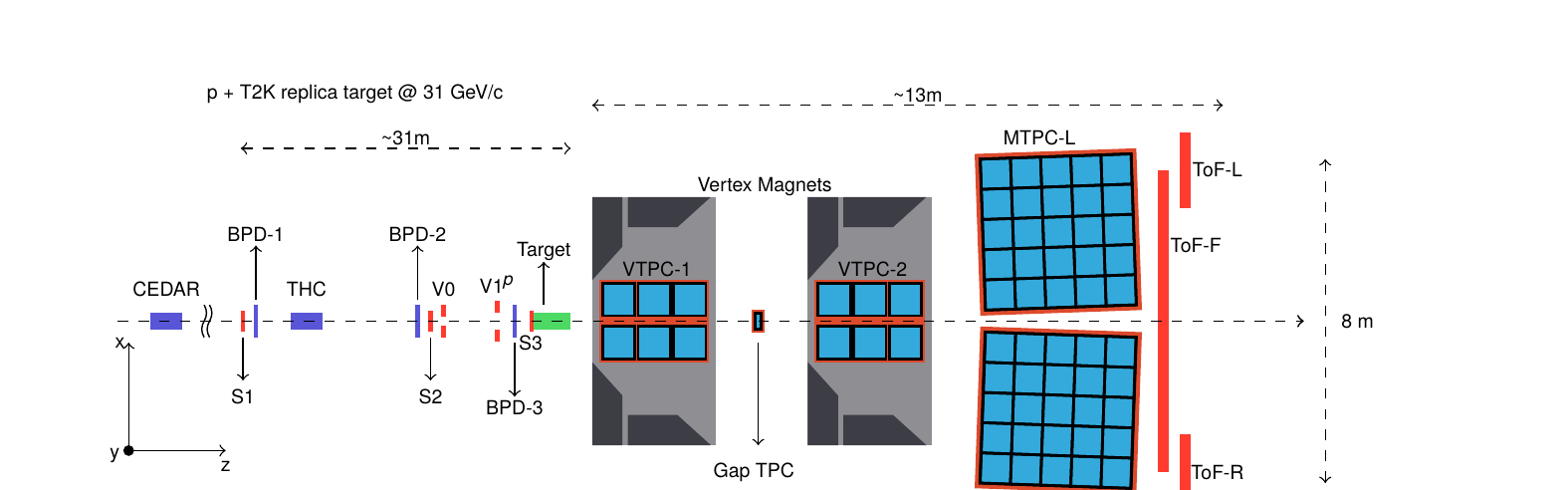}
                        \caption{}
                \end{subfigure}
                \begin{subfigure}[t]{0.98\textwidth}
                        \includegraphics[trim={0cm, 1cm, 0cm, 6cm},clip,width=1\textwidth]{figures/setup/na61setup.pdf}
                        \caption{}
                \end{subfigure}
                \caption{A top view (a) and a side view (b) of the \NASixtyOne experimental setup used in 2010 for the T2K replica target data-taking. The beam comes from the left. The orientation of the coordinate system is shown in the bottom left corner, while its origin is located at the centre of VTPC-2.}\label{fig:NA61Setup}
    \end{center}
    
\end{figure*}

The \NASixtyOne setup during the T2K replica target data-taking in 2010 consisted of five Time Projection Chambers (TPCs), three Time-of-Flight walls (ToF-F, ToF-L and ToF-R), three Beam Position Detectors (BPD-1, BPD-2, BPD-3), a $90$-\cm-long graphite target, five scintillator counters and two Cherenkov detectors. Two of the vertex TPCs (VTPC-1 and VTPC-2) are inside the magnetic field created by two superconducting magnets. For the study presented here the magnetic field of the dipole magnets was set to a bending power of 1.14~T\,m (standard magnetic field), while a small subset of data was also taken with the full magnetic field of 9~T\,m. A schematic overview of the setup is presented in Figure~\ref{fig:NA61Setup}. More details can be found in Ref.~\cite{NA61detector_paper}. The coordinate system is defined as follows: the $z$-axis is in the nominal direction of the beam, the $x$-axis in the horizontal plane is such that positively charged particles are bent in the positive $x$-direction, and the $y$-axis is perpendicular to the horizontal plane and points upward. The origin is located in the centre of the VTPC-2. 


\subsection{Beamline}

The \NASixtyOne spectrometer~\cite{NA61detector_paper} is served by the so-called H2 beamline at the north area of the SPS. The beamline is designed to transport both primary and secondary hadrons and ions from the maximum SPS energy ($400\:$\GeVc) down to $13\:$\AGeVc. Secondary hadron beams of various momenta are produced by impinging a primary $400\:$\GeVc proton beam on a beryllium target. Produced hadrons are selected by two spectrometers and a set of collimators according to their rigidity (momentum to charge ratio). The secondary beam is then transported towards  \NASixtyOne. The beam is defined by three scintillator counters used in coincidence ($S_1$, $S_2$ and $S_3$) and two scintillator counters with holes used in anticoincidence ($V_0$ and $V_1^p$). The $S_3$ counter has a radius equal to the target radius, and it was placed $0.5\:$\cm upstream of the target. The $V_0$ counter has a $1\:$\cm diameter hole centered on the beam axis, which allows the selection of a narrower beam, if necessary. A Cherenkov Differential Counter with Achromatic Ring Focus~\cite{CEDAR} (CEDAR) and Threshold Cherenkov Counter (THC) were used in coincidence and anti-coincidence respectively to identify beam particles. By changing the gas pressure in these detectors, it is possible to estimate the beam composition. The $30.92\:$\GeVc beam contains around $76.3$\% $\pi^+$, $1.6$\%  $K^+$, and $12$\% protons~\cite{PavinThesis}. The estimated purity of the selected proton sample is better than $99.9$\%. Finally, the beam position is measured by a set of six multi-wire proportional chambers. Two chambers, one measuring position along the $x$-axis and another measuring the position along the $y$-axis, are part of one BPD. The precision of the position measurement in one BPD is around $200\:$\micron~\cite{NA61detector_paper}.

The third BPD was mounted on the same support as the target, just upstream of the $S_3$ counter.

\subsection{Target}
\begin{figure*}[tb]
        \includegraphics[trim= 0 0 0 0, clip, width=0.85\textwidth]{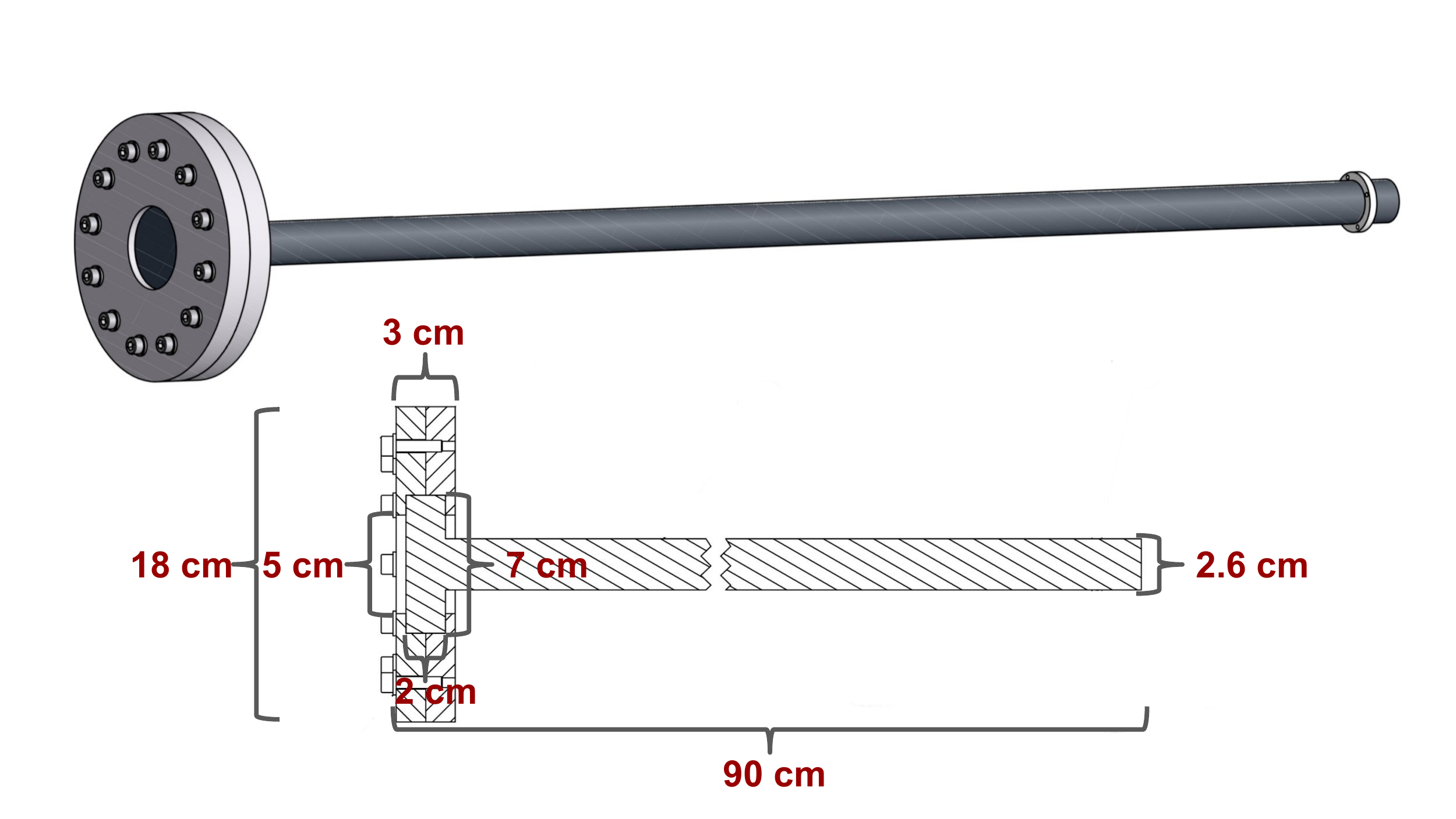}

\caption{The T2K replica target with aluminium flange attached to the upstream disk. The hole at the upstream side of the flange was used for the $S_3$ counter. A technical drawing of the T2K replica target with the upstream flange zoomed in is shown on the bottom of the figure.}\label{fig:target}
\end{figure*}

The T2K target~\cite{T2K} has a modular structure which includes a graphite core, a thin titanium case, cooling pipes filled with helium. The replica target design matches dimensions and material of the T2K graphite core. It was made of Toyo Tanso IG-43 graphite~\cite{toyotanso}. The target density was estimated by measuring its volume and mass, and it is equal to $1.83\pm0.03\:$g~cm$^{-3}$. This value is $1.5$\% higher than the density of the T2K target, but well within the measurement uncertainty. The target shape was machined to form a $2\:$\cm thick disk with an $88\:$\cm long rod coming from its centre. The radii of the disk and rod are $3.5$ and $1.3\:$\cm, respectively. The disk was used to mount an aluminium flange which was, in turn, tightened to a target holder. The upstream side of the flange has a hole to minimize beam interactions with aluminium and provide space for the $S_3$ counter used in the trigger. A schematic overview of the target can be seen in Figure~\ref{fig:target}. The position and tilt of the target were adjusted by the screws on the target holder.
The downstream face of the target was placed around $67\:$\cm from the upstream side of VTPC-1. 

\subsection{Tracking system}
The tracking system consists of two Vertex TPCs located in the magnetic field with an additional small TPC called the Gap TPC (GTPC) located between them, and two Main TPCs (MTPC-L and MTPC-R), located downstream symmetrically with respect to the beamline. All TPCs were inherited from the previous NA49 experiment~\cite{NA49-NIM}. The GTPC was introduced in order to measure forward-going particles with small polar angles, $\theta < 10\:$\mrad, in the laboratory frame. The magnetic field was set so that the total bending power of the magnets was $1.14$~T\,m. This allowed a momentum resolution of $\sigma_p/p^2 = 5 \times 10^{-3}\:$(\GeVc)$^{-1}$, except for very forward tracks that only passed through the GTPC and MTPCs and had a momentum resolution of $22 \times 10^{-3}\:$(\GeVc)$^{-1}$~\cite{PavinThesis}.  
The angular resolution is around 3--4~\mrad and 
it does not change much as a function of angle. 
The longitudinal position resolution changes with angle. For angles 
around 20~\mrad it is around 15~\cm. For larger angles, it goes down 
to around 2.5~\cm. Particles emitted from the downstream face of the target 
behave differently. Forward going particles are usually emitted from 
the center of the downstream target face. Therefore, extrapolation usually 
works quite well in this case. However, higher-angle particles coming from 
the downstream face of the target are usually emitted closer to the edge. 
In that case, migration effects increase.

The measured energy loss (\dedx) in the TPCs provided excellent particle identification capabilities; the achieved resolution is around $4$\%. In the region where the energy loss distributions for different particles cross, time-of-flight measurements ($\tof$) were used for particle identification.

\subsection{Time-of-flight walls}
Although the \NASixtyOne spectrometer has three different Time-of-Flight (ToF) detectors, for the T2K replica target measurements only the forward wall (ToF-F) was used for particle identification. The left and right walls (ToF-L and ToF-R) have superior granularity, but their coverage is insufficient. The ToF-F wall was constructed in 2007 and upgraded in 2009 for the T2K hadron production measurements. It consists of $80$ scintillator bars with dimensions W$\times$H$\times$L = $10\times120\times2.5\:$\cm arranged in ten separate modules. The signal is read out by two PMTs placed at both ends of the bars. The estimated $\tof$ resolution was $115\:$\ps~\cite{PavinThesis}.

\subsection{Triggers}

\begin{figure}[ht]
    \begin{center}
    \includegraphics[width=0.5\textwidth]{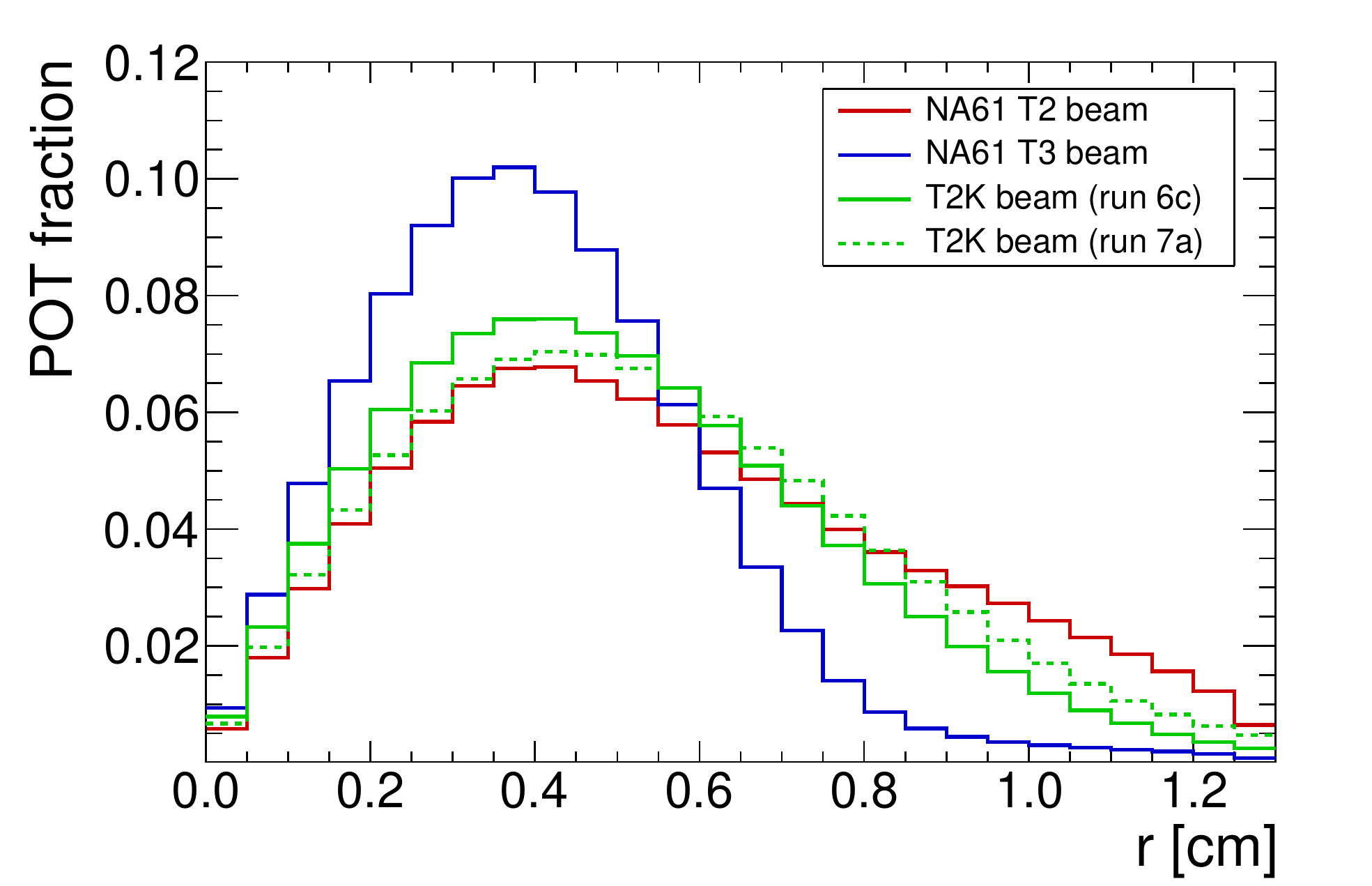}
    \caption{Radial distributions of the incoming proton beam with respect to the center of the target: $T_3$ beam profile (blue), $T_2$ beam profile (red) and T2K beam profiles for two different runs (green). Beam profiles in T2K change on a run-by-run basis and perfect agreement is not possible. All distributions are normalised to the corresponding total number of protons on target.\label{fig:beamrad}}
    \end{center}
\end{figure}

The proton beam profile on target at NA61/SHINE is wider than that at J-PARC. Previous studies~\cite{LTpaper2009} have shown that longitudinal ($z$) distribution of the hadrons emitted from the replica target surface depends on the width and position of the incoming proton beam at the target upstream face. In the simplest case where the beam has some radial distribution and where beam divergence and possible re-interactions in the target are neglected, a hadron, produced at position $z_i$ along the target with polar angle $\theta$ and radial position $\Delta r$ with respect to the target longitudinal axis, would escape the target at a minimal longitudinal position,  $z$, given by:
\begin{equation}
	z = z_i + \frac{\Delta r}{\tan \theta}.
\end{equation}
Evidently, for narrower beams, the longitudinal ($z$) distribution of the hadrons exiting the target surface is pushed more downstream. With a sufficiently narrow beam, one can suppress the number of hadrons coming from the first $18\:$cm of the target by an order of magnitude or more. Although this effect seems to be purely geometrical, a Monte Carlo simulation suggest that different beam widths and positions can lead to slightly different numbers of low momentum pions produced in re-interactions. The difference in the number of pions is around $1$\%. The \NASixtyOne proton beam achieved in 2009~\cite{LTpaper2009} is much wider than the J-PARC beam. To avoid large differences between the \NASixtyOne beam and the J-PARC beam, four special triggers were used during the 2010 data-taking (see Figure~\ref{fig:NA61Setup}):

\begin{subequations}
	\begin{align}
	T_1 = S_1 \cdot S_2 \cdot \:\:\quad\: \overline{V}_0 \cdot \overline{V}_1^p \cdot CEDAR \cdot \overline{THC},\\
	T_2 = S_1 \cdot S_2 \cdot S_3 \cdot \:\:\:\:\:\quad \overline{V}_1^p \cdot CEDAR \cdot \overline{THC},\\
	T_3 = S_1 \cdot S_2 \cdot S_3 \cdot \overline{V}_0 \cdot \overline{V}_1^p \cdot CEDAR \cdot \overline{THC},\\
	T_4 = S_1 \cdot S_2 \cdot S_3 \cdot \:\:\:\quad\: \:\overline{V}_1^p. \qquad \qquad \qquad 
	\end{align}
\end{subequations}
The $T_1$ trigger is the only trigger that selects beam protons that did not necessarily hit the target or when a hit in the $S_3$ scintillator was not detected. The $T_2$ and $T_3$ triggers select only beam protons that hit the target. The only difference is that $T_3$ selects a narrower beam because the $V_0$ counter has a smaller hole. The fourth trigger selects all beam particles that hit the target. It is important to note that $T_1$, $T_2$ and $T_4$ were prescaled - only a fraction of them was recorded. The prescaling was applied to $T_2$ to achieve the desired radial distribution of the beam. If an event is a $T_2$ but not a $T_3$, this means that the triggered beam particle is in the tail of the radial distribution on the upstream target face. To reduce the beam tail, only one out of two such events was registered. The achieved radial distribution is shown in Figure~\ref{fig:beamrad}. Perfect agreement with the T2K beam profile is not possible since in T2K the beam profile changes on a run-by-run basis. Any differences must be studied before using these measurements in the T2K neutrino beam simulation. Such study is presented in Section~\ref{sec:ReWeightFactors}.

In total, $10.2 \times 10^6$ triggers were recorded. Around $11.7$\% of the triggers were recorded with the maximum magnetic field of 9~T\,m. This allowed the recording of beam protons that pass through the target without interacting and the measurements of their parameters in the TPCs. However, all secondary hadrons below $6\:$\GeVc were bent out of the TPCs and did not reach the ToF-F wall. The analysis of data recorded with the maximum magnetic field is a subject of an independent publication. In this paper, the data taken with the standard magnetic field configuration ($89.3$\% of triggers) were analysed, while the high magnetic field subset was used for alignment between the BPDs and TPCs (see Subsection~\ref{subsec:targetpos}).

\section{Analysis} \label{sec:Analysis}

The reconstruction and simulation procedures used for the replica target
analysis are described in earlier publications~\cite{LTpaper,LTpaper2009}.

The analysis of the data taken with the standard magnetic field configuration was performed with the so-called $\tof$-\dedx method, which makes use of the TPC measurements of the specific energy loss (\dedx) and the time-of-flight measurements~\cite{LTpaper,LTpaper2009}. The specific energy loss of a track is calculated as a truncated mean~\cite{NA49-NIM} of the charges of the clusters (points) on the track traversing the TPCs. After event and track selection, selected tracks were binned in momentum ($p$), polar angle ($\theta$), both measured in the laboratory frame, and longitudinal position of the exit point on the target surface ($z$), as required by T2K. The different hadron yields were estimated by fitting a two-dimensional \dedx - \mtof distribution for every bin. These raw yields were then corrected for all inefficiencies by applying Monte Carlo and data-based correction factors. 
Before any event or track selection is applied, it was necessary to determine the target position with respect to the BPDs and TPCs in order to refine the original surveyor measurements.

\subsection{Target position and alignment}\label{subsec:targetpos}

The target upstream face $x$ and $y$ positions with respect to the BPDs can be determined just by plotting the $T_2$ beam profile at the target upstream $z$ position. The position of the target centre was obtained by maximising the number of entries in the circle with radius equal to $1.3\:$\cm. This was only possible because $S_3$ counter, which was part of the trigger, had the same radius as the target and it was placed just upstream of the target. The target upstream $z$ position was measured by surveyors before the data-taking. Once the relative alignment of the target and BPDs was known, the alignment of the TPCs with respect to the BPDs could also be determined. To do so, the
data taken with the high magnetic field configuration were used to select events with only one high momentum track corresponding to the beam track. Those tracks were extrapolated to the surveyed z position of the target using an analytical method (see Ref.~\cite{Gorbunov:2006pe}) with error propagation that accounts for multiple scattering (see Ref.~\cite{Wolin:1992ti}). The positions of the beam tracks extrapolated forward from the BPDs and backward from the TPCs were compared and the mean value
of the differences was taken to be the misalignment of the TPCs with respect to the BPDs.

\begin{figure*}[tb]
  \begin{subfigure}[t]{0.33\textwidth}
        \includegraphics[trim= 0 0 0 0, clip, width=1\textwidth]{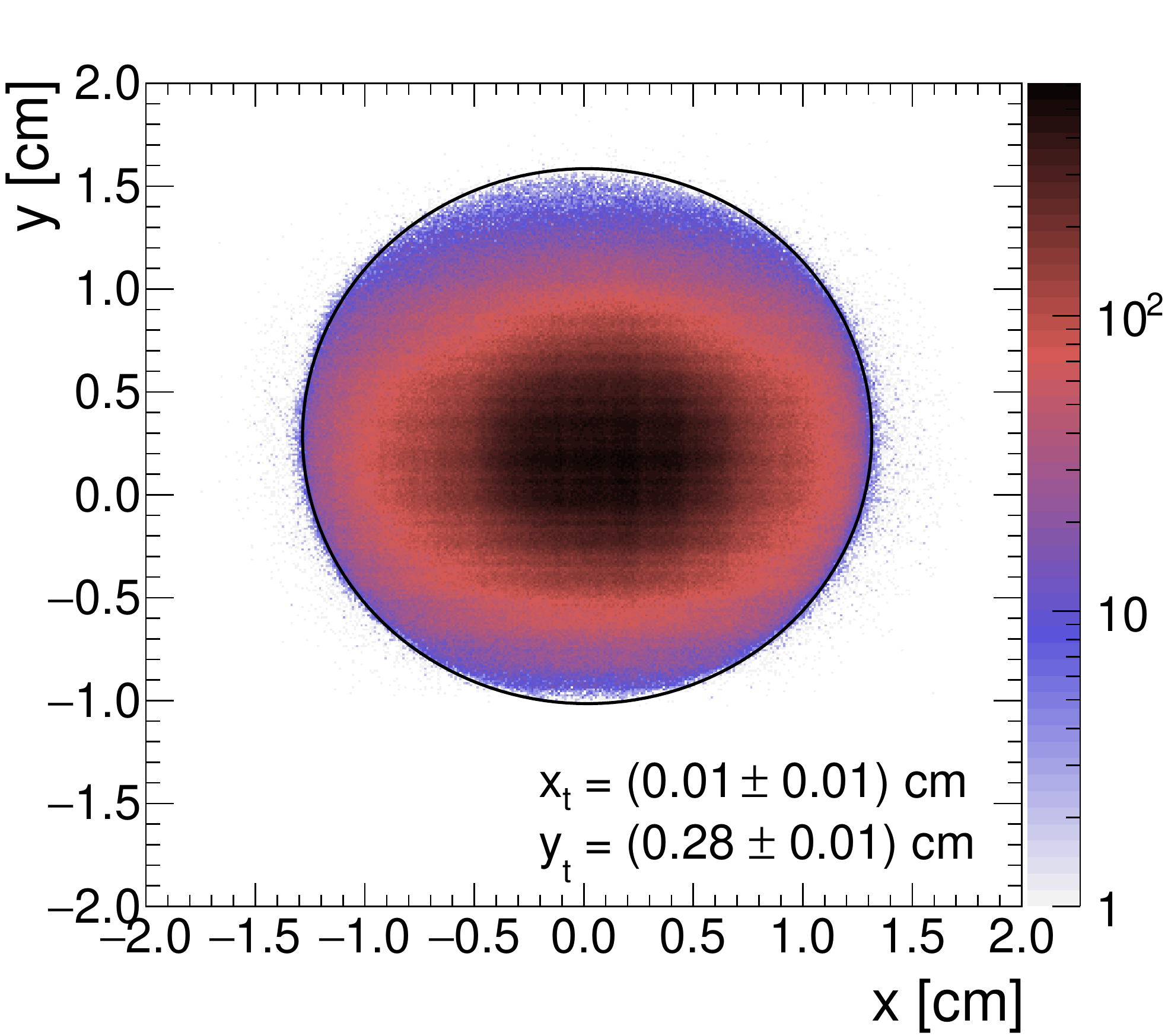}
    \caption{}
  \end{subfigure}
  \begin{subfigure}[t]{0.33\textwidth}
        \includegraphics[trim= 0 0 0 0, clip, width=1\textwidth]{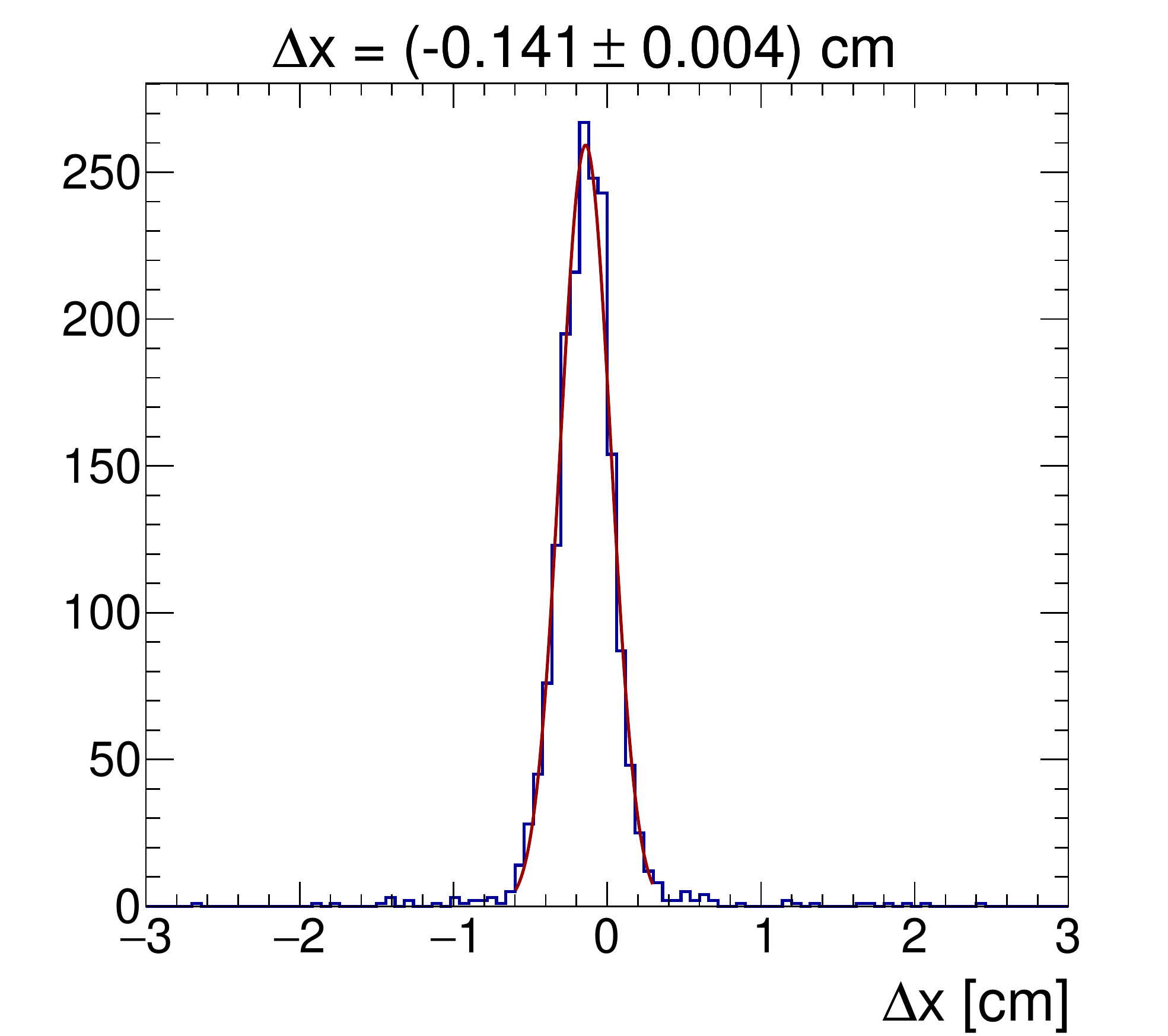}
    \caption{}
  \end{subfigure}
  \begin{subfigure}[t]{0.33\textwidth}
        \includegraphics[trim= 0 0 0 0, clip, width=1\textwidth]{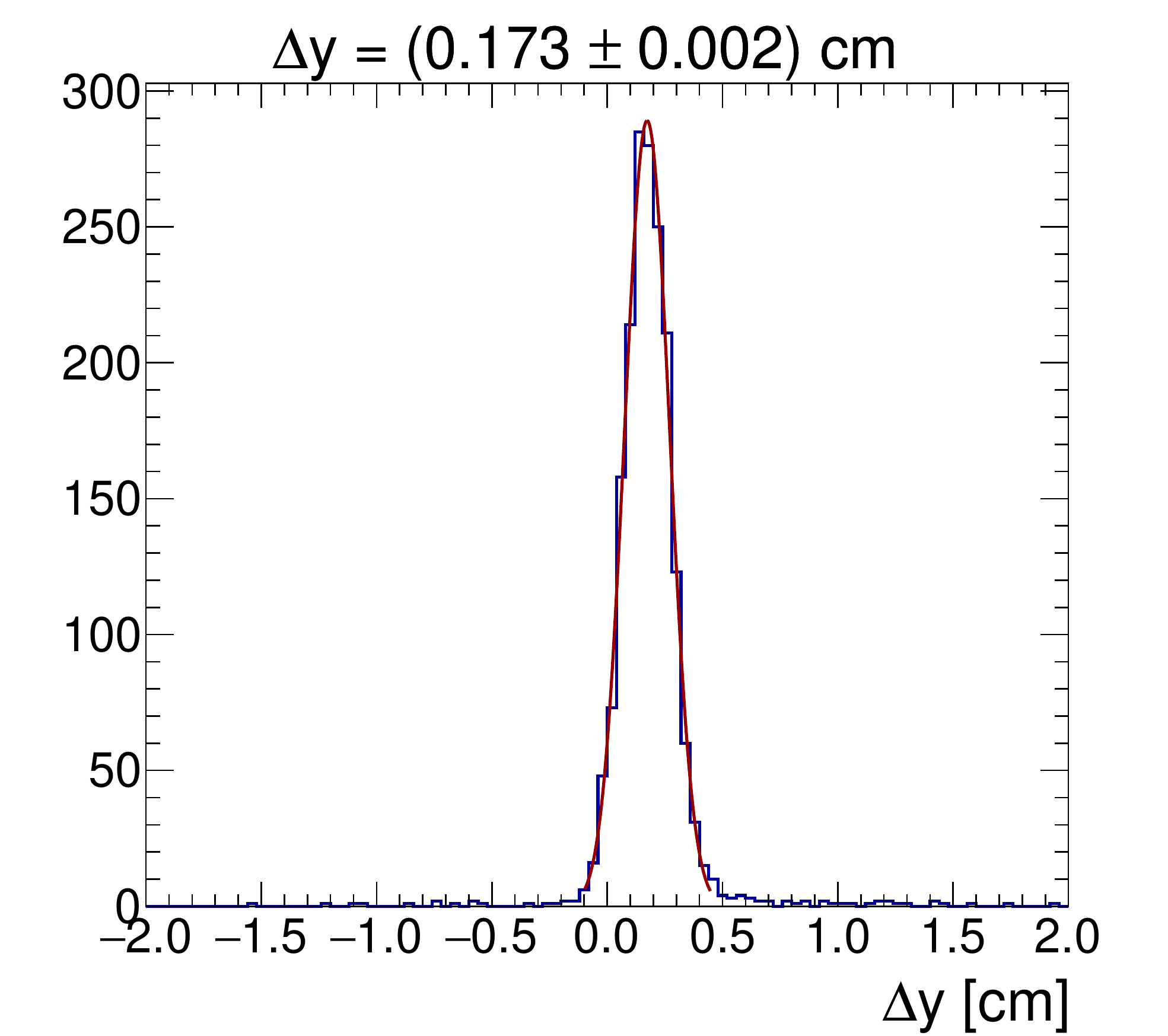}
    \caption{}
  \end{subfigure}
	\caption{Target position with respect to the BPDs: distribution of beam particles in $x-y$ at the target upsteam face (a), BPD-TPC alignment in $x$ (b) and BPD-TPC alignment in $y$ (c).}\label{fig:targetxy}
\end{figure*}

\begin{figure}[ht]
    \begin{center}
    \includegraphics[width=0.5\textwidth]{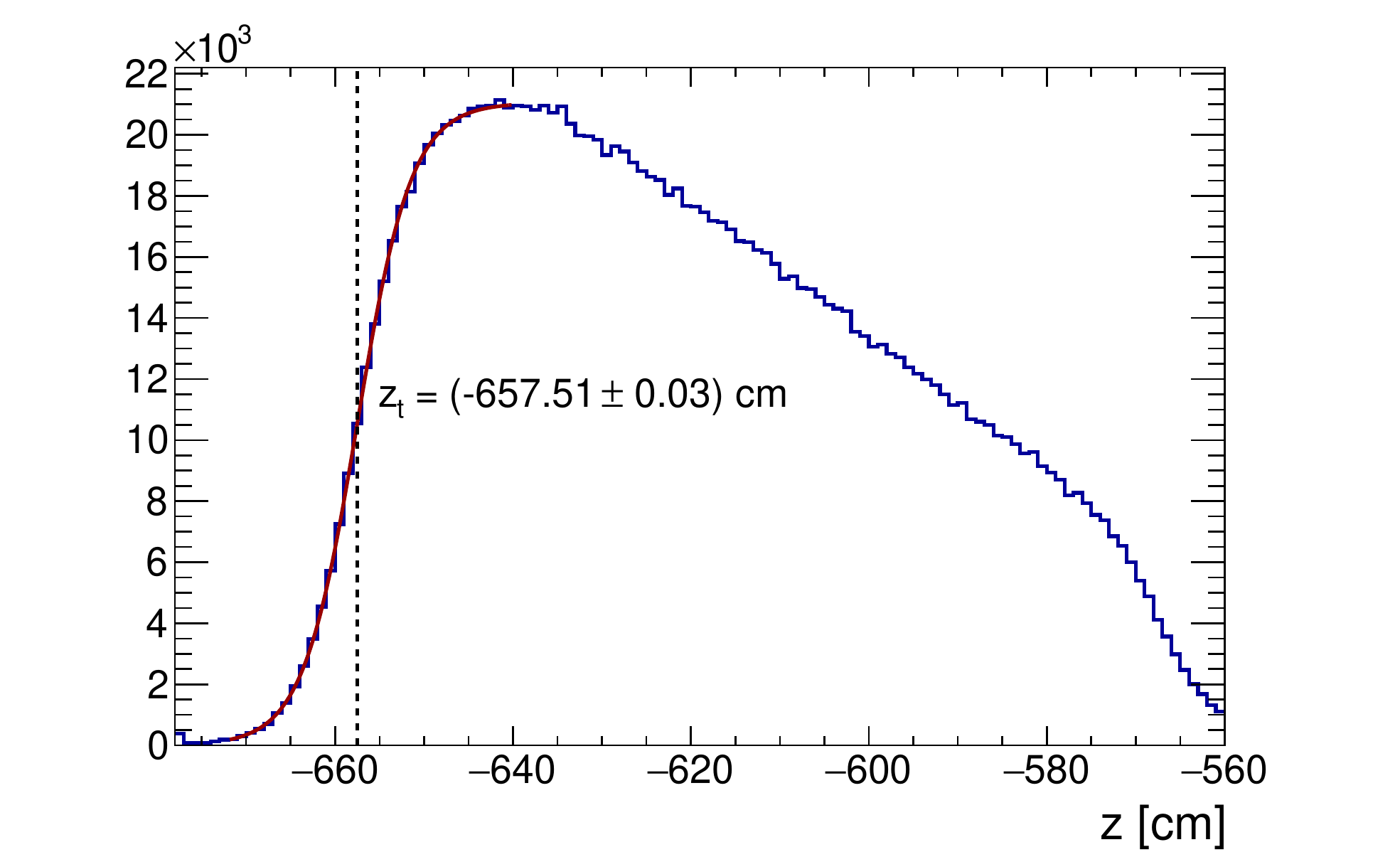}
    \caption{The distribution of $z$ positions of closest approach for TPC tracks extrapolated to the beam track. Only tracks whose distance to the beam track is within extrapolation uncertainty were selected. The red line shows fitted rising edge and the position of the half maximum was taken to be the target upstream $z$ position.}\label{fig:targetz}
    \end{center}
\end{figure}

Because of the non-negligible probability of re-interactions and long extrapolation distance, it is difficult to reconstruct interaction vertices in the target. However, by extrapolating TPC tracks and selecting only those for which the point of closest approach to the beam track is within the extrapolation uncertainty, it was possible to check the surveyors' measurement of the target $z$ position. The $z$ distribution of the points of closest approach was plotted and a rising edge could be seen as shown in Fig.~\ref{fig:targetz}. The $z$ position at the half-max of this rising edge was taken to be the upstream target face. Full agreement with the surveyor position measurement was obtained. 

The final step
was to determine the tilt of the target in $x-z$ and $y-z$ planes. First, the target was assumed to be parallel to the $z$-axis. Then, only events with beam tracks passing through the whole length of the target were selected. The TPC tracks were again extrapolated backwards until the minimum distance from the beam track was reached. By plotting the $x$ and $y$ distributions for $18$ slices in $z$ and checking the width and position of the distributions, it is straightforward to extract any possible tilt of the target. For example, if the target is tilted in positive $x$ direction, edges of the $x$ distributions would be shifted towards the centre because the target is moved from the beam. The shift would increase with $z$. No appreciable tilt was found. In Table~\ref{tab:targpos} the measured position of the target is summarized with uncertainties for all parameters.
\begin{table}[!ht]
	\begin{center}
		\begin{tabular}{cccccc} \toprule
			 & $x\:$[\cm] & $y\:$[\cm] & $z\:$[\cm] & $t_x\:$[\mrad] & $t_y\:$[\mrad]\\
			\midrule 
			 Value  & $0.15$ & $0.12$ & $-657.5$ & $0.0$ & $0.0$ \\
			 Uncertainty & $0.03$ & $0.02$ & $0.1$ & $0.3$ & $0.3$ \\ 
			\bottomrule
		\end{tabular}
	\end{center}
	\caption{Coordinates of the upsteam face of the T2K replica target (in the NA61/SHINE coordinate system, see Fig.~\ref{fig:NA61Setup}) and their uncertainty as well as the tilt of the target. 
}\label{tab:targpos}
\end{table}

\subsection{Event selection}
Three different criteria were imposed on the recorded events. Beam particles were required to hit the target, in other words, an event must correspond to $T_2$ or $T_3$ trigger. Also, the position of each beam particle must have been measured by all three BPDs. And finally, the reconstructed path of the beam particle must have passed through the whole length of the target. 

Only $75$\% of the events passed this selection. Events were mostly discarded at the second requirement due to inefficiencies in the BPD measurements. 
The event selection should also ensure that events containing off-time TPC tracks were not selected as these tracks can potentially bias the measured hadron yields. The beam intensity during the data-taking was around $8.3\:$\kHz which means that the mean time difference between beam particles was around $120\:$\mus. However, the ToF-F wall has a time acquisition window of $100\:$\ns, and because of the requirement that all tracks must have a $\tof$ hit for particle identification, most of the off-time TPC tracks were automatically discarded. The fraction of accepted off-time tracks was negligible.

\subsection{Track selection}
\begin{figure}[ht]
    \begin{center}
    \includegraphics[width=0.5\textwidth]{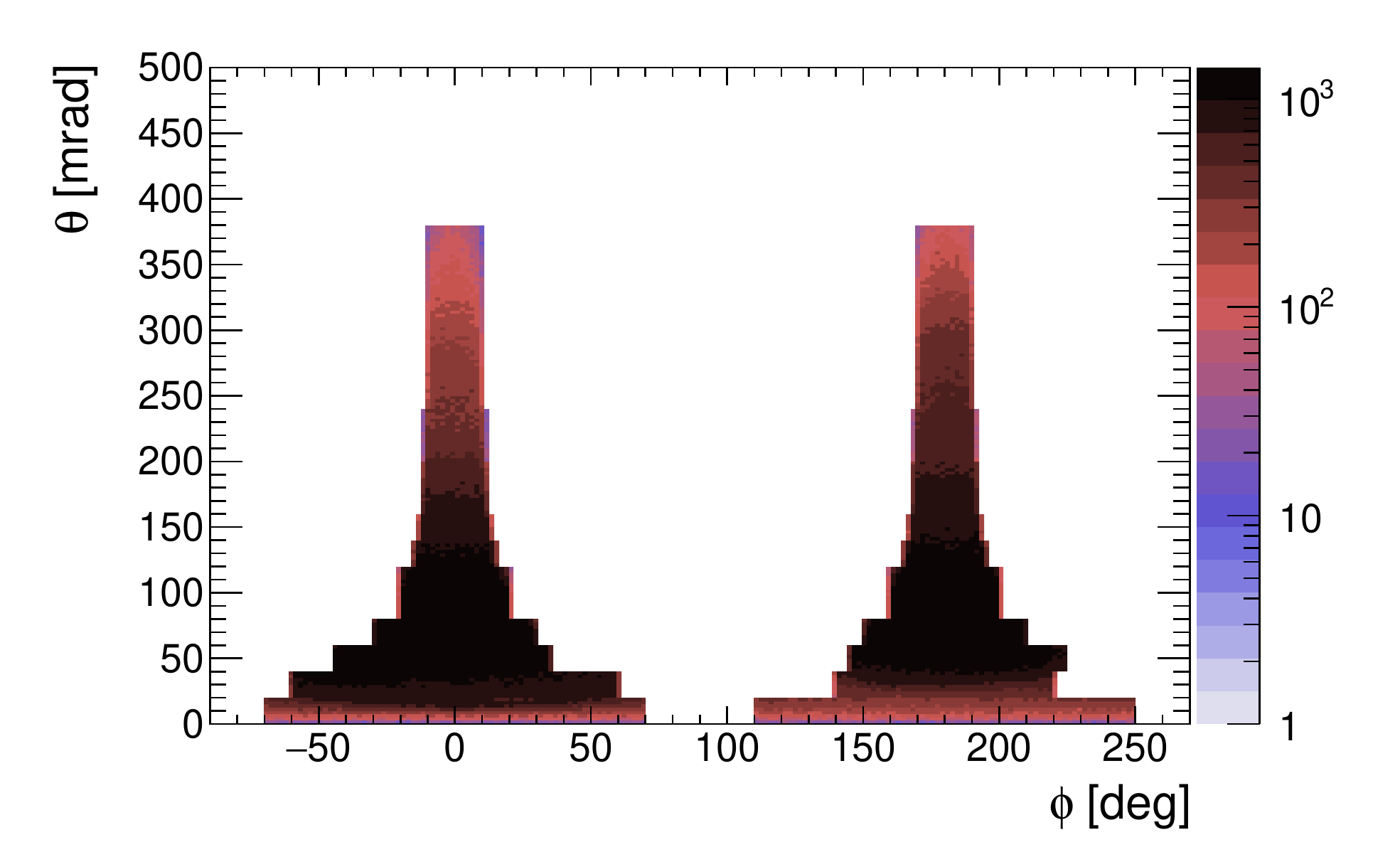}
    \caption{The $\theta - \phi$ distribution of selected tracks: regions of flat acceptance are selected. The selected regions in azimuthal angle $\phi$ depend on the polar angle $\theta$ interval.}\label{fig:phiCut}
    \end{center}
\end{figure}

Track selection was rather simple and can be divided into two parts. First, a track was required to have fitted momentum, energy loss, and time-of-flight measurements. Afterwards, cuts were applied to increase the quality of the selected track sample. Tracks can have segments in different TPCs and therefore, can be divided into different topologies. For each topology, a different requirement on the number of clusters was applied. For example, if a track passes through the VTPCs, at least $25$ points in the VTPCs were required. A cut on the number of clusters in MTPCs was not applied to such tracks. However, tracks with momentum measured only in the GTPC have between four and seven GTPC points. At least six points in the GTPC were required as well as additional $30$ points in the MTPCs. This ensured that the track parameters were measured with sufficient precision and that the track did not pass very close to the MTPC walls where possible field distortions are the largest. Afterwards, the azimuthal angle distributions were plotted for polar angle intervals of $20\:$\mrad or $40\:$\mrad. Two different types of azimuthal regions were removed: the regions with rapidly changing acceptance and the regions in which the Monte Carlo acceptance does not correspond to the data. This may be caused by slightly different sizes of the TPCs or distortions of the magnetic field which are not present in the magnetic field maps used in the Monte Carlo simulation.  
The selected regions in ($\phi, \theta$) space are presented in Fig.~\ref{fig:phiCut}.  This reduces the possibility of bias once the Monte Carlo corrections were applied to the data.

 Finally, the selected TPC tracks were extrapolated towards the target surface. The track parameters and covariance matrices were saved for the positions where tracks hit the target surface or the minimum distance from the target surface was less than $3\sigma_R$, where $\sigma_R$ is the extrapolated radial uncertainty. This cut reduced the number of selected tracks that were not coming from the target, but were created in decays or interactions outside of the target. 

\subsection{Phase space}

Selected tracks were binned in ($p, \theta, z$) space. As required by T2K~\cite{LTpaper2009}, six longitudinal bins in total were used: the 90-\cm-long graphite rod was divided into five bins of 18~\cm each and the target downstream face was considered as an additional sixth bin. Different ($p, \theta$) binning was applied for extracting $\pi^\pm$, $K^\pm$ and \prot yields. The reason for this is due to statistics: several times fewer kaons than pions are expected, so coarser binning must be used for kaons. Unequal bin sizes were used to reduce the large variability in the statistical uncertainty for low and high momenta. The appropriate size of the bins was estimated from \FlukaEleven.$2$c.$5$ simulations~\cite{Fluka,Fluka_CERN,Fluka_new}. Also, the starting value for the momentum binning was carefully adjusted. The necessity for this comes from the fact that the ToF-F response was not simulated in the Monte Carlo. Instead, in the reconstruction chain applied to simulated events the $\tof$ measurements were just assigned to the tracks that hit the ToF-F wall, and later on the inefficiency of the ToF-F wall 
was corrected with a data-based correction. Actually, very low momentum particles, depending on their mass, cannot reach the ToF-F wall within the ToF-F acquisition window. Therefore, Monte Carlo corrections in this region would be heavily biased. For example, momentum bins for protons must start at $0.5\:$\GeVc.  

\begin{figure*}[tb]
    \includegraphics[trim= 0 0 0 0, clip, width=1\textwidth]{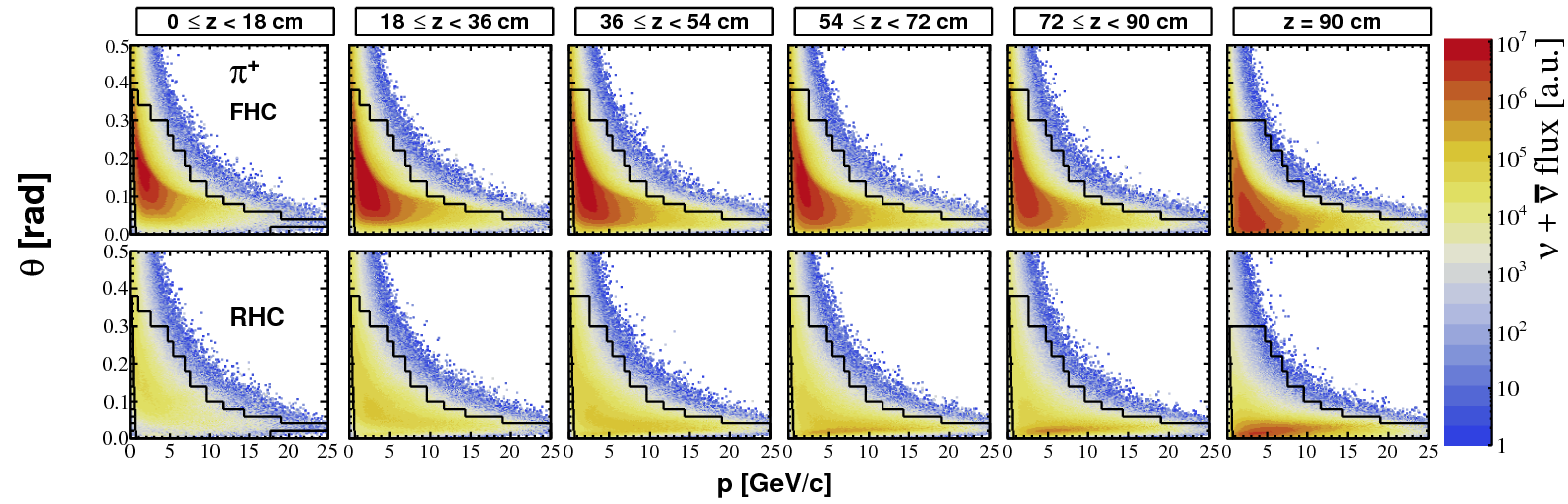}
    \caption{The $p-\theta$ distribution of positively charged pions emitted from the surface of the T2K target which contribute to the (anti)neutrino flux in \sk. The top row shows the phase space plots in the forward horn current configuration, while bottom plots show the phase space for the reverse horn current configuration. The \NASixtyOne coverage from the 2010 replica target measurement is overlaid on top of the plots.}\label{fig:covpip}
\end{figure*}

\begin{figure*}[tb]
    \includegraphics[trim= 0 0 0 0, clip, width=1\textwidth]{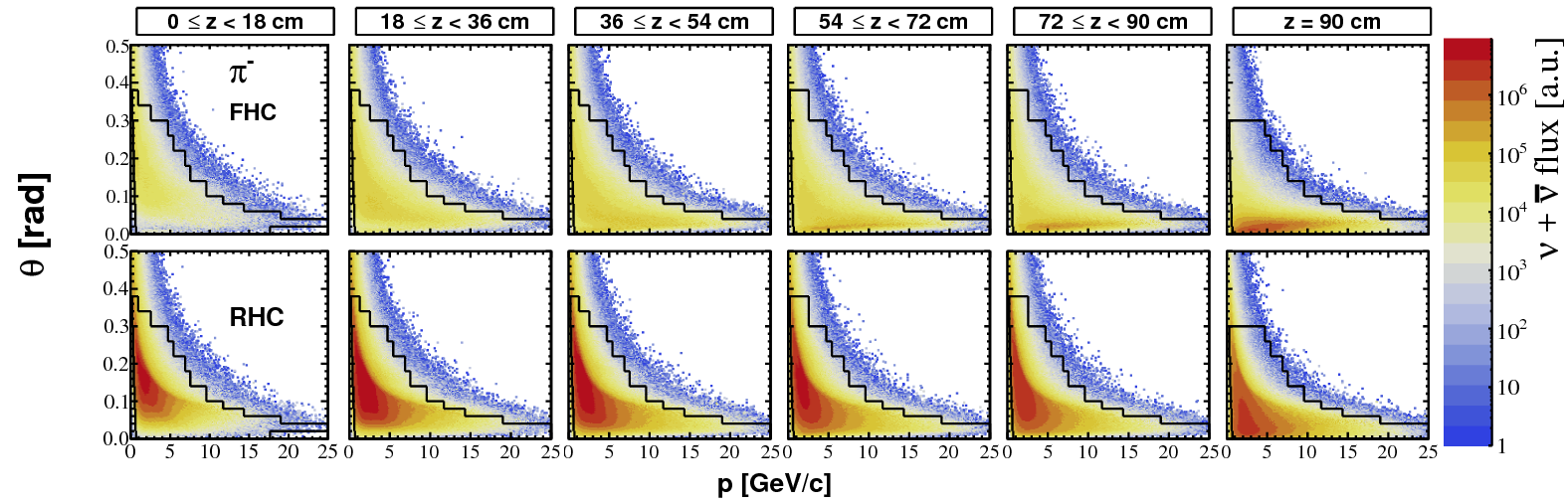}
    \caption{The $p-\theta$ distribution of negatively charged pions emitted from the surface of the T2K target which contribute to the (anti)neutrino flux in \sk. The top row shows the phase space plots in the forward horn current configuration, while bottom plots show the phase space for the reverse horn current configuration. The \NASixtyOne coverage from the 2010 replica target measurement is overlaid on top of the plots.}\label{fig:covpim}
\end{figure*}

\begin{figure*}[tb]
    \includegraphics[trim= 0 0 0 0, clip, width=1\textwidth]{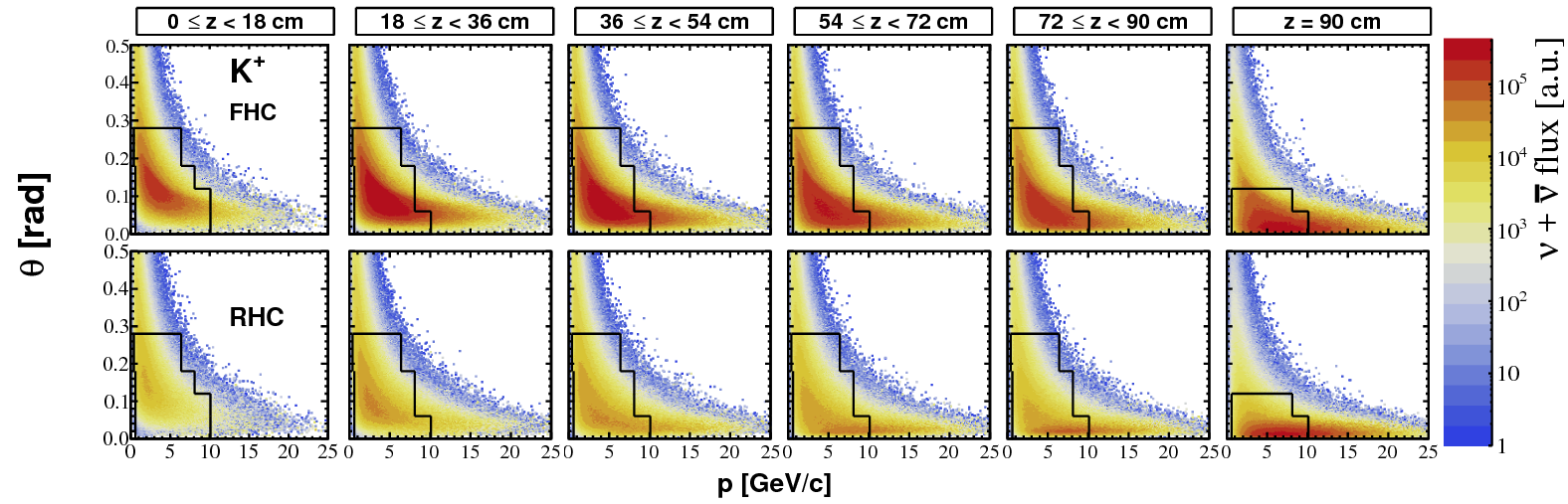}
    \caption{The $p-\theta$ distribution of positively charged kaons emitted from the surface of the T2K target which contribute to the (anti)neutrino flux in \sk. The top row shows the phase space plots in the forward horn current configuration, while bottom plots show the phase space for the reverse horn current configuration. The \NASixtyOne coverage from the 2010 replica target measurement is overlaid on top of the plots.}\label{fig:covkp}
\end{figure*}

\begin{figure*}[tb]
    \includegraphics[trim= 0 0 0 0, clip, width=1\textwidth]{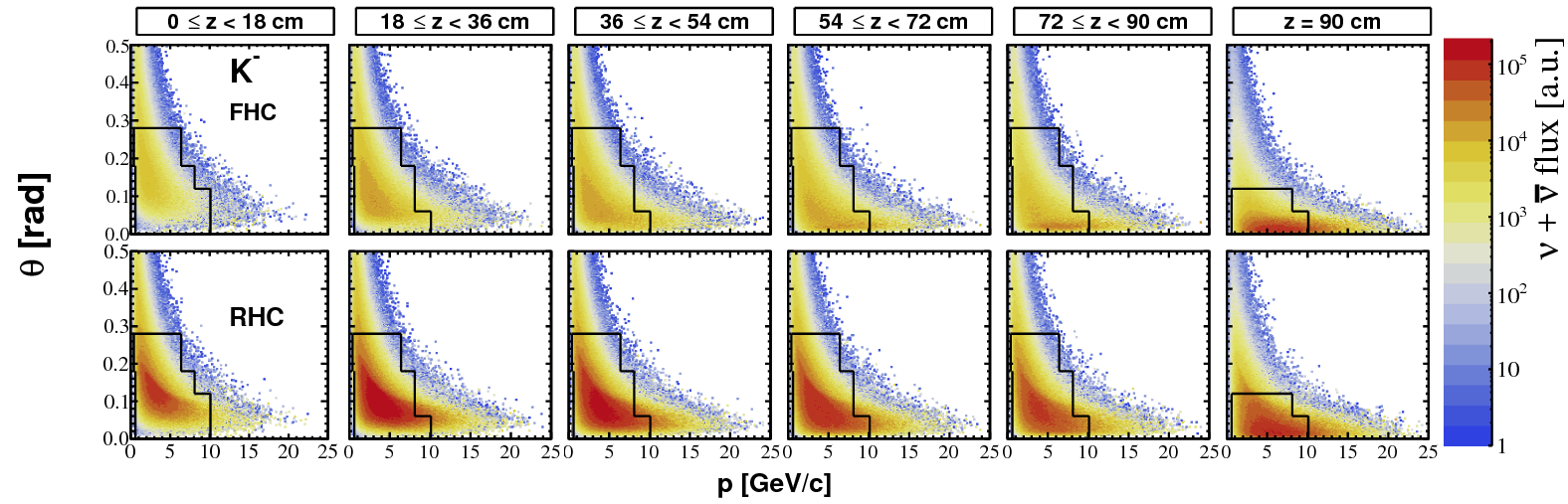}
    \caption{The $p-\theta$ distribution of negatively charged kaons emitted from the surface of the T2K target which contribute to the (anti)neutrino flux in \sk. The top row shows the phase space plots in the forward horn current configuration, while bottom plots show the phase space for the reverse horn current configuration. The \NASixtyOne coverage from the 2010 replica target measurement is overlaid on top of the plots.}\label{fig:covkm}
\end{figure*}

\begin{figure*}[tb]
    \includegraphics[trim= 0 0 0 0, clip, width=1\textwidth]{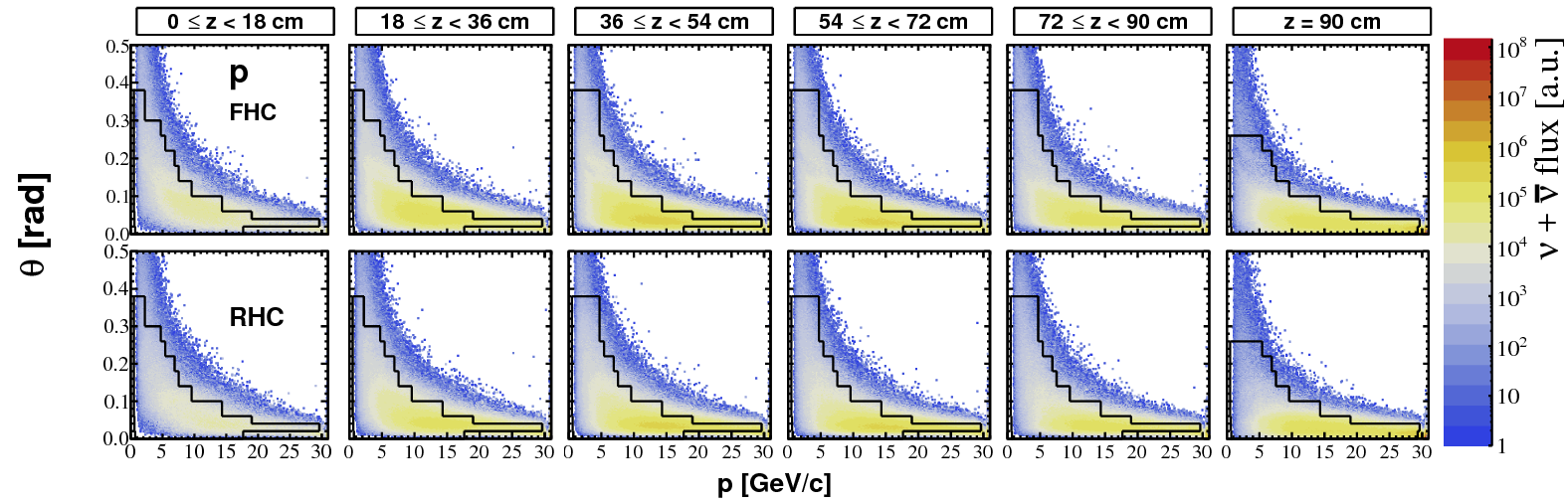}
    \caption{The $p-\theta$ distribution of protons emitted from the surface of the T2K target which contribute to the (anti)neutrino flux in \sk. The top row shows the phase space plots in the forward horn current configuration, while bottom plots show the phase space for the reverse horn current configuration. The \NASixtyOne coverage from the 2010 replica target measurement is overlaid on top of the plots.}\label{fig:covp}
\end{figure*}

\begin{table}[!ht]
	\begin{center}
		\begin{tabular}{ccccccc} \toprule
			 & \pip [\%] & \pim [\%] & \kp [\%] & \km [\%] & \prot [\%] & Total [\%]\\
			\midrule 
			 FHC  & $99.22$ & $97.47$ & $84.50$ & $83.08$ & $71.65$ & $96.92$ \\
			 RHC & $97.03$ & $98.89$ & $72.56$ & $89.61$ & $69.66$ & $96.62$ \\ 
			\bottomrule
		\end{tabular}
	\end{center}
	\caption{Fractions of (anti)neutrino fluxes at \sk produced by \pip, \pim, \kp, \km, and \prot emitted from the surface of the T2K target (estimated using \FlukaEleven.$2$c.$5$ simulations) and covered by the measurements presented in this paper. The fractions are presented for the forward and reverse horn current configurations. The last column is a sum of the coverage for five hadrons presented in this paper and does not include other particles contributing to the neutrino flux (\kz, \lm, $\mu$).}\label{tab:cov}
\end{table}

The overall phase space binning covers more than $96\%$ of (anti)\-neutrinos crossing the T2K far detector (\sk) which are produced by the charged hadrons emitted from the target. More than $97\%$ of (anti)neutrinos produced by pions are covered, while the coverage drops for kaons and protons. This is summarised in Table~\ref{tab:cov}. 
The phase space of the charged hadrons coming from the T2K target surface that contribute to the (anti)neutrino fluxes in \sk are plotted in Figs.~\ref{fig:covpip} -~\ref{fig:covp}. The top row in each figure represents the coverage for the forward horn current (FHC) corresponding to the neutrino mode, while the bottom row represents the coverage for the reverse horn current (RHC) corresponding to the anti-neutrino mode. Each column represents a different longitudinal bin. The phase space coverage of the measurements presented in this paper is overlaid on top of the figures as a black line. Since \NASixtyOne added new forward TPCs in 2017~\cite{Status_Report_2017}, possible future measurements could improve coverage for the forward-going high momentum $K^\pm$ and \prot~\cite{Addendum2020+}.

\subsection{Particle identification}
\begin{figure*}[ht]
    \begin{center}
    \includegraphics[width=1\textwidth]{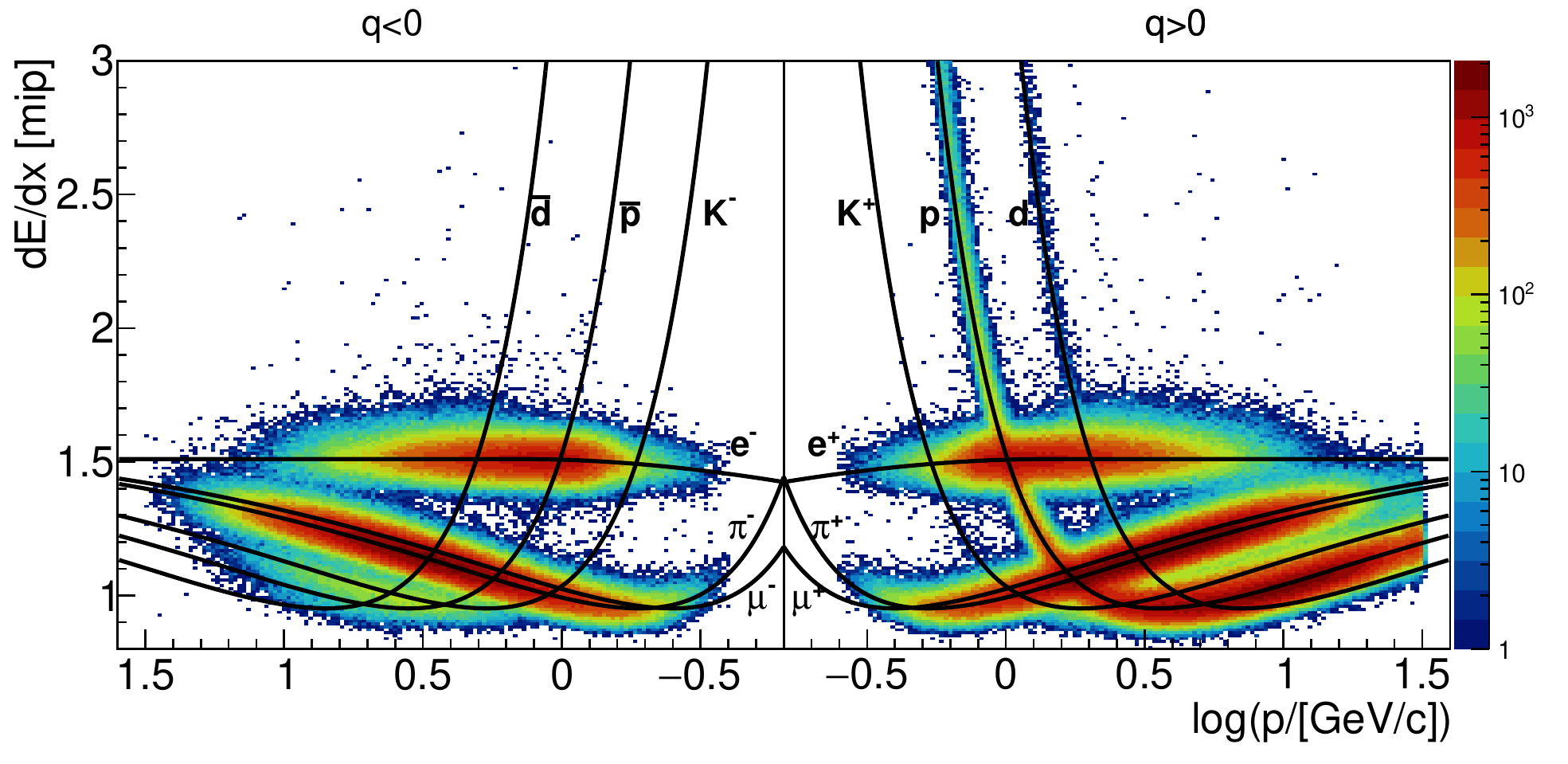}
    \caption{Distribution of specific energy loss in the TPCs as a function of particle momentum for negatively (left panel) and positively (right panel) charged particles.}\label{fig:dEdx}
    \end{center}
\end{figure*}

\begin{figure}[ht]
    \begin{center}
    \includegraphics[width=0.5\textwidth]{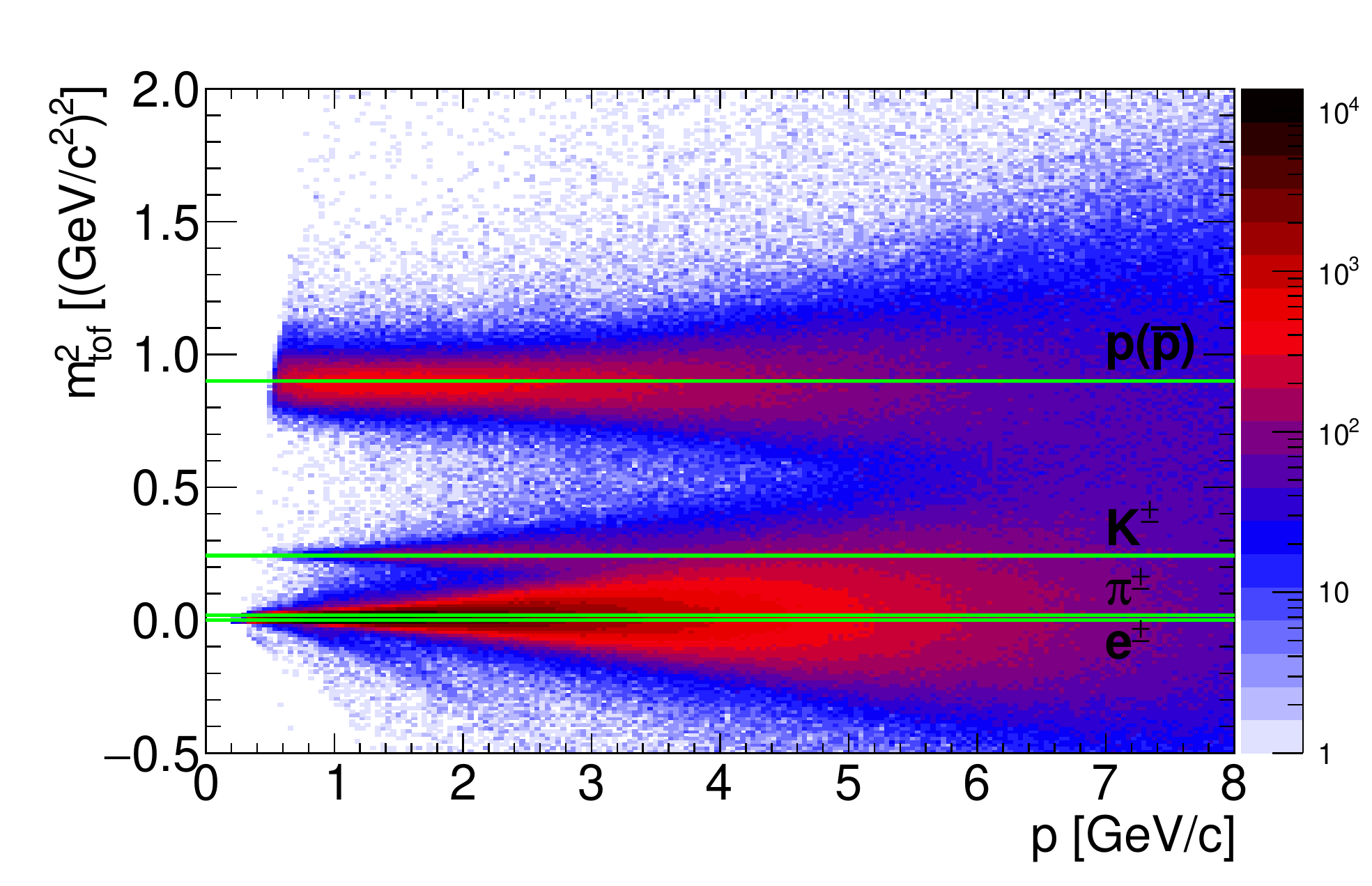}
    \caption{A mass squared distribution of selected tracks, calculated from the $\tof$ measurements, as a function of particle momentum. Known squared masses of $\pi^\pm$, $K^\pm$, $e^\pm$ and \prot($\overline{\prot}$) are overlaid on top. }\label{fig:m2tof}
    \end{center}
\end{figure}

\begin{figure*}[ht]
    \begin{center}
    \includegraphics[width=1\textwidth]{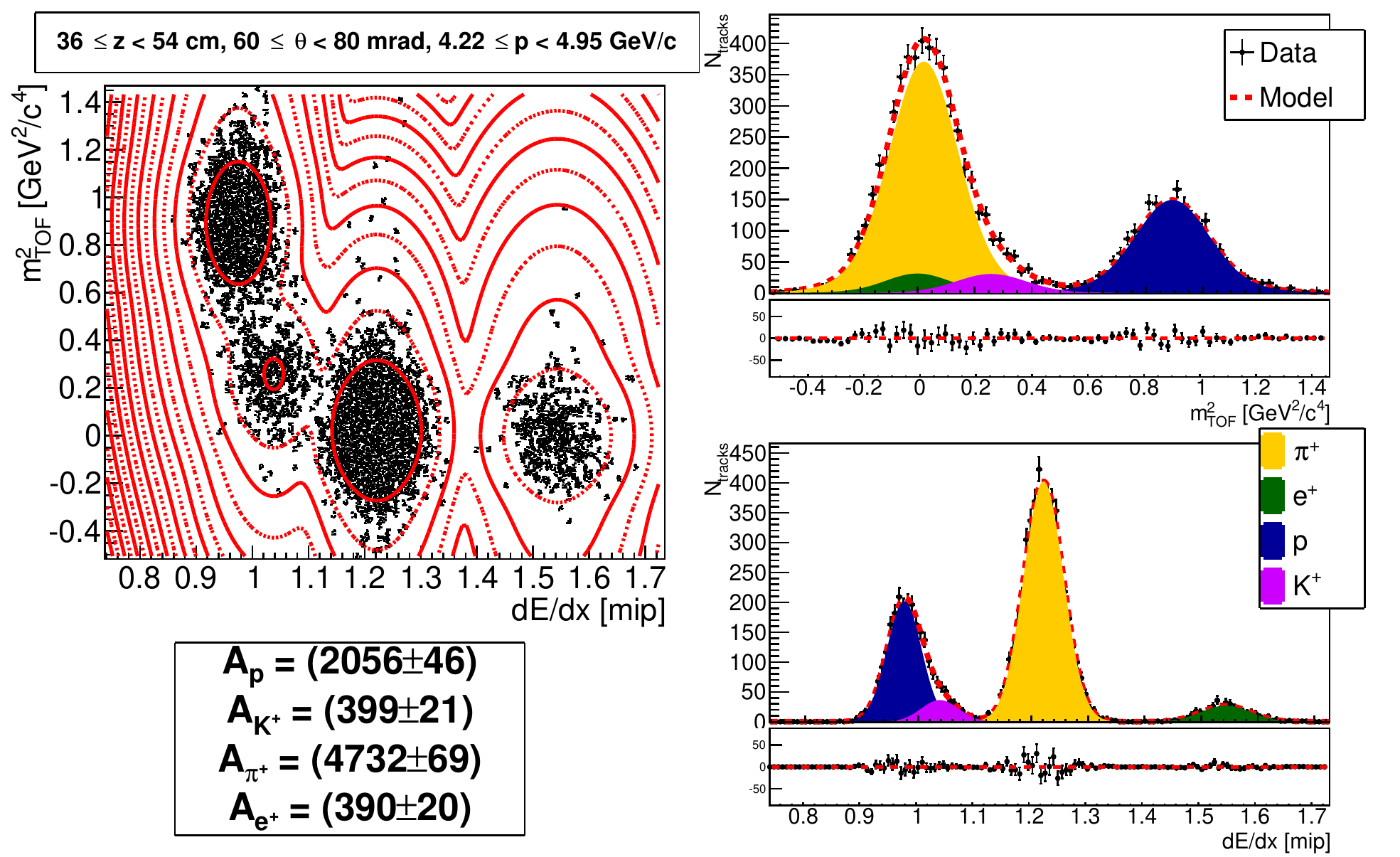}
    \caption{An example of the joint \mtof - \dedx fit for a single ($p, \theta, z$) bin. The top left panel shows the 2D distribution of the data with a contour plot of the fitted function in red. The bottom left panel shows the extracted number of \pip, \kp, \prot and \ep. The top right and bottom right panels show projections of the data and fitted function to \mtof and \dedx axis, respectively.}\label{fig:fitpos}
    \end{center}
\end{figure*}

Since large fractions of the phase space are covered, a robust particle identification method is needed. For the momentum range between $1\:$\GeVc to $3\:$\GeVc (see Fig.~\ref{fig:dEdx}), the energy loss distributions cross, hence particle identification based on the energy loss alone is not possible. However, in these regions, $\tof$ measurements can be used to distinguish between particles. Kaons can be separated easily up to $3\:$\GeVc, while protons can be separated up to $8\:$\GeVc, as can be seen in Fig.~\ref{fig:m2tof}. For higher momentum, the $\tof$ resolution becomes too poor, while energy loss measurements allow for better identification. It is clear that both approaches are complementary and therefore they can be combined to cover all the bins. The $\tof$ measurement was used to calculate the particle mass squared (\mtof) and it was combined with the energy loss. Particles were represented by islands in the \mtof-\dedx space. Therefore, raw hadron yields could be obtained by fitting an appropriate function to the distribution for each phase space bin. Both, \mtof and \dedx were assumed to be normally distributed. This assumption for \dedx must be closely examined. In general, \dedx is not normally distributed, as its distribution is similar to the Landau distribution. The measured \dedx in NA61/SHINE is calculated as a truncated mean (see Ref.~\cite{NA49-NIM}) of all energy depositions for all clusters. Therefore, we assume that truncated mean is normally distributed. This is only valid if all tracks have the same number of TPC clusters. The selected tracks can have a different number of clusters which can significantly affect the \dedx resolution if the number of TPC clusters is small. However, the \dedx resolution in \NASixtyOne saturates at around $3.5$\% for tracks with more than $70$ clusters. Around $98.5$\% of the selected tracks have more than $70$ clusters. This is a good justification for using only a single Gaussian for describing energy loss in a single phase space bin for one particle species. Our previous measurements prove this assumption (see Refs.~\cite{thin2009paper, LTpaper2009}). The total fitting function was constructed from four two-dimensional Gaussians, one for each particle species ($e^\pm$, $\pi^\pm$, $K^\pm$, \prot($\overline{\prot}$)). The fitting was done in the RooFit framework~\cite{Verkerke:2003ir} by using extended log-likelihood minimization which treats the number of observed particles as a Poisson random variable. An example of a fit is shown in Fig.~\ref{fig:fitpos}. In total, there were $20$ parameters in the fit: eight mean values, eight standard deviations, and four particle multiplicities. The obtained raw multiplicities must be corrected for various inefficiencies with Monte Carlo and data-based corrections.

\subsection{Monte Carlo correction factors}
\begin{figure*}[ht]
	\begin{center}
  \begin{subfigure}[t]{0.48\textwidth}
        \includegraphics[trim= 0 0 0 0, clip, width=1\textwidth]{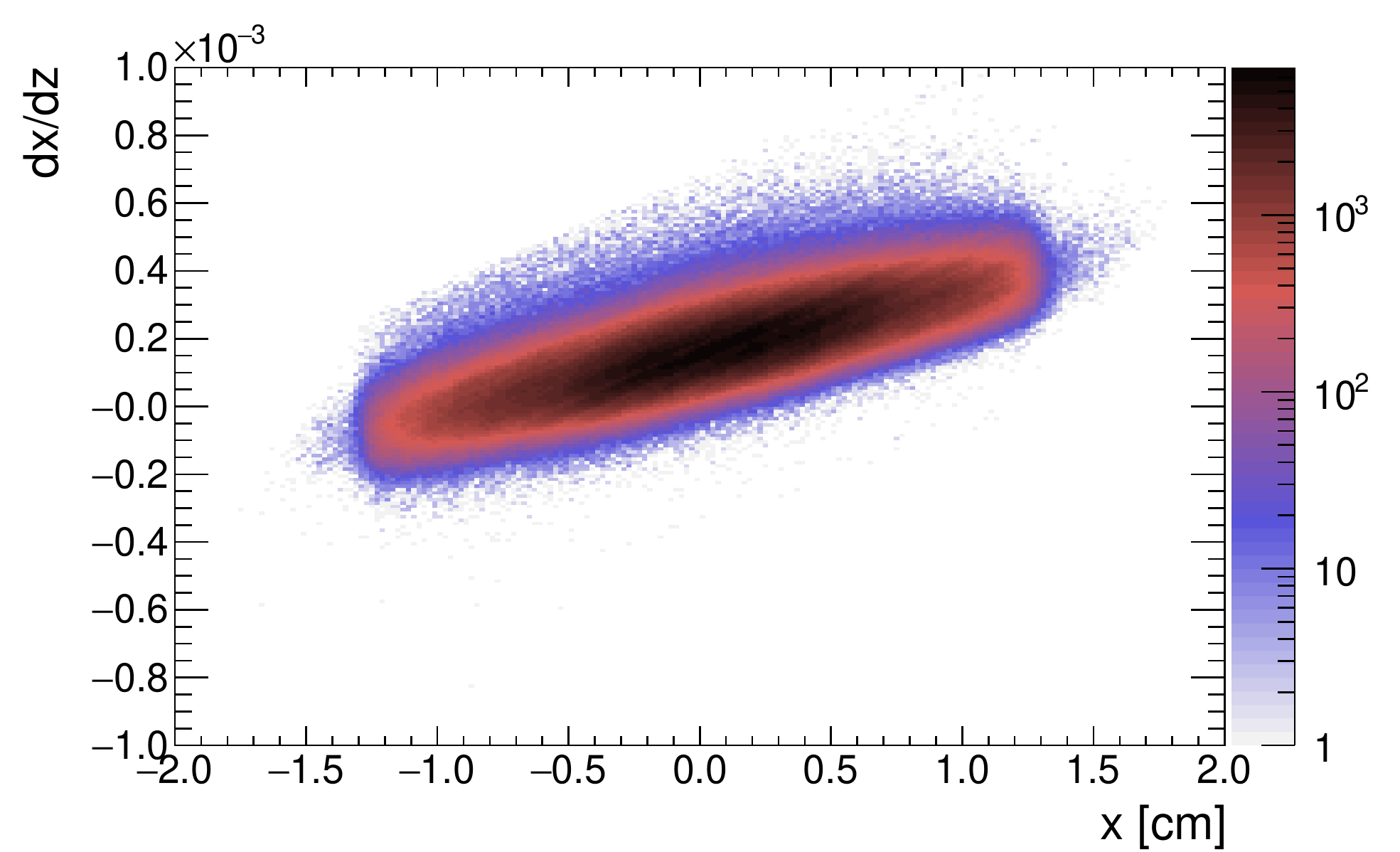}
    \caption{}
  \end{subfigure}	
  \begin{subfigure}[t]{0.48\textwidth}
        \includegraphics[trim= 0 0 0 0, clip, width=1\textwidth]{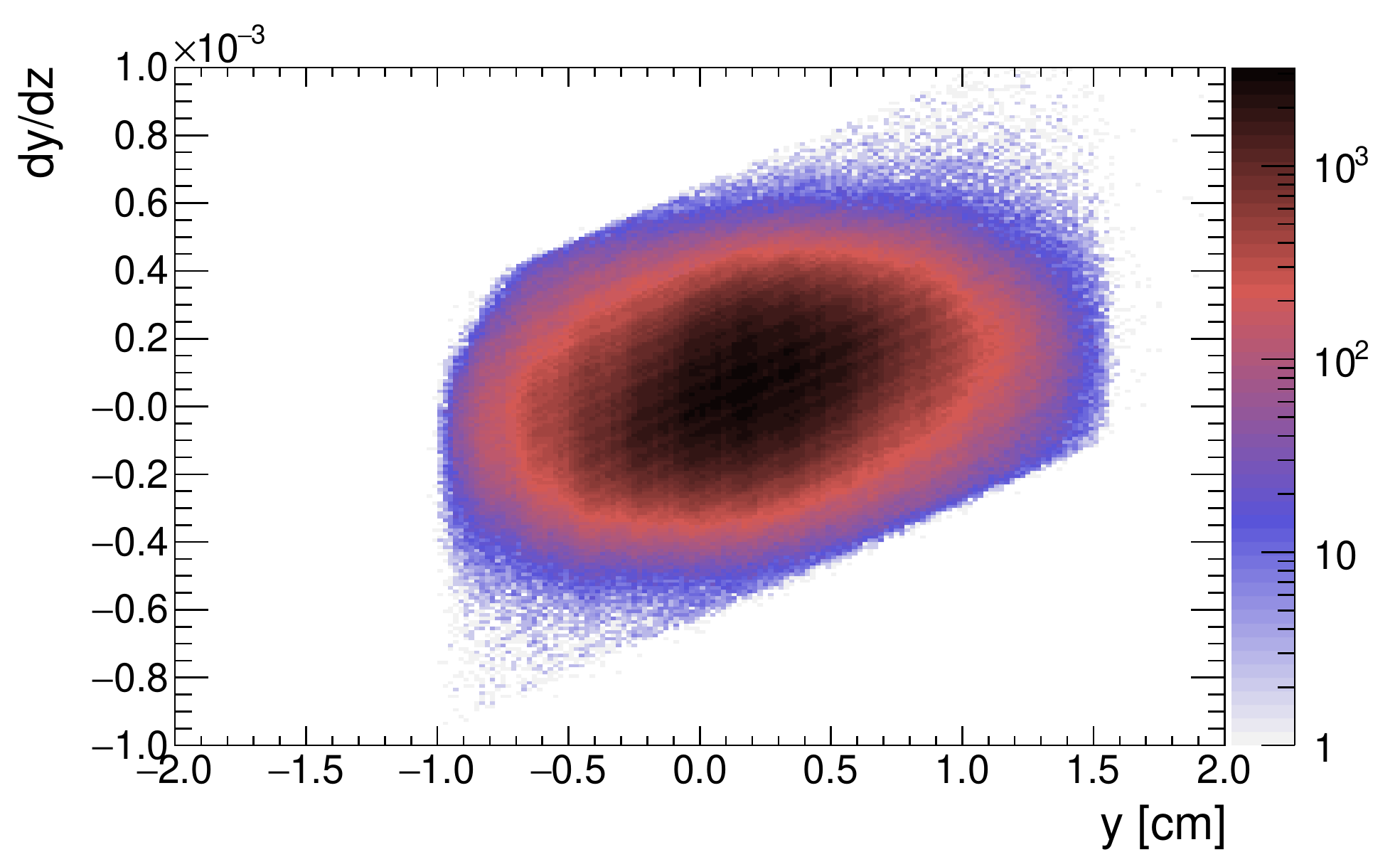}
    \caption{}
  \end{subfigure}	

  \end{center}
	\caption{Beam slope $\text{d}x/\text{d}z$ vs. $x$ position (a) and $\textrm{d}y/\text{d}z$ vs. $y$ (b) for beam particles selected by the $T_2$ trigger. The distributions are used to randomly generate beam particles in the \NASixtyOne Monte Carlo simulation.}\label{fig:beamdiv}
\end{figure*}

Interactions in the target were simulated by \FlukaEleven.$2$c.$5$~\cite{Fluka, Fluka_CERN, Fluka_new}. The kinematics of the particles at their exit position on the surface of the target was then fed to \GeantThree$/$\GCALOR~\cite{Brun:1987ma} for propagation through the \NASixtyOne detector.
The incoming proton beam information from the data was used to generate the Monte Carlo beam profile. The beam track parameters for each simulated proton were thrown randomly according to the beam divergence and position distributions shown in Fig.~\ref{fig:beamdiv}. The $x$- and $y$- positions and divergences were considered to be independent.
In total, nearly $40 \times 10^6$ protons on target (POT) were simulated.

A simple Monte Carlo correction was applied to the raw yields:
\begin{equation}
	C_{ijk} = \frac{N_\text{sim}^{ijk}}{N_\text{rec,sim}^{ijk}},
\end{equation} 
where $N_\text{sim}^{ijk}$ is the number of simulated hadrons in the bin ($p,\theta, z$)=\\($i,j,k$) and $N_\text{rec,sim}^{ijk}$ is the number of simulated hadrons reconstructed in the same bin. This correction can be separated into several factors, which allow for a better determination of the associated systematic errors:
\begin{equation}
\frac{1}{C_{ijk}} = \epsilon_{ijk}^\text{feed-down}\cdot \epsilon_{ijk}^\text{mig}\cdot \epsilon^\text{sel}_{ijk} \cdot \epsilon^\text{rec}_{ijk} \cdot \epsilon^\text{loss}_{ijk}\cdot \epsilon^{\phi}_{ijk},
\end{equation}
where $\epsilon^{\phi}_{ijk}$ is the geometrical efficiency (because of the $\phi$ cut), $\epsilon^\text{loss}_{ijk}$ is the hadron loss efficiency due to decays and interactions in the detector (hadrons which are produced in re-interactions and by chance extrapolated to the target surface are included in this correction), $\epsilon^\text{rec}_{ijk}$ is the reconstruction efficiency, $\epsilon^\text{sel}_{ijk}$ is the selection efficiency, $\epsilon_{ijk}^\text{mig}$ is the bin migration efficiency and $\epsilon_{ijk}^\text{feed-down}$ is the feed-down efficiency. The geometrical efficiency gives the percentage of $4\pi$ solid angle covered by the \NASixtyOne detector after the selection. Some of the produced hadrons inside the detector coverage can decay or re-interact in the detector and therefore, they do not hit the TOF-F wall. The percentage of surviving hadrons is given by the hadron loss efficiency. The reconstruction efficiency is a convolution of the efficiency of software algorithms used for the reconstruction and the efficiency of the detector. The selection efficiency corrects for the hadrons lost in the quality selection. The values of momentum ($p$), polar angle ($\theta$) and longitudinal position along the target surface ($z$) for each hadron are reconstructed with finite precision. Consequently, some of the hadrons will be placed in a wrong $(p, \theta, z$) bin. The ratio of the number of hadrons which are reconstructed in a given bin to the number of hadrons which are simulated in the same bin represents the migration efficiency. The feed-down efficiency corrects for weak decays outside of the target whose decay products are extrapolated to the target surface.

Although only the total Monte Carlo correction was applied to raw yields, it was useful to study each contribution separately in order to estimate possible biases. The systematic uncertainties for each correction are presented in the next section. As previously described, the efficiency of the ToF-F detector was not included in the Monte Carlo correction.
  
\subsection{ToF-F efficiency correction factors}
\begin{figure}[ht]
    \begin{center}
    \includegraphics[width=0.5\textwidth]{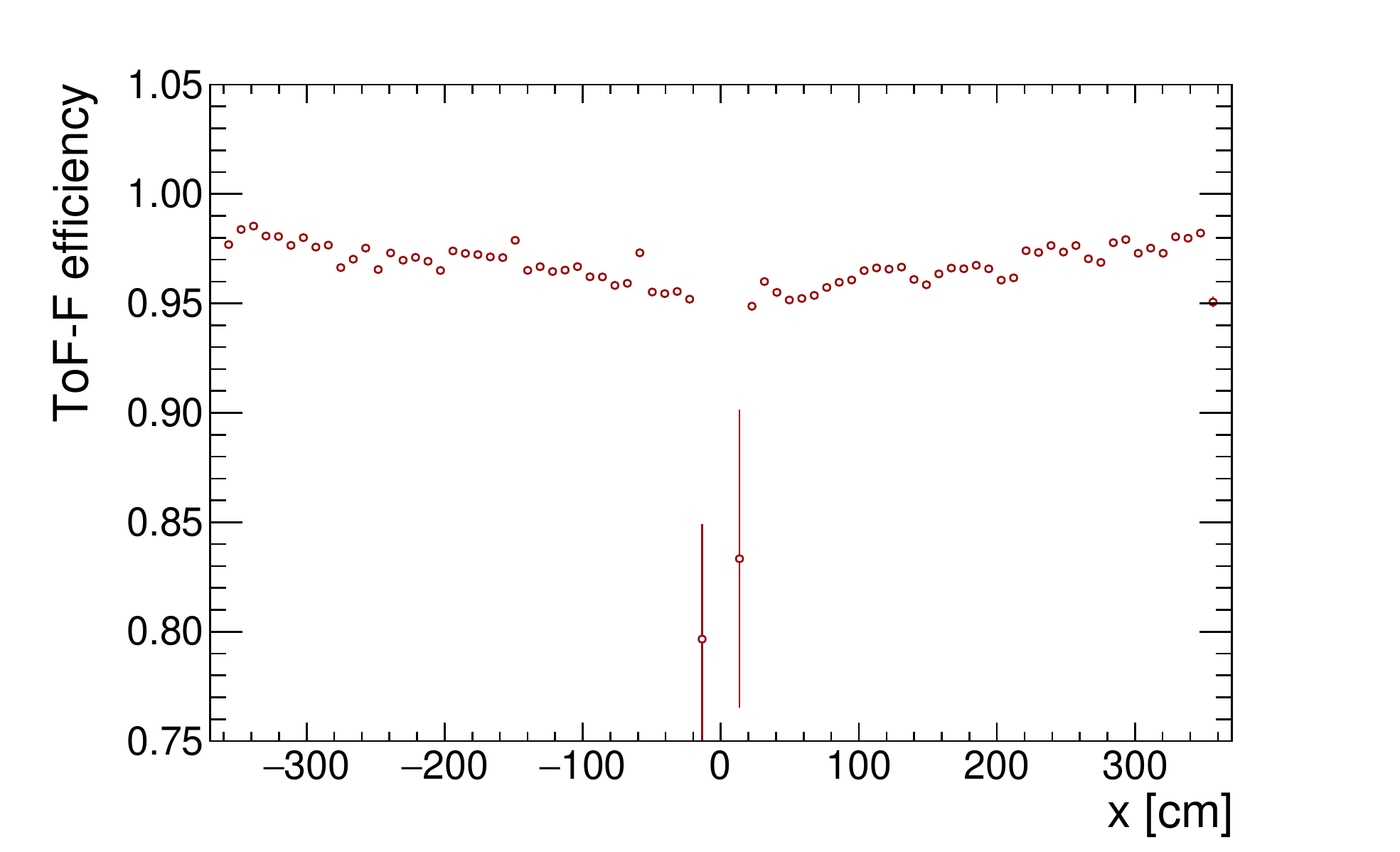}
    \caption{Efficiency of the scintillator bars in the ToF-F wall as a function of the $x$-coordinate. It is clear that the  efficiency drops if the bar is closer to the beamline ($x=0\:$\cm). This is due to an increasing track density with decreasing distance to the beam. }\label{fig:tofEff}
    \end{center}
\end{figure}

A different track selection was used to estimate the ToF-F efficiency from the data. Only tracks that reached the end of the MTPCs were selected. Since the ToF-F wall was placed just downstream of the MTPCs, it was necessary to make sure that tracks did not re-interact or decay before hitting the ToF-F wall. The efficiency of each scintillator bar was simply calculated as the number of tracks with hits in the ToF-F wall divided by the total number of selected tracks. Inefficiency was caused mainly by two effects. First, if two tracks hit a scintillator bar in the same event, the hit was discarded during reconstruction since it was not possible to distinguish between these tracks (close to the beamline there is a 4\% probability another track will be matched to the same \mbox{ToF-F} hit, while this probability is below 1\% further away from the beamline). Second, a hit is also discarded if the difference in time of flight measured by the top and bottom PMTs in a bar was larger than $2000\:$\ps. During the data-taking period, four scintillator bars had only one PMT working or there was a DAQ problem for these channels. These scintillators did not suffer from the inefficiency caused by the quality cut during reconstruction. However, these bars had a worse time resolution by a factor of $\sqrt{2}$. The efficiency of each bar is shown in Fig.~\ref{fig:tofEff}. Efficiency uncertainties were calculated as binomial errors. This information was then used to estimate the ToF-F efficiency for each phase space bin. Tracks in a single bin usually contain hits in several scintillators (usually around three or four). The error on the efficiency for each bin was taken as a ToF-F systematic uncertainty.   

\subsection{Ad hoc correction factors}
A 50\% drop in the number of ToF-F hits in the data was observed for a couple of upstream longitudinal bins and high polar angle ($\theta \approx 300\:$\mrad). However, such a drop was not seen in the MC simulation and could not be related to the ToF-F inefficiency. Under further inspection, it was discovered that 
the corresponding tracks actually did not reach the MTPCs but were bent out or absorbed while passing very close to the edge of the magnetic field map used in the simulation.
It is thus possible that the field map in this region is biased. It appears to be biased asymmetrically as negatively-charged tracks were more affected than positively-charged ones, pointing out
to asymmetric $E \times B$ corrections in the TPCs.
Nevertheless, the affected bins represent only around $1\%$ of the total $\pi^\pm$ bins and other hadrons were not affected, since they were not measured in the affected area because of the low statistics. To account for this inefficiency, an additional ad hoc correction factor was applied:
\begin{equation}
	C^{adhoc}_{ijk} = \left(\frac{n^{MC}_{sel, \tof}}{n^{MC}_{sel}}\right)_{ijk}/ \left(\frac{C_{ijk}^{\tof} n^{data}_{sel, \tof}}{n^{data}_{sel}}\right)_{ijk},
\end{equation}
where $n^{MC(data)}_{sel, \tof}$ is the number of selected tracks with $\tof$ hit in the Monte Carlo (data), $n^{MC(data)}_{sel}$ is the number of selected tracks in the Monte Carlo(data) without the $\tof$ requirement, and $C^{\tof}_{ijk}$ is the ToF-F efficiency for the given phase space bin. It is clear that this correction can be model-dependent and therefore this point must be addressed for a possible systematic bias.

\section{Systematic Uncertainties} \label{sec:Syst}

The systematic uncertainties of this analysis were carefully studied.
A brief summary is given below, while more detailed information
can be found in Ref.~\cite{PavinThesis}.

\begin{figure*}[ht]
    \begin{center}
    \includegraphics[width=1\textwidth]{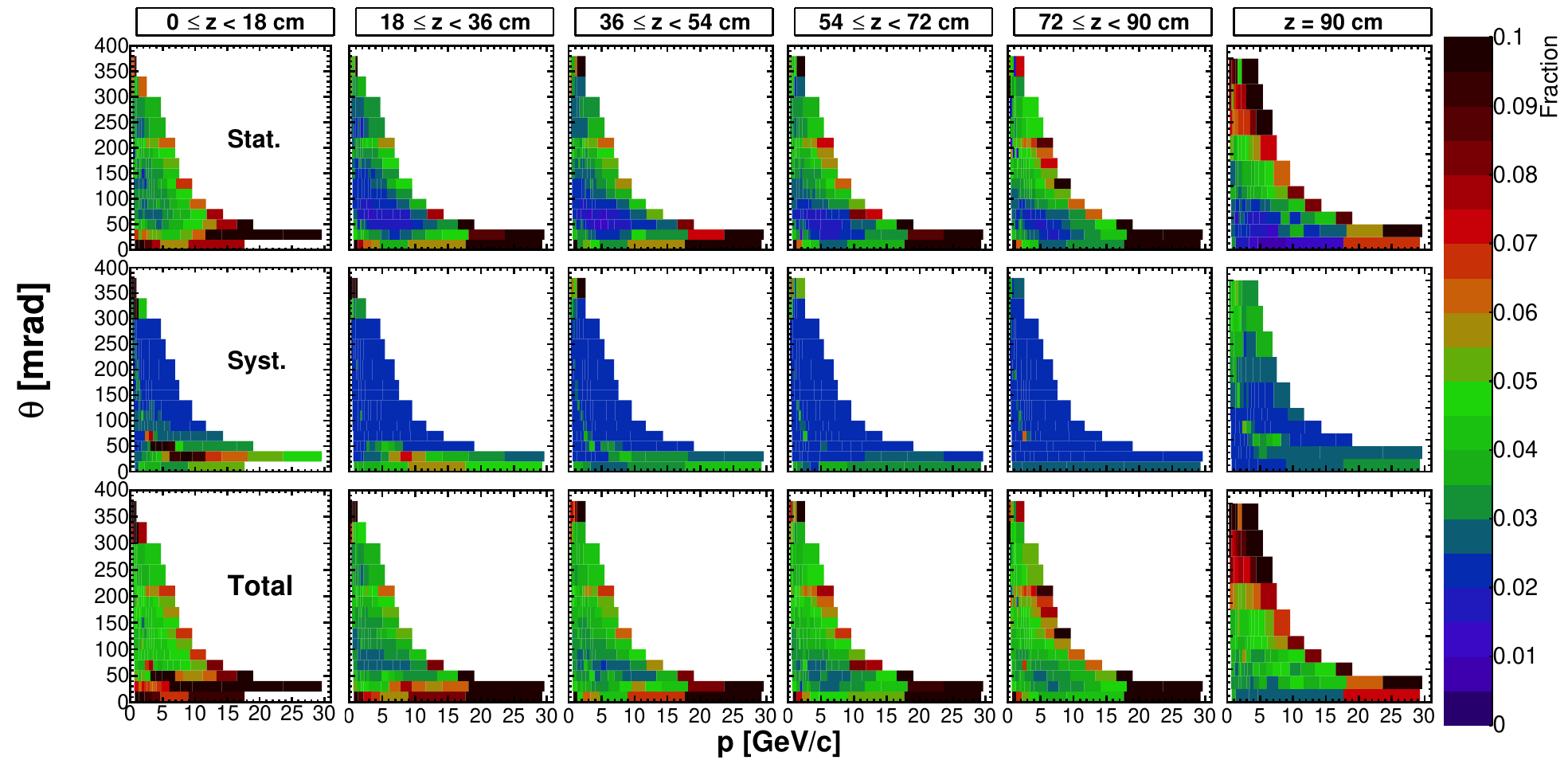}
    \caption{Uncertainties on \pip yields shown as a function of $p$ and $\theta$: statistical uncertainties (top row), total systematic uncertainties (middle row) and total uncertainties (bottom row). Each column corresponds to a different longitudinal bin.}\label{fig:piperrors}
    \end{center}
    
\end{figure*}

\begin{figure*}[ht]
    \begin{center}
    \includegraphics[width=1\textwidth]{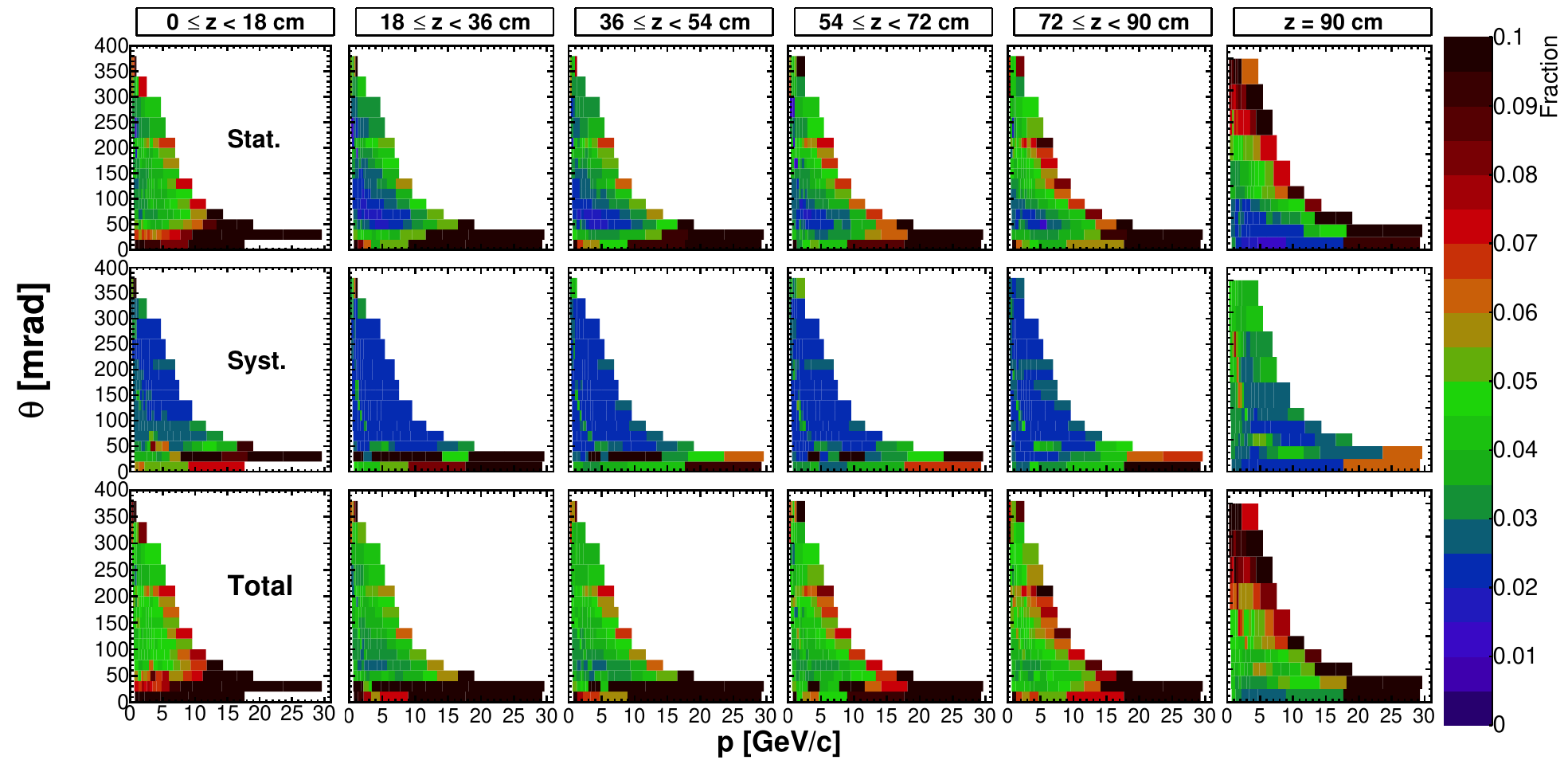}
    \caption{Uncertainties on \pim yields shown as a function of $p$ and $\theta$: statistical uncertainties (top row), total systematic uncertainties (middle row) and total uncertainties (bottom row). Each column corresponds to a different longitudinal bin.}\label{fig:pimerrors}
    \end{center}
    
\end{figure*}

\begin{figure*}[ht]
    \begin{center}
    \includegraphics[width=1\textwidth]{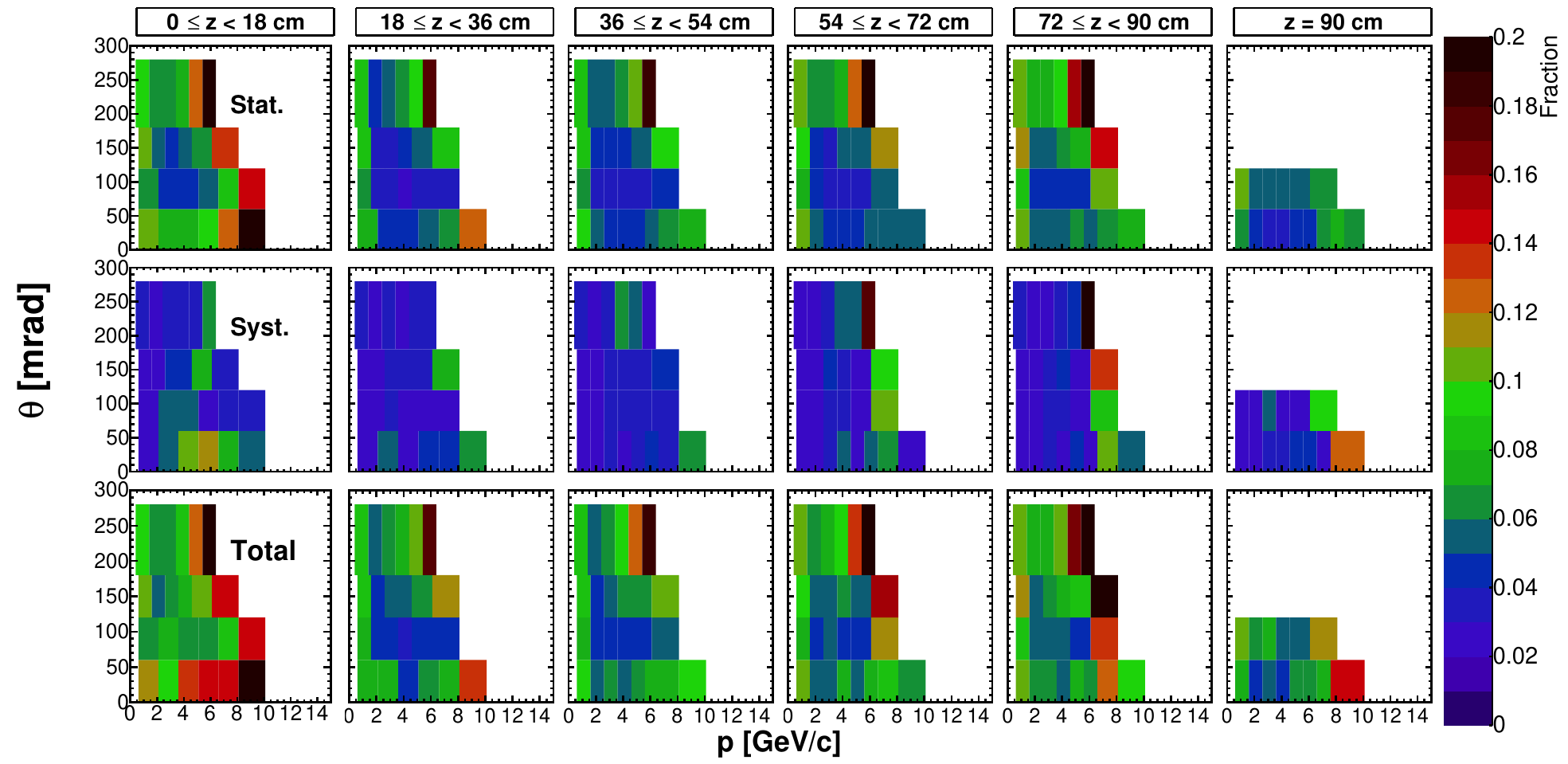}
    \caption{Uncertainties on \kp yields shown as a function of $p$ and $\theta$: statistical uncertainties (top row), total systematic uncertainties (middle row) and total uncertainties (bottom row). Each column corresponds to a different longitudinal bin.}\label{fig:kperrors}
    \end{center}
    
\end{figure*}

\begin{figure*}[ht]
    \begin{center}
    \includegraphics[width=1\textwidth]{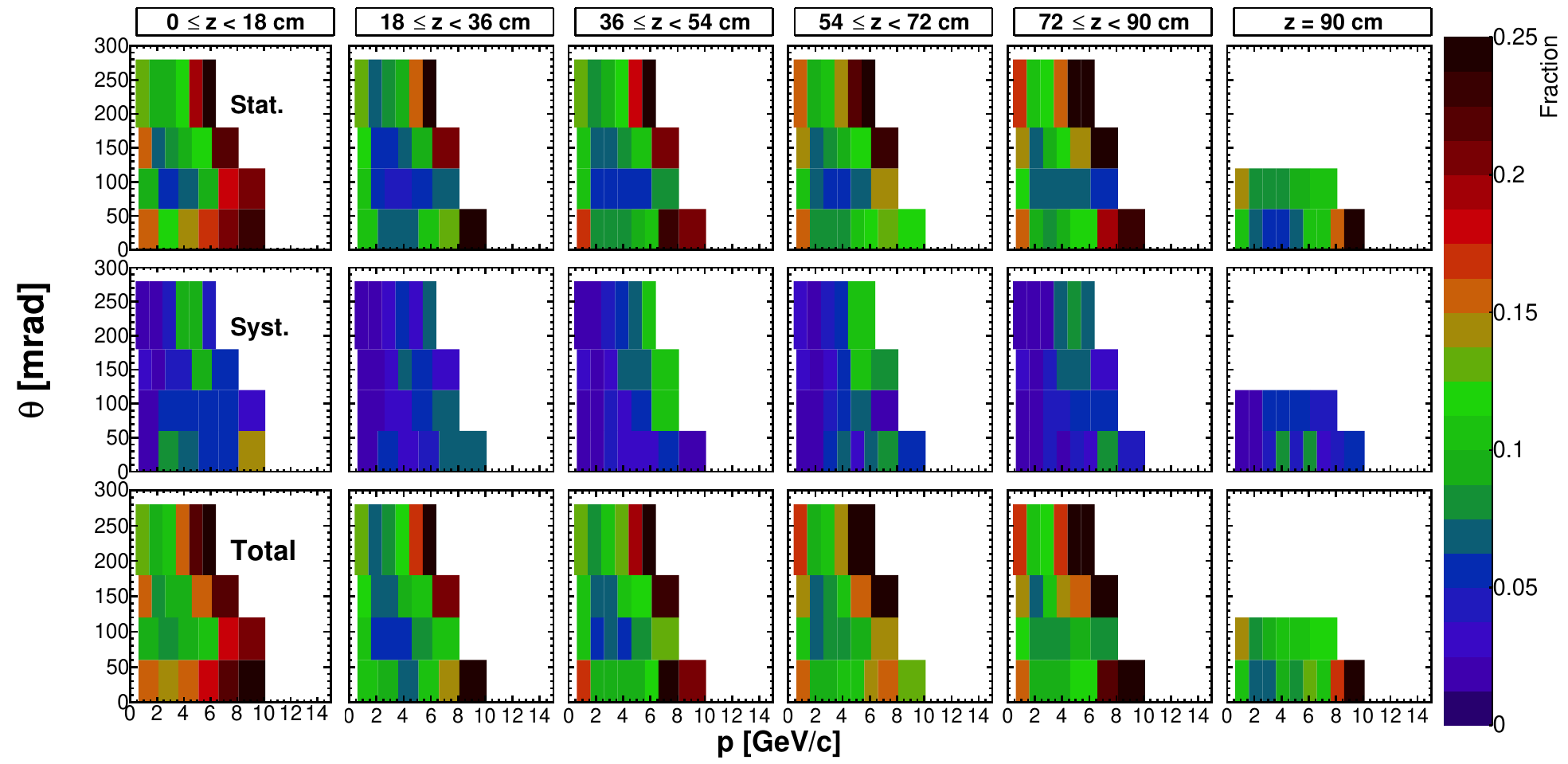}
    \caption{Uncertainties on \km yields shown as a function of $p$ and $\theta$: statistical uncertainties (top row), total systematic uncertainties (middle row) and total uncertainties (bottom row). Each column corresponds to a different longitudinal bin.}\label{fig:kmerrors}
    \end{center}
    
\end{figure*}

\begin{figure*}[ht]
    \begin{center}
    \includegraphics[width=1\textwidth]{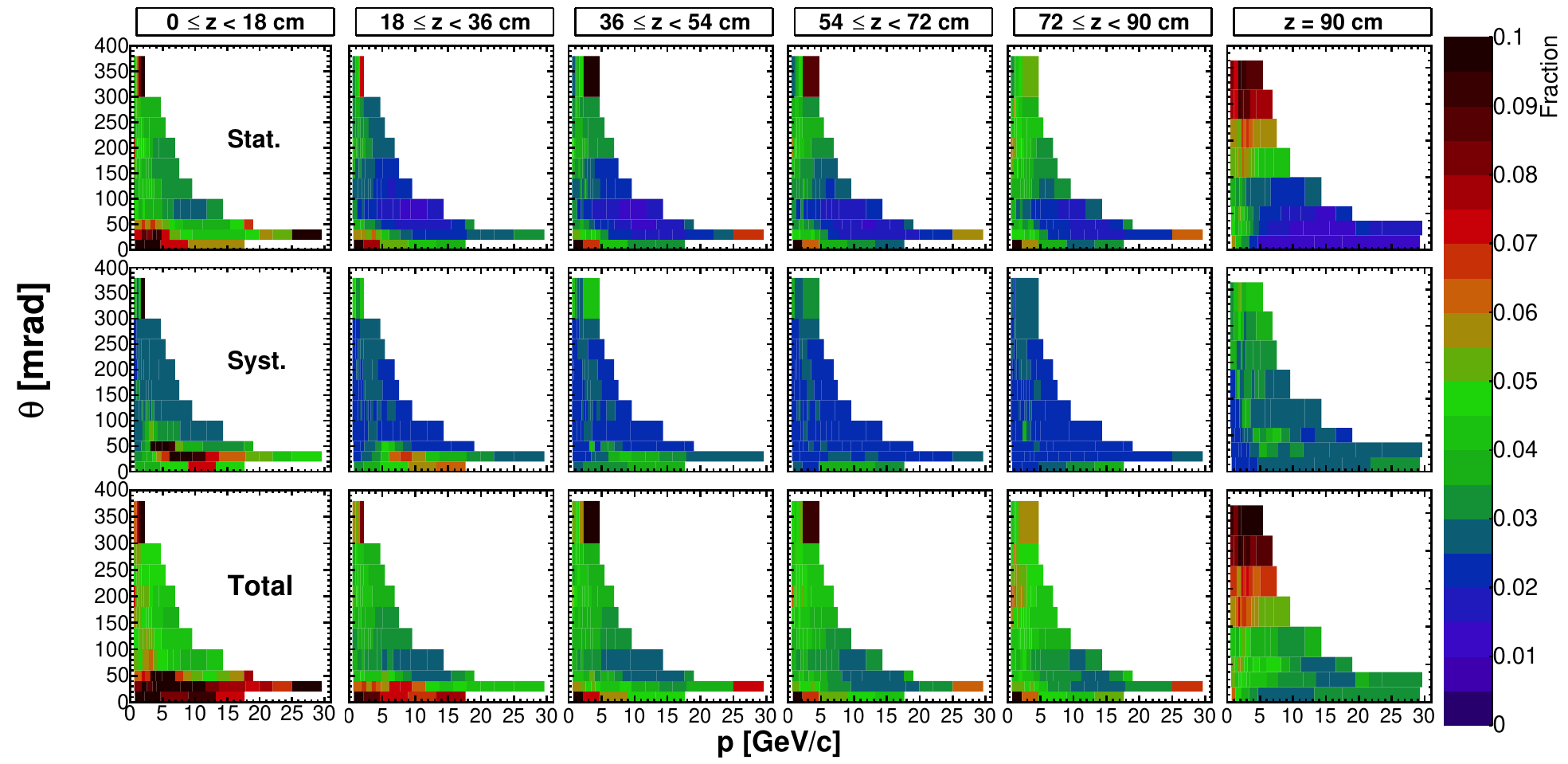}
    \caption{Uncertainties on \prot yields shown as a function of $p$ and $\theta$: statistical uncertainties (top row), total systematic uncertainties (middle row) and total uncertainties (bottom row). Each column corresponds to a different longitudinal bin.}\label{fig:pperrors}
    \end{center}
    
\end{figure*}

The following systematic uncertainties were considered:
\begin{enumerate}[(i)]
	\item Hadron loss uncertainty. Apart from the ToF-F inefficiency, tracks can miss the ToF-F wall, re-interact in the detector, or decay before reaching the end of the MTPCs. Possible imperfections in the Monte Carlo description of the detector can create biases while correcting for the previously described effects. To check for potential biases, an attempt to impose more strict cuts and to select only tracks with a long segment in the MTPCs was made. After re-calculating and comparing the yields to the standard results, the difference was taken as a systematic uncertainty. In the majority of bins, this uncertainty is well below $1\%$. However, for tracks with a momentum of several \GeVc in the $20-40\:$\mrad region in the most upstream longitudinal bin, the hadron loss uncertainty goes up to $20\%$ (for a single bin it reached $50\%$). These tracks cross the beamline and pass very close to the walls of the MTPCs.
	\item Backward track extrapolation (bin migration) uncertainty. The uncertainty on the target position can induce systematic bin migration while extrapolating tracks backwards to the surface of the target. The position of the target was changed in the Monte Carlo and hadron yields were recalculated. Differences were taken as systematic uncertainties. For tracks with a very small angle originating from the first two longitudinal bins, this uncertainty goes up to $8\%$. Also, high angle tracks extrapolated toward the downstream face of the target are sensitive to the $x$ and $y$ shifts of the target, and for them the bin migration uncertainty is up to $5\%$. For all other regions of the phase space, the bin migration uncertainty drops below $1\%$.
	\item Reconstruction efficiency uncertainty. This uncertainty was estimated in the previous \NASixtyOne publications~\cite{thin2009paper, LTpaper2009}. Since the experimental setup and the reconstruction software did not change, the same value was used for the measurements presented in this paper. A conservative value of $2\%$ is assigned for all bins. 
	\item Particle identification (PID) uncertainty. The width of the energy loss distribution for a single bin depends on the number of clusters in tracks. The energy loss distribution should be a sum of Gaussians, where each Gaussian represents tracks with a distinct number of clusters. Selected tracks, in this case, are very long and the width of the energy loss distribution becomes nearly constant. For this reason, a single Gaussian was used to describe the energy loss distributions. However, at higher momenta and, thus, for larger momentum bins, the mean values will also shift and possibly create an  asymmetry in the distribution. To check for potential biases, a sum of two Gaussians was used and the obtained yields were compared. The mean values of Gaussians were kept independent, while the width of the second Gaussian was forced to be larger than the width of the first Gaussian. For low momentum, no difference was found, however, for higher momentum and larger bins, a difference of up to $1\%$ was observed for $\pi^\pm$ and \prot. A constant value of $1\%$ was assigned to all of these bins. Positively charged kaons are located under the proton peak in the energy loss distribution and the pion peak in the \mtof distribution for momentum larger than $3\:$\GeVc. For this reason, the uncertainty for \kp can reach $15\%$ for the higher momentum bins. In case of \km, the obtained uncertainties are similar to the ones for \kp.   
	\item Feed-down uncertainty. Produced \kz and \lm in the target can decay before reaching the VTPC-1 and some of the daughter particles reconstructed in the TPCs can be extrapolated to the target surface. Since the correction for this effect is calculated with the Monte Carlo simulation, the number of produced \kz and \lm is model dependent. A systematic uncertainty of $30\%$ on the feed-down correction factor was assigned, following the standard approach in Refs.~\cite{thin2009paper, LTpaper2009}. The systematic uncertainty is up to $2\%$ for $\pi^\pm$ and protons. Mostly low momentum bins in the first and last longitudinal bins are affected.
	\item The ToF-F efficiency uncertainty. The uncertainty of the ToF-F efficiency was taken as a systematic uncertainty. It goes up to $2\%$, but for most of the bins it is well below $1\%$.
	\item Ad-hoc uncertainty. A constant conservative uncertainty of $25\%$ was assigned to the bins that are corrected by the ad-hoc correction factor. The value was based on half the size of the typical ad hoc correction ($50\%$).
\end{enumerate}
Total systematic uncertainties for individual analysis bins and for each particle type are shown in Figs.~\ref{fig:piperrors} -~\ref{fig:pperrors} (middle row).

\begin{figure*}[ht]
    \begin{center}
    \includegraphics[width=1\textwidth]{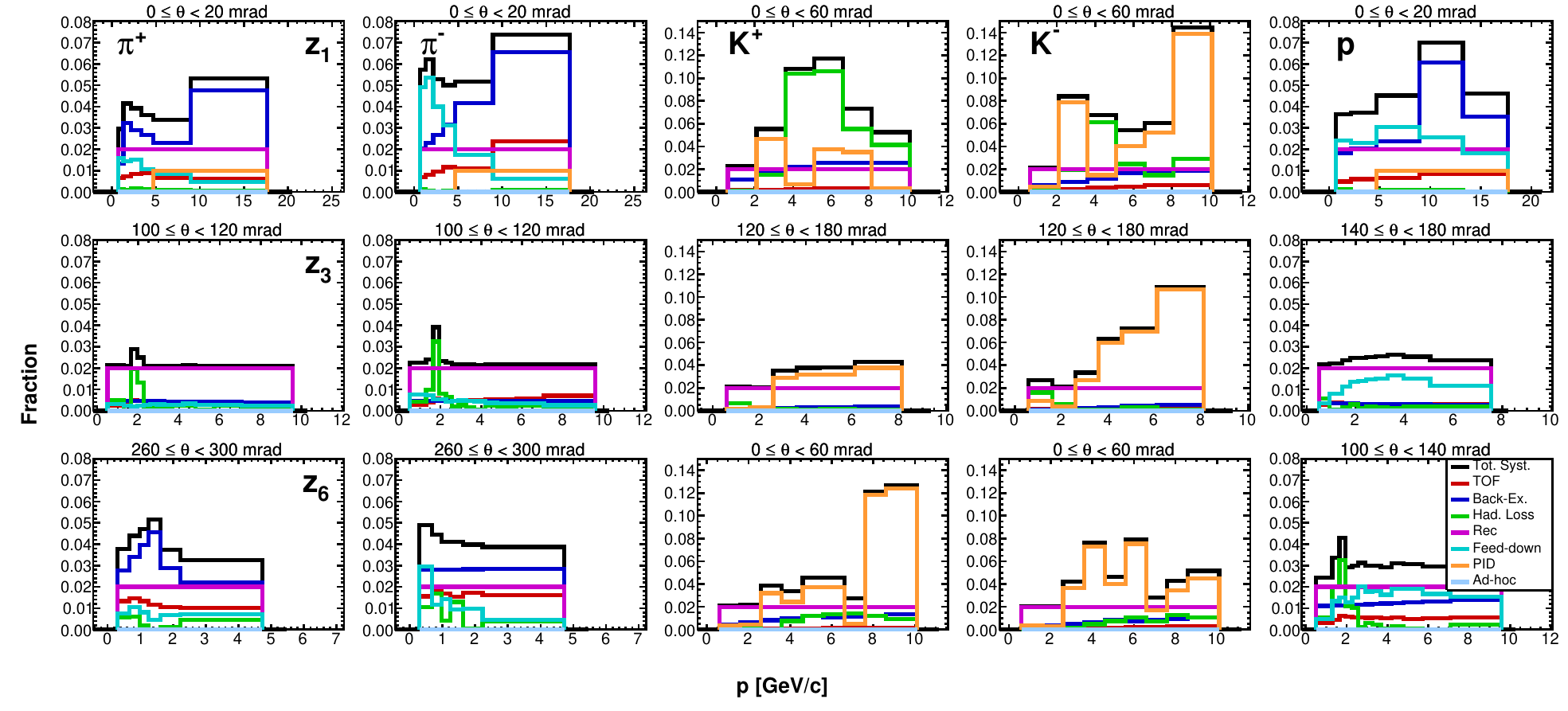}
    \caption{Systematic uncertainties for the measured yields as a function of momentum. Each column corresponds to a different particle type, from left to right: \pip, \pim, \kp, \km, and \prot. 
    Each row corresponds to a different longitudinal bin. Specific theta ranges are given on top of each plot. Presented ($p,\theta, z$) bins are selected to show typical sizes of the systematic uncertainties.}\label{fig:syst}
    \end{center}
    
\end{figure*}

Respective contributions to the total systematic uncertainties for different particle types are shown in Fig.~\ref{fig:syst}. Different regions of the phase space have different dominant contributions. For example, the dominant contribution in the third longitudinal bin ($z_3$) for $\pi^{\pm}$ yields comes from the reconstruction efficiency uncertainty, while the dominant contribution for $K^\pm$ comes from the PID uncertainty.

\section{Results and comparison with hadron production models} \label{sec:Results}

Results are presented in the form of double differential yields normalised by the total number of protons on target:
\begin{equation}
	 \frac{1}{N_{POT}}\left(\frac{d^2 n}{dp \cdot d\theta}\right)_{ijk} = \frac{1}{N_{POT}}\frac{N^{\alpha}_{ijk}C_{ijk}}{\Delta p_{ijk}\Delta \theta_{ij}},
\end{equation}
where ($i,j,k$) are the ($p, \theta, z$) bin numbers, $N_{POT}$ is the number of protons on target, $\alpha = \pip, \pim, \kp, \km, \prot$ is the hadron species, $N^{\alpha}_{ijk}$ is the number of extracted hadrons in a given bin, $C_{ijk}$ is the correction factor, $\Delta p_{ijk}$ is the momentum bin size, and $\Delta \theta_{ij}$ is the polar angle bin size. Uncertainties on the hadron yields are shown in Figs.~\ref{fig:piperrors} -~\ref{fig:pperrors}. The top rows show statistical uncertainties, the middle rows show systematic uncertainty and the bottom rows show total uncertainties. For \pip, \pim and \prot statistical uncertainties dominate in the low ($<40\:$\mrad) and high ($>200\:$\mrad) polar angle region. While for $40 < \theta < 200\:$\mrad, the systematic uncertainties are comparable to or larger than the statistical uncertainties. For kaons, the total uncertainty is mostly dominated by the statistical uncertainty. Tables with numerical results are presented in Ref.~\cite{numtables}.

\subsection{Comparisons with models}
\begin{figure*}[ht]
    \begin{center}
    \includegraphics[width=1\textwidth]{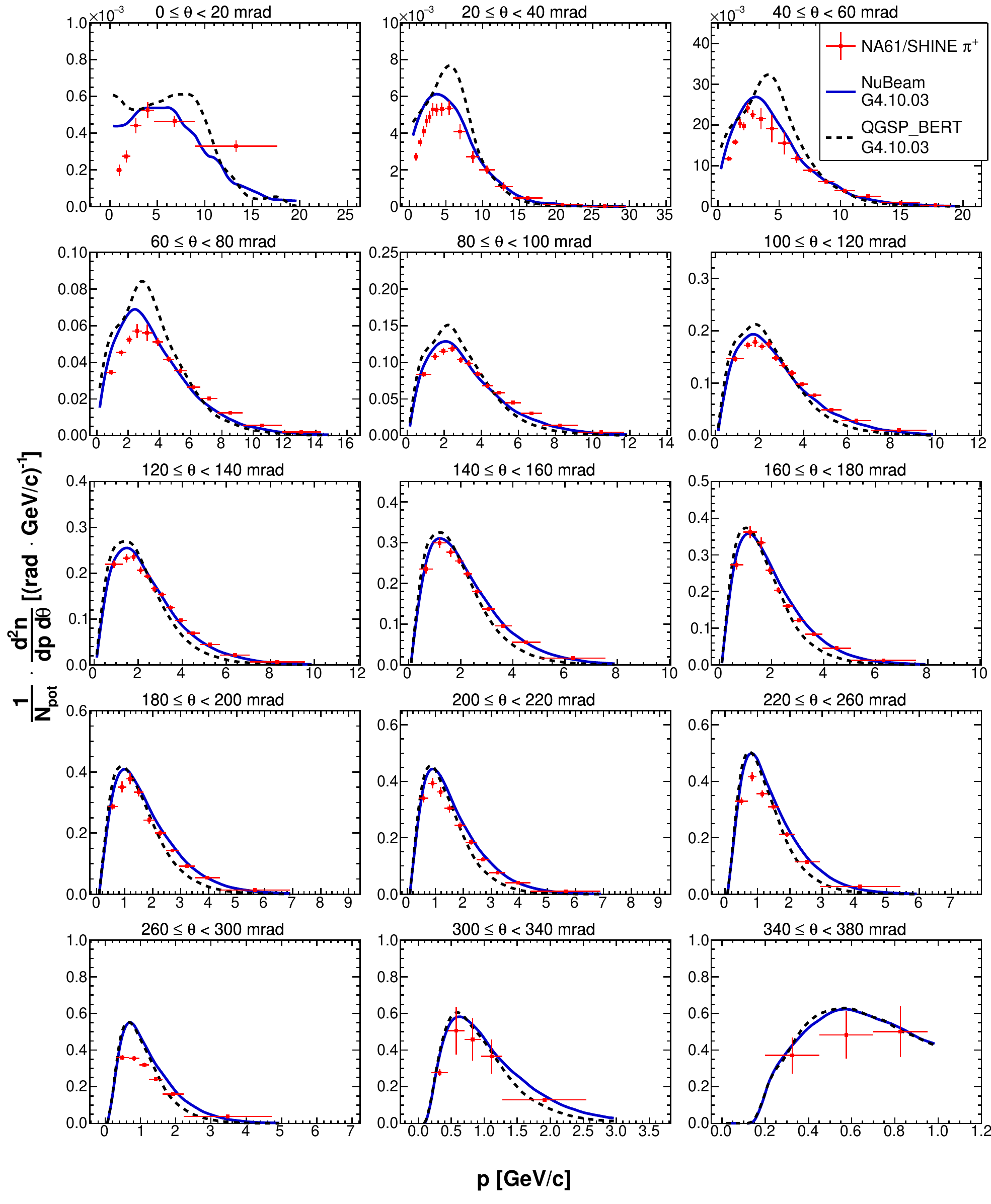}
    \caption{Double differential yields of positively charged pions for the most upstream longitudinal bin ($0\leq z < 18\:$\cm). Vertical bars represent the total uncertanties. Predictions from the \NuBeam (solid blue line) and \QGSP (dashed black line) physics lists from \GeantFT.03~\cite{GEANT4, GEANT4bis} are overlaid on top of the data.}\label{fig:piPlus1}
    \end{center}
\end{figure*}

\begin{figure*}[ht]
    \begin{center}
    \includegraphics[width=1\textwidth]{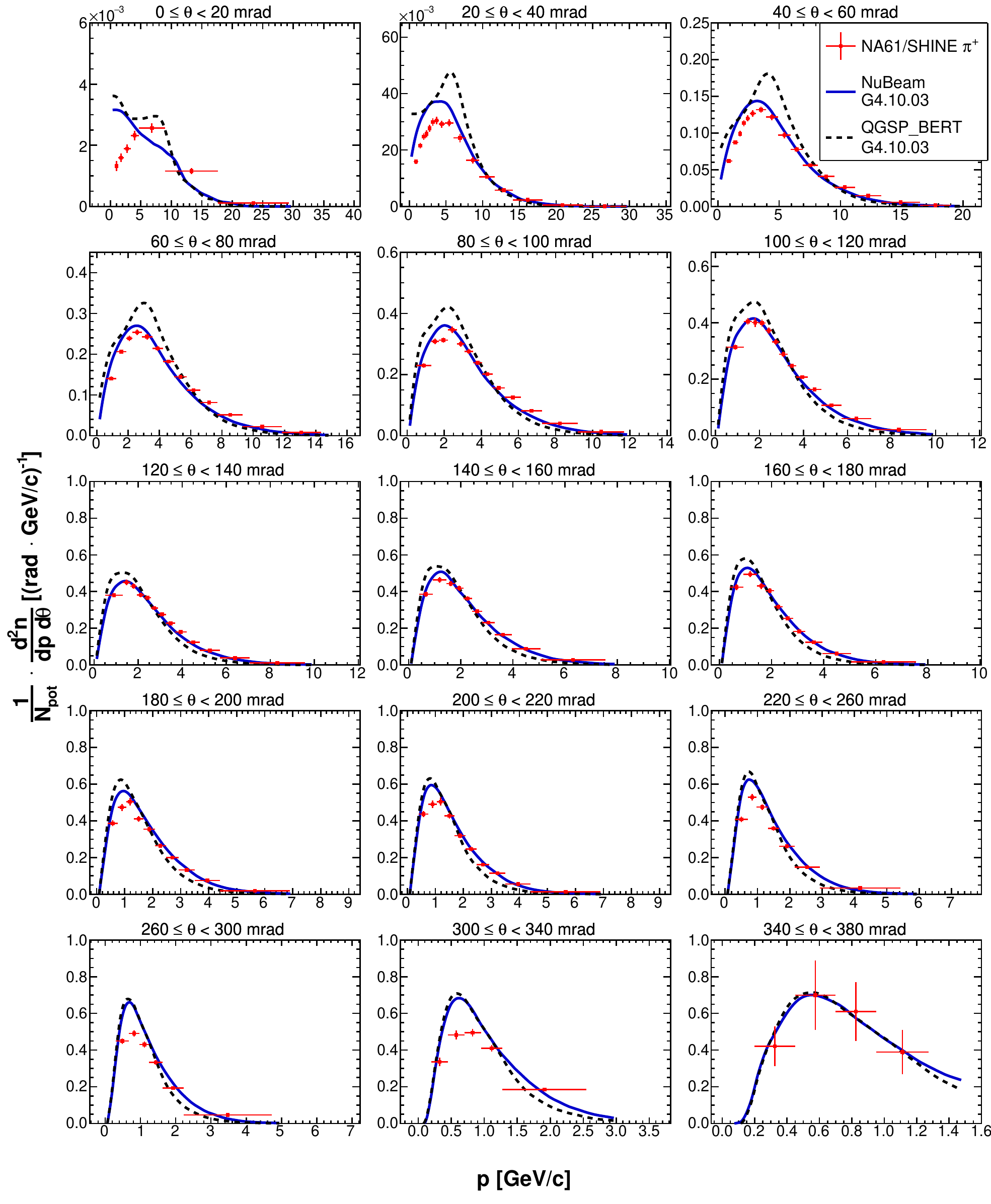}
    \caption{Double differential yields of positively charged pions for the second upstream longitudinal bin ($18\leq z < 36\:$\cm). Vertical bars represent the total uncertanties. Predictions from the \NuBeam (red) and \QGSP (dashed black line) physics lists from \GeantFT.03~\cite{GEANT4, GEANT4bis} are overlaid on top of the data.}\label{fig:piPlus2}
    \end{center}
\end{figure*}

\begin{figure*}[ht]
    \begin{center}
    \includegraphics[width=1\textwidth]{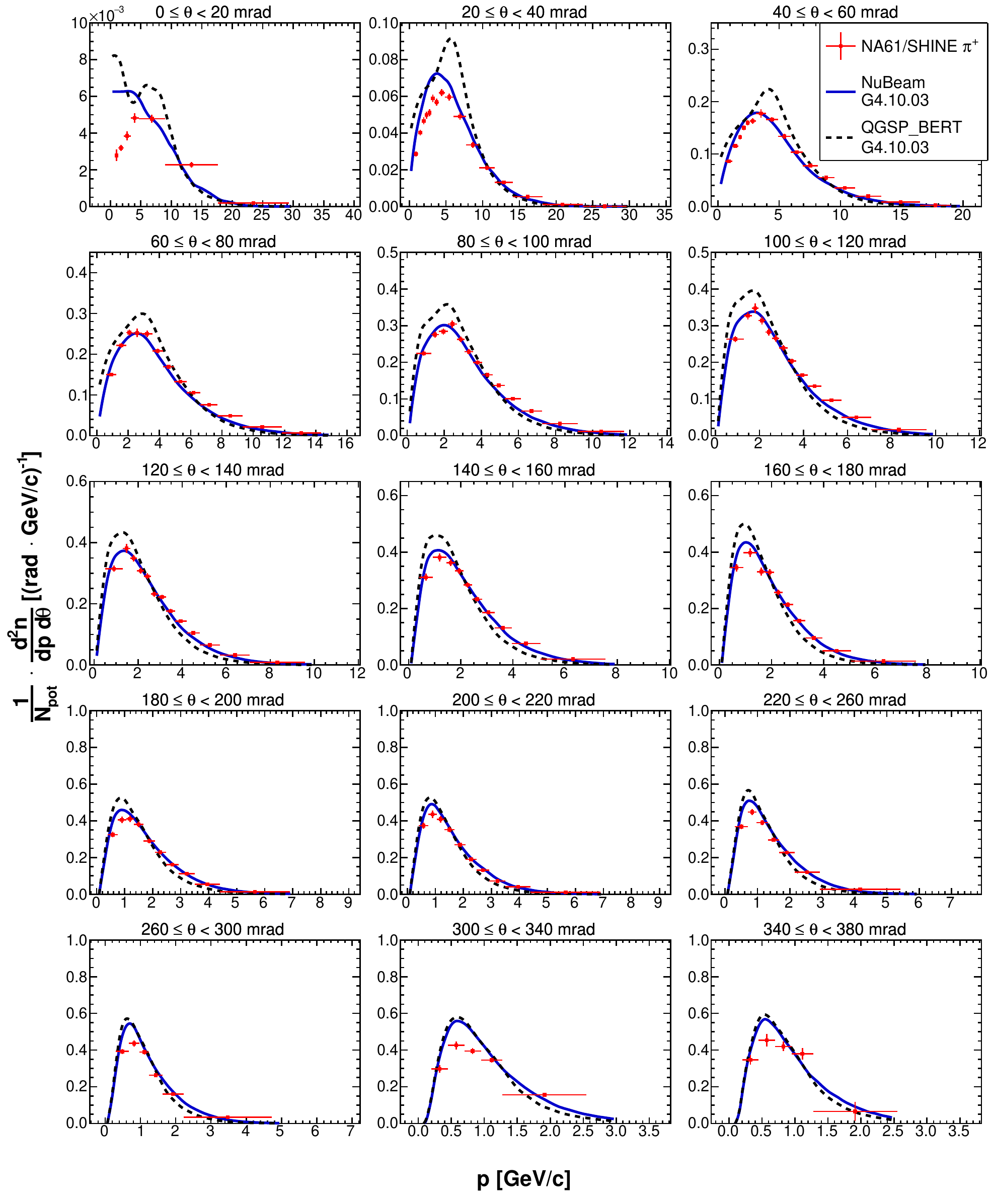}
    \caption{Double differential yields of positively charged pions for the third upstream longitudinal bin ($36\leq z < 54\:$\cm). Vertical bars represent the total uncertanties. Predictions from the \NuBeam (solid blue line) and \QGSP (dashed black line) physics lists from \GeantFT.03~\cite{GEANT4, GEANT4bis} are overlaid on top of the data.}\label{fig:piPlus3}
    \end{center}
\end{figure*}

\begin{figure*}[ht]
    \begin{center}
    \includegraphics[width=1\textwidth]{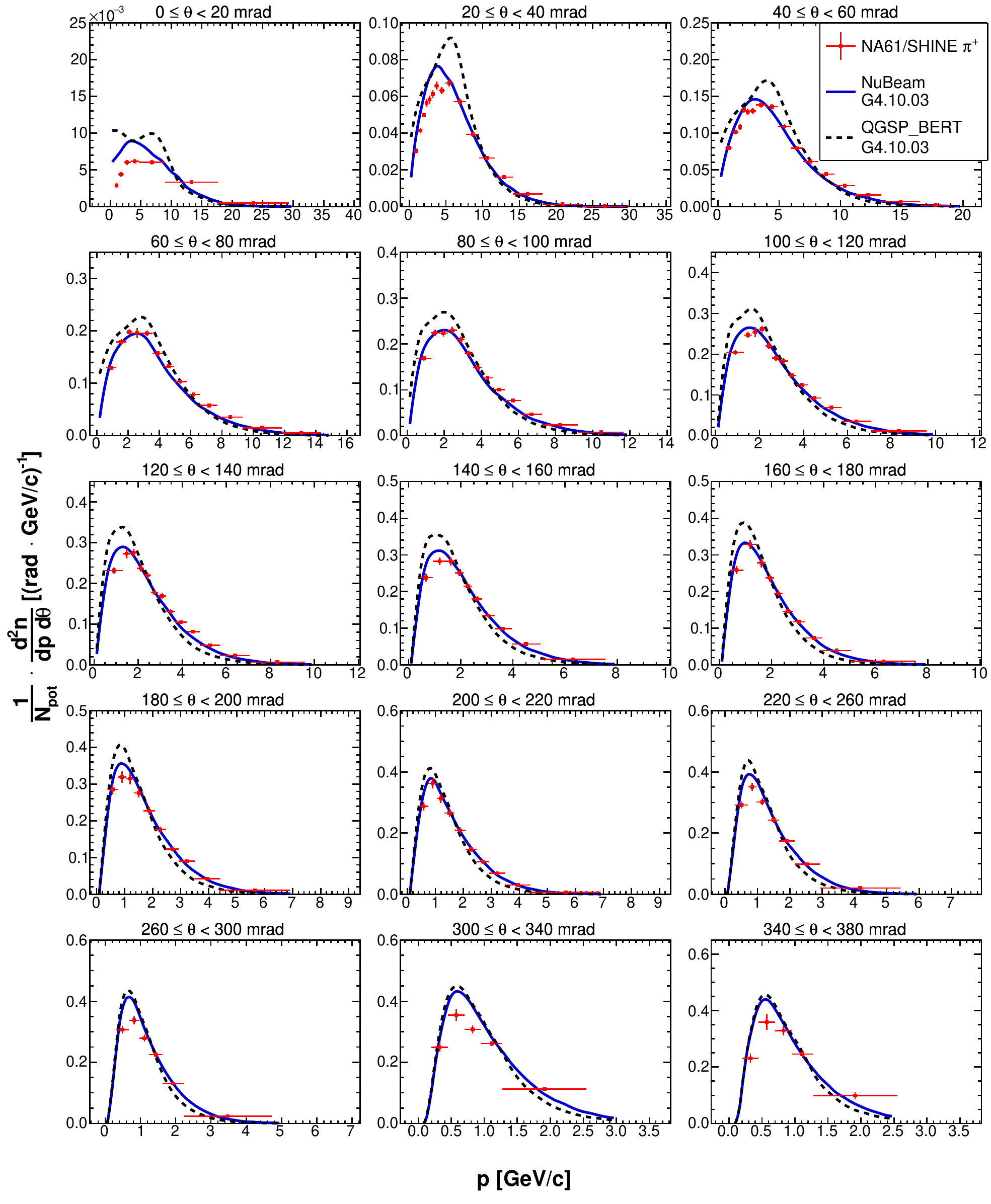}
    \caption{Double differential yields of positively charged pions for the fourth upstream longitudinal bin ($54\leq z < 72\:$\cm). Vertical bars represent the total uncertanties. Predictions from the \NuBeam (solid blue line) and \QGSP (dashed black line) physics lists from \GeantFT.03~\cite{GEANT4, GEANT4bis} are overlaid on top of the data.}\label{fig:piPlus4}
    \end{center}
\end{figure*}

\begin{figure*}[ht]
    \begin{center}
    \includegraphics[width=1\textwidth]{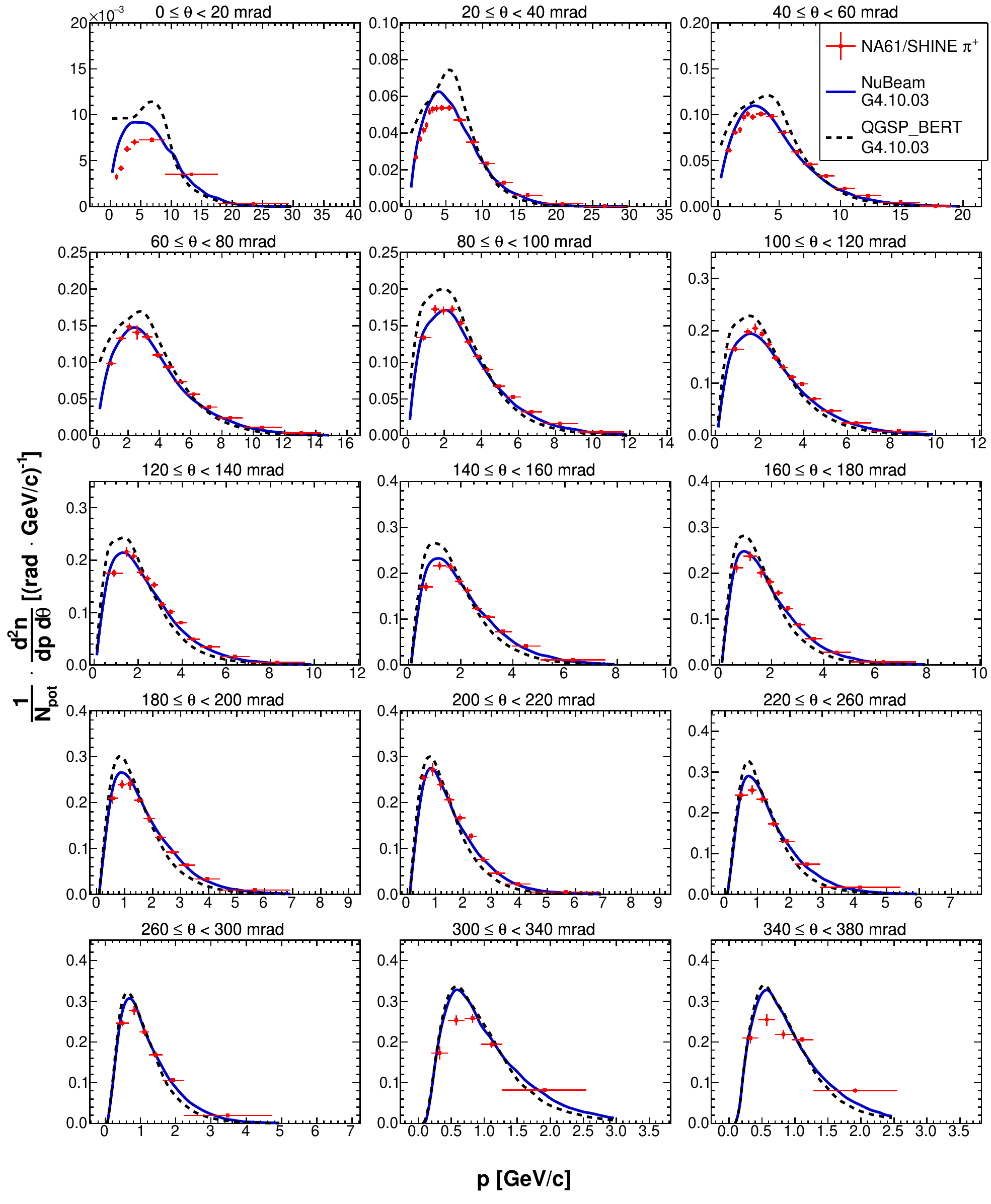}
    \caption{Double differential yields of positively charged pions for the fifth upstream longitudinal bin ($72\leq z < 90\:$\cm). Vertical bars represent the total uncertanties. Predictions from the \NuBeam (solid blue line) and \QGSP (dashed black line) physics lists from \GeantFT.03~\cite{GEANT4, GEANT4bis} are overlaid on top of the data.}\label{fig:piPlus5}
    \end{center}
\end{figure*}

\begin{figure*}[ht]
    \begin{center}
    \includegraphics[width=1\textwidth]{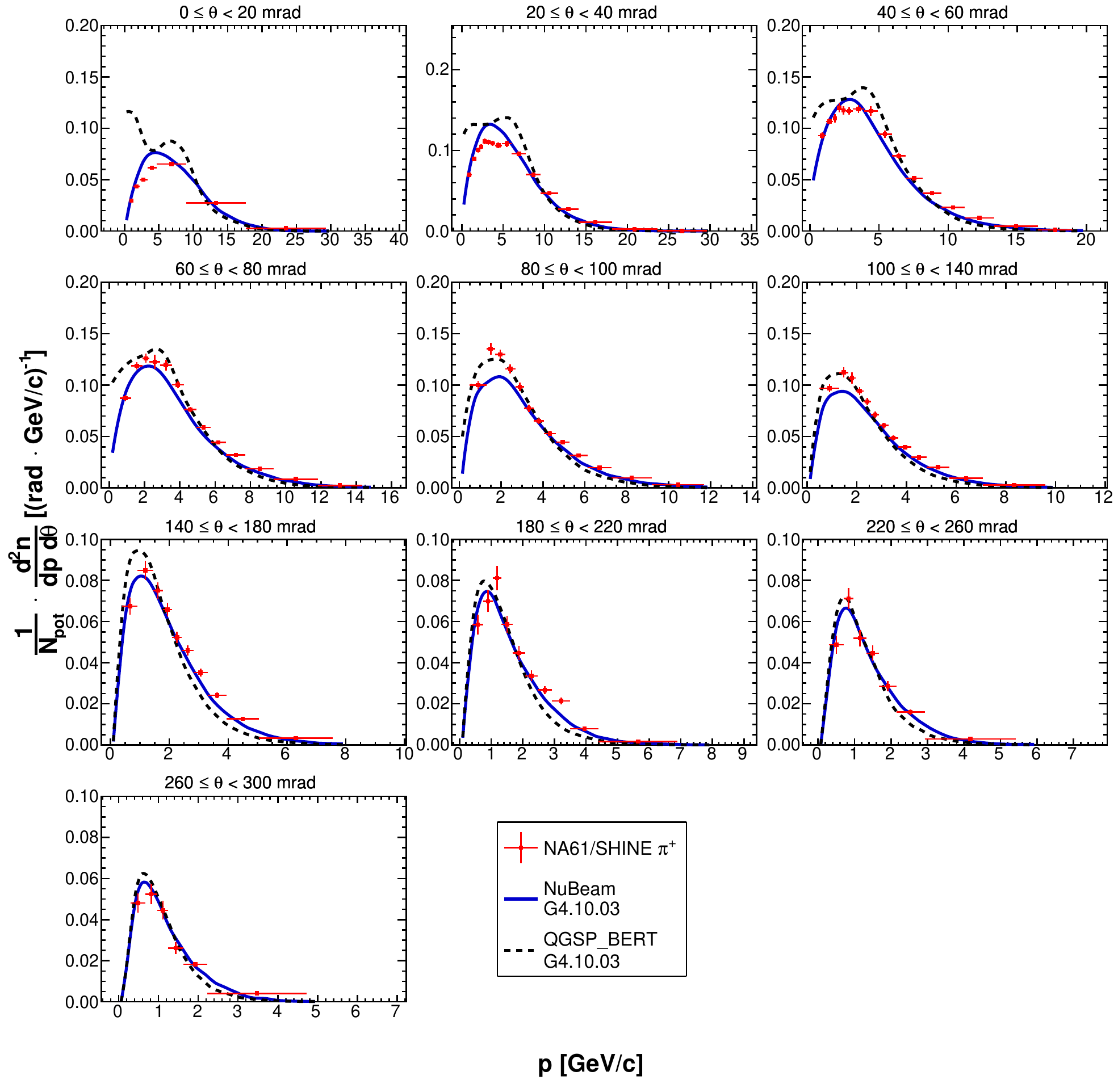}
    \caption{Double differential yields of positively charged pions for the downstream target face ($z=90\:$\cm). Vertical bars represent the total uncertanties. Predictions from the \NuBeam (solid blue line) and \QGSP (blue) physics lists from \GeantFT.03~\cite{GEANT4, GEANT4bis} are overlaid on top of the data.}\label{fig:piPlus6}
    \end{center}
\end{figure*}

\begin{figure*}[ht]
    \begin{center}
    \includegraphics[width=1\textwidth]{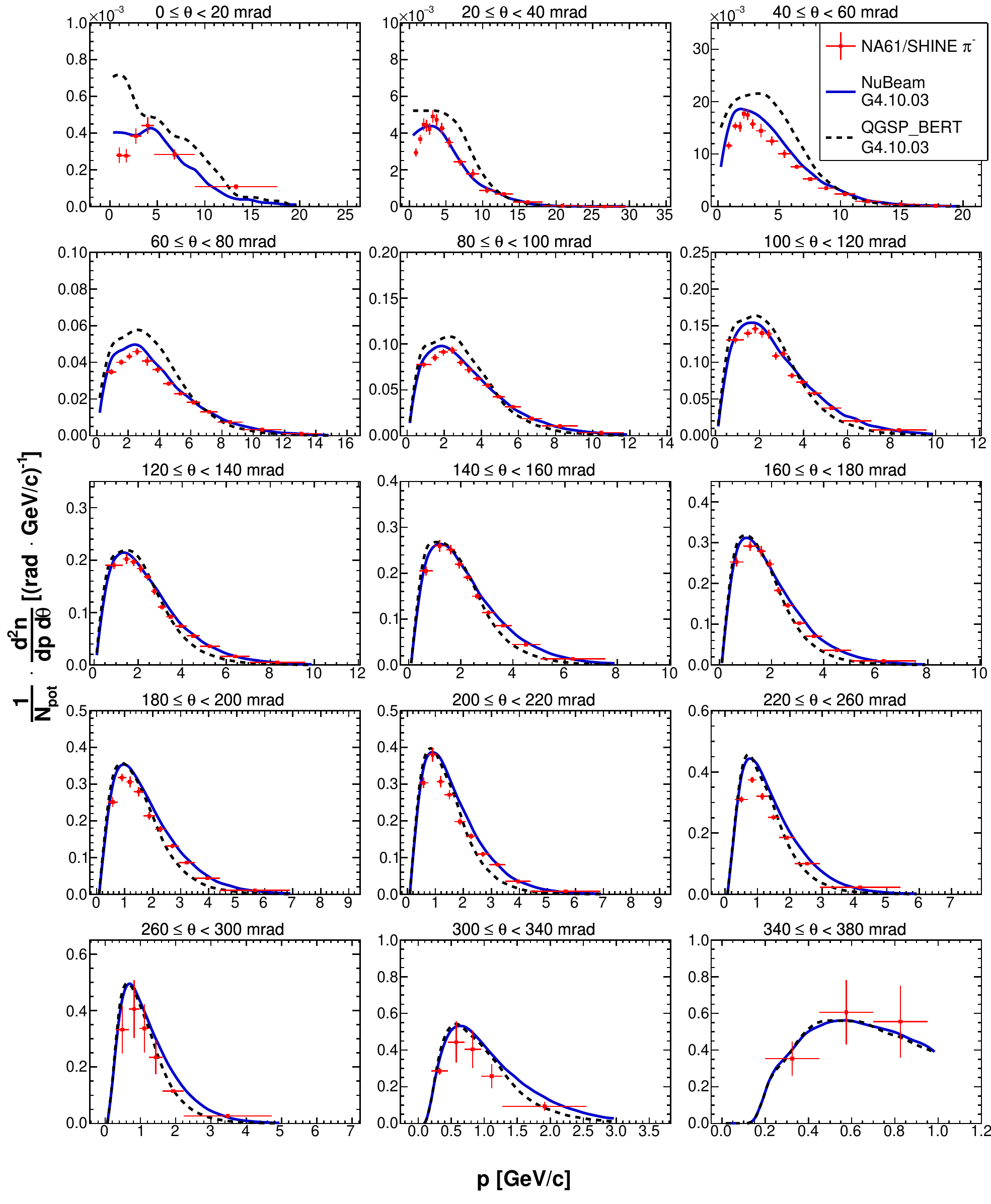}
    \caption{Double differential yields of negatively charged pions for the most upstream longitudinal bin ($0\leq z < 18\:$\cm). Vertical bars represent the total uncertanties. Predictions from the \NuBeam (solid blue line) and \QGSP (blue) physics lists from \GeantFT.03~\cite{GEANT4, GEANT4bis} are overlaid on top of the data.}\label{fig:piMinus1}
    \end{center}
\end{figure*}

\begin{figure*}[ht]
    \begin{center}
    \includegraphics[width=1\textwidth]{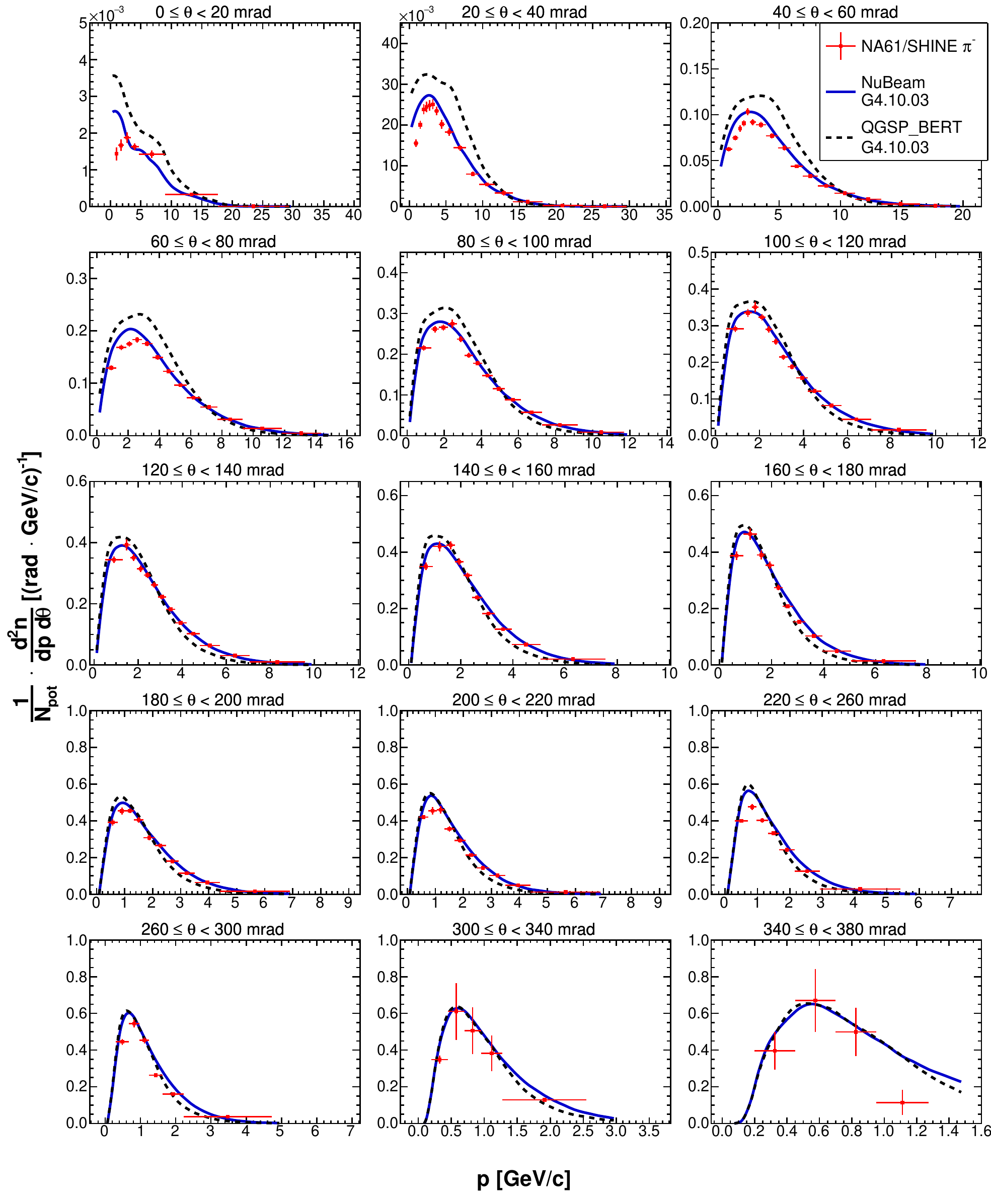}
    \caption{Double differential yields of negatively charged pions for the second upstream longitudinal bin ($18\leq z < 36\:$\cm). Vertical bars represent the total uncertanties. Predictions from  the \NuBeam (solid blue line) and \QGSP (dashed black line) physics lists from \GeantFT.03~\cite{GEANT4, GEANT4bis} are overlaid on top of the data.}\label{fig:piMinus2}
    \end{center}
\end{figure*}

\begin{figure*}[ht]
    \begin{center}
    \includegraphics[width=1\textwidth]{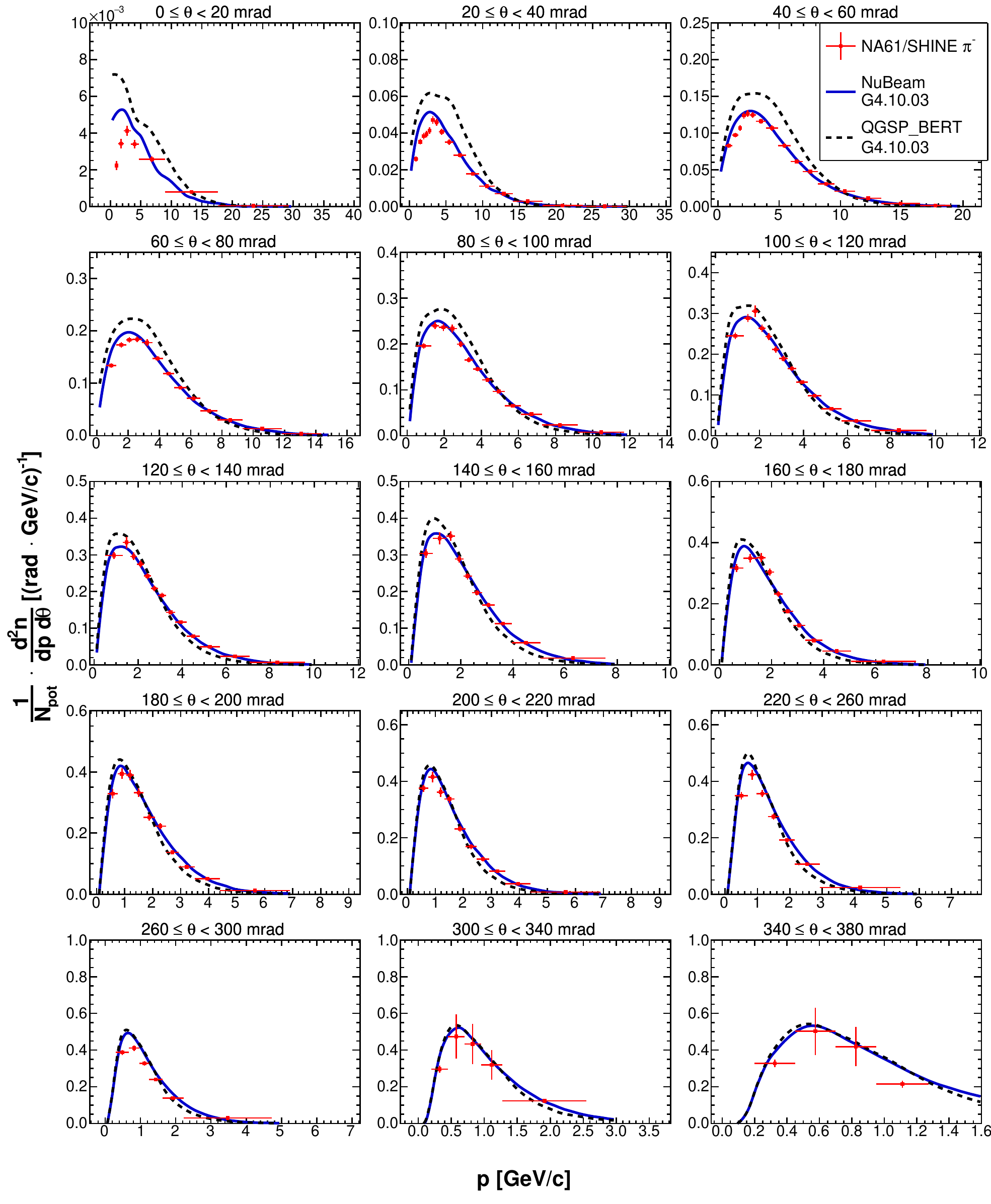}
    \caption{Double differential yields of negatively charged pions for the third upstream longitudinal bin ($36\leq z < 54\:$\cm). Vertical bars represent the total uncertanties. Predictions from the \NuBeam (solid blue line) and \QGSP (dashed black line) physics lists from \GeantFT.03~\cite{GEANT4, GEANT4bis} are overlaid on top of the data.}\label{fig:piMinus3}
    \end{center}
\end{figure*}

\begin{figure*}[ht]
    \begin{center}
    \includegraphics[width=1\textwidth]{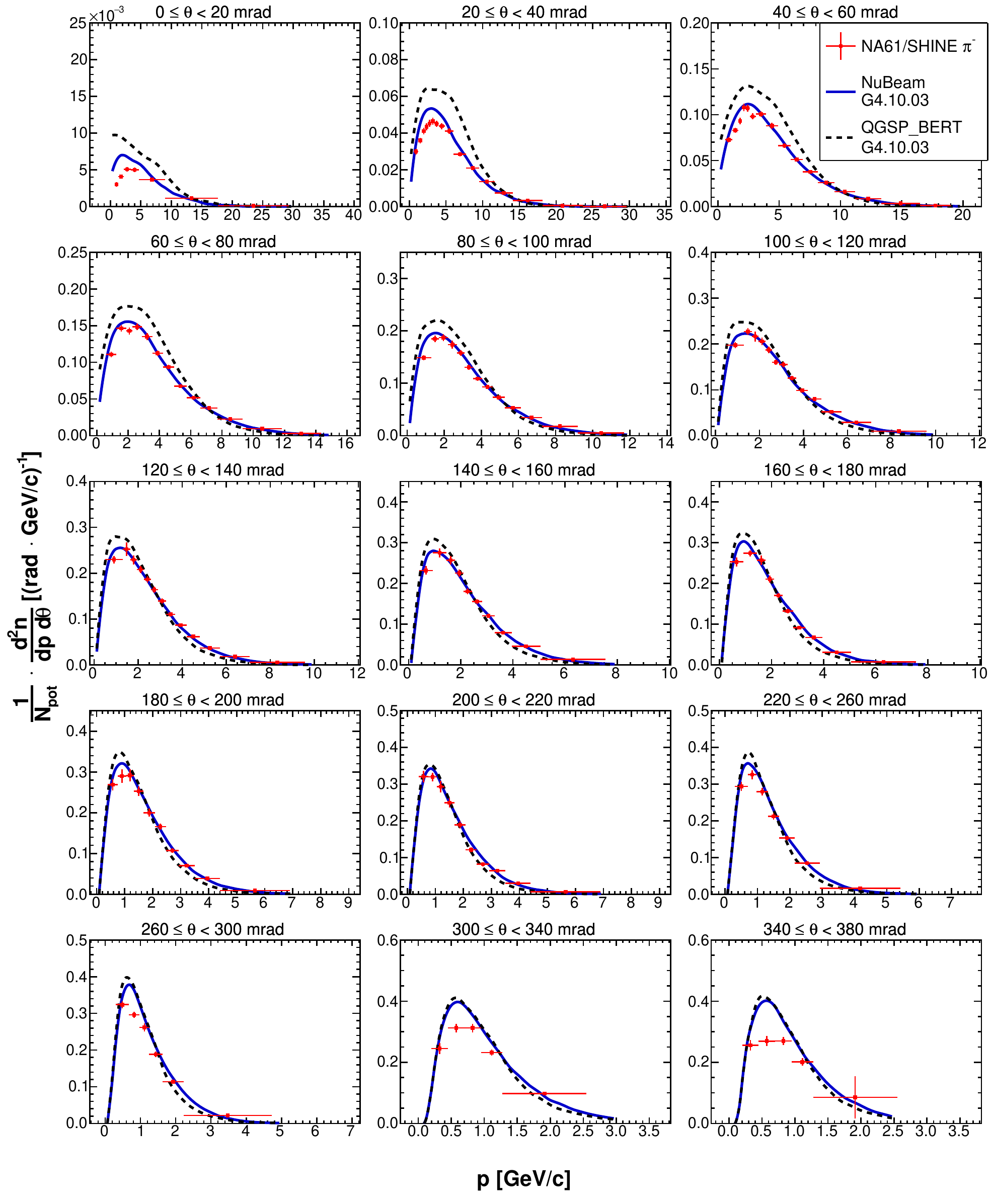}
    \caption{Double differential yields of negatively charged pions for the fourth upstream longitudinal bin ($54\leq z < 72\:$\cm). Vertical bars represent the total uncertanties. Predictions from the \NuBeam (solid blue line) and \QGSP (dashed black line) physics lists from \GeantFT.03~\cite{GEANT4, GEANT4bis} are overlaid on top of the data.}\label{fig:piMinus4}
    \end{center}
\end{figure*}

\begin{figure*}[ht]
    \begin{center}
    \includegraphics[width=1\textwidth]{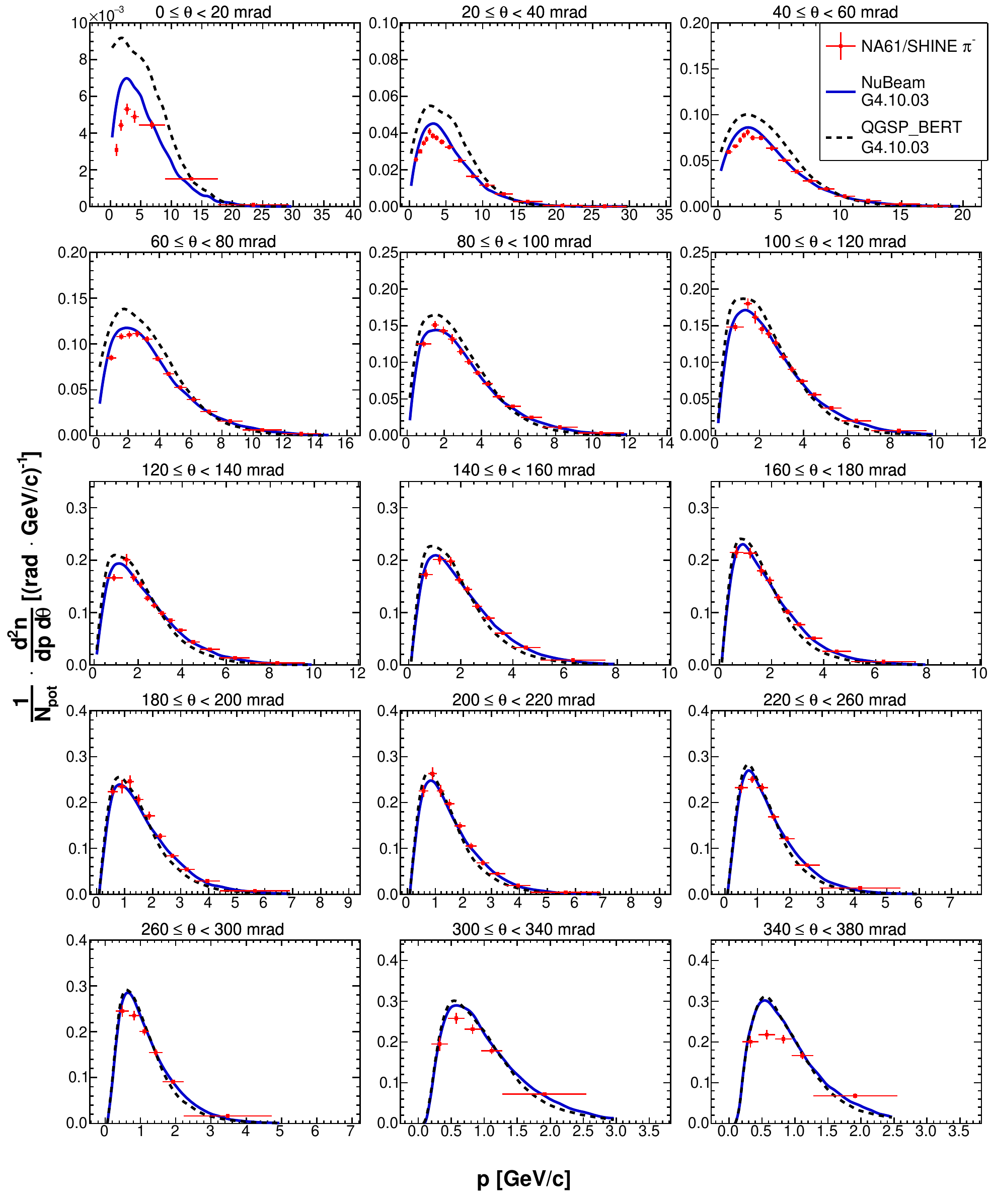}
    \caption{Double differential yields of negatively charged pions for the fifth upstream longitudinal bin ($72\leq z < 90\:$\cm). Vertical bars represent the total uncertanties. Predictions from the \NuBeam (solid blue line) and \QGSP (dashed black line) physics lists from \GeantFT.03~\cite{GEANT4, GEANT4bis} are overlaid on top of the data.}\label{fig:piMinus5}
    \end{center}
\end{figure*}

\begin{figure*}[ht]
    \begin{center}
    \includegraphics[width=1\textwidth]{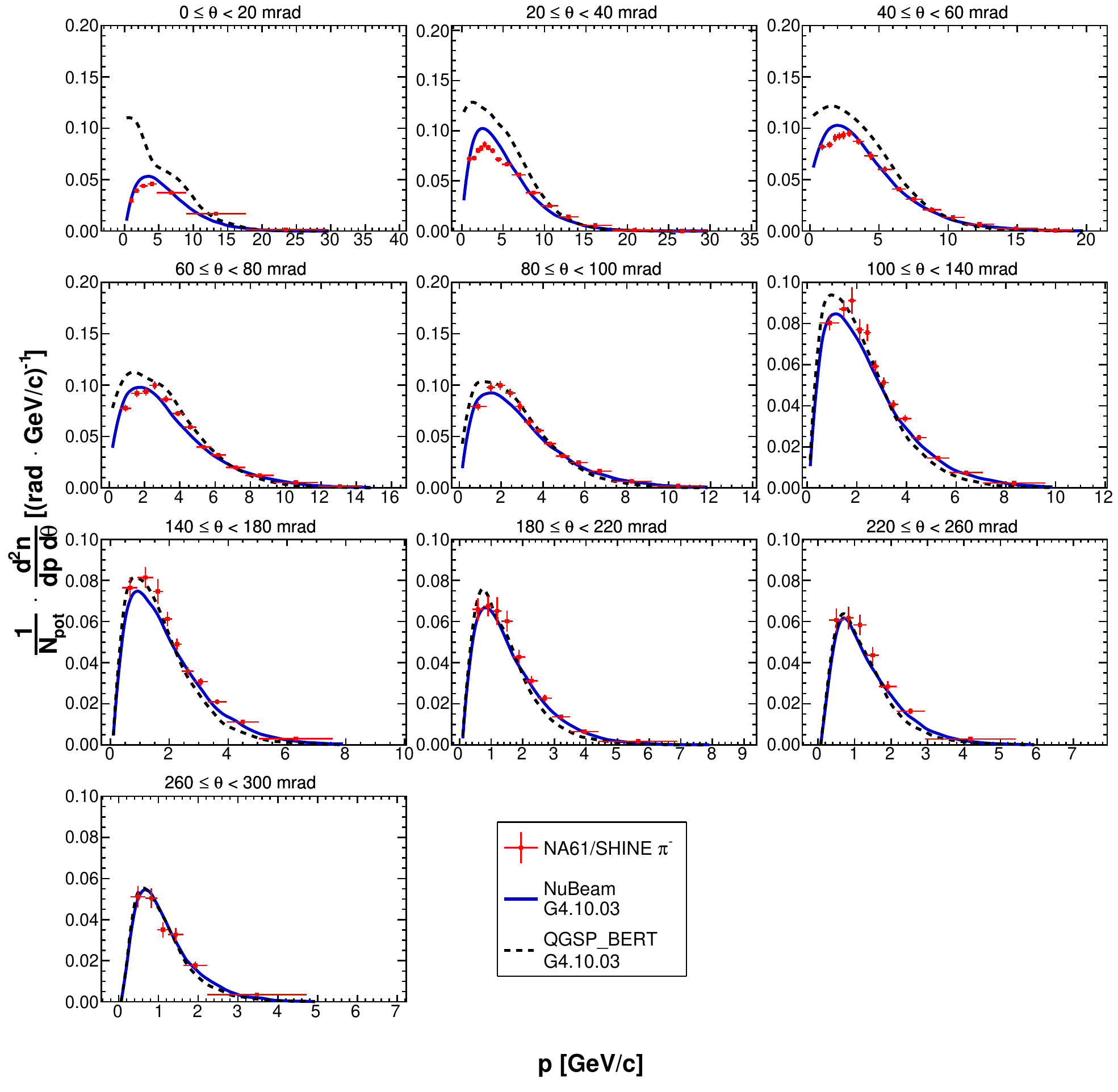}
    \caption{Double differential yields of negatively charged pions for the downstream target face ($z=90\:$\cm). Vertical bars represent the total uncertanties. Predictions from the \NuBeam (solid blue line) and \QGSP (dashed black line) physics lists from \GeantFT.03~\cite{GEANT4, GEANT4bis} are overlaid on top of the data.}\label{fig:piMinus6}
    \end{center}
\end{figure*}

\begin{figure*}[ht]
    \begin{center}
    \includegraphics[width=1\textwidth]{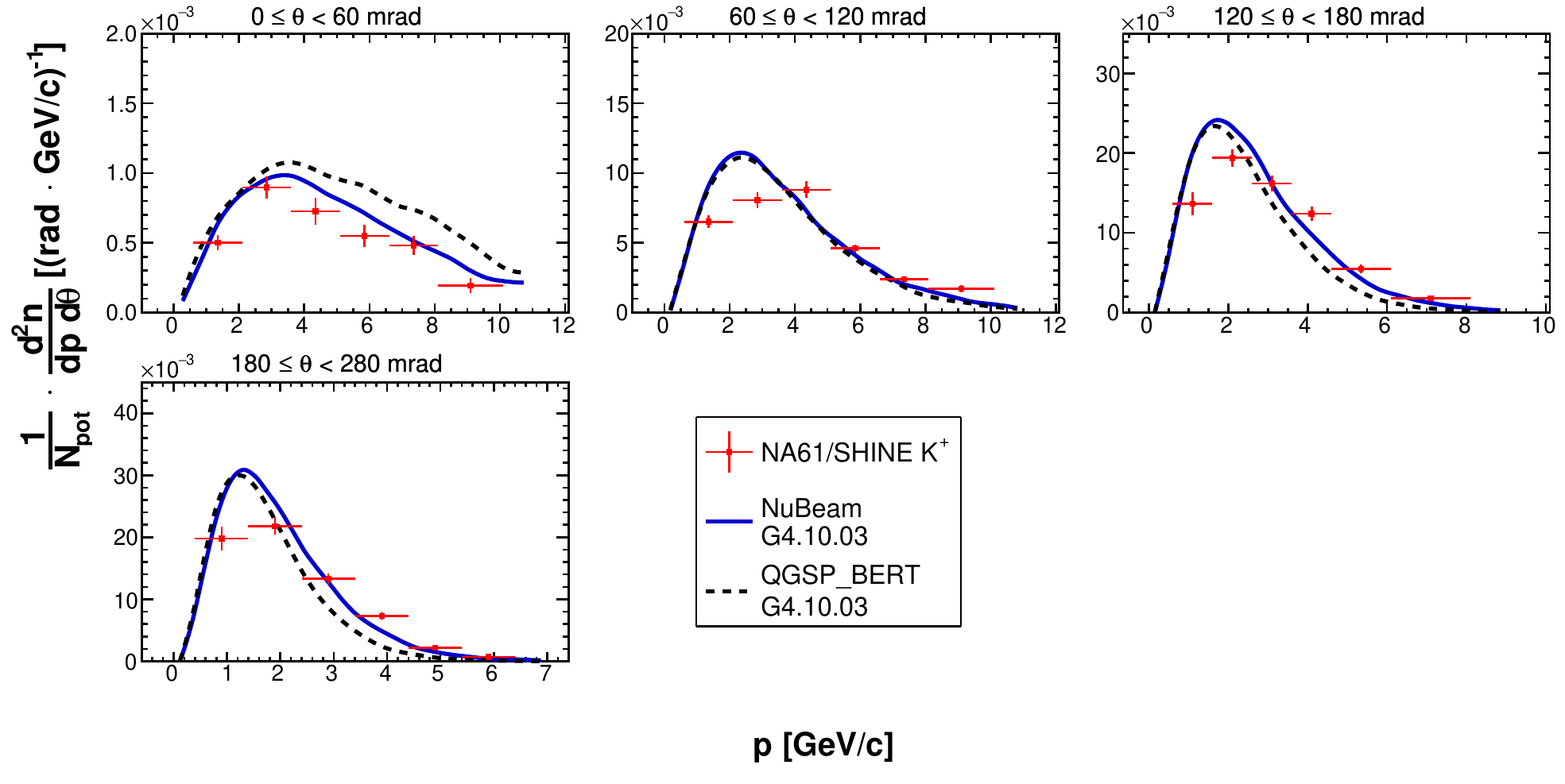}
    \caption{Double differential yields of positively charged kaons for the most upstream longitudinal bin ($0\leq z < 18\:$\cm). Vertical bars represent the total uncertanties. Predictions from the \NuBeam (solid blue line) and \QGSP (dashed black line) physics lists from \GeantFT.03~\cite{GEANT4, GEANT4bis} are overlaid on top of the data.}\label{fig:Kplus1}
    \end{center}
\end{figure*}

\begin{figure*}[ht]
    \begin{center}
    \includegraphics[width=1\textwidth]{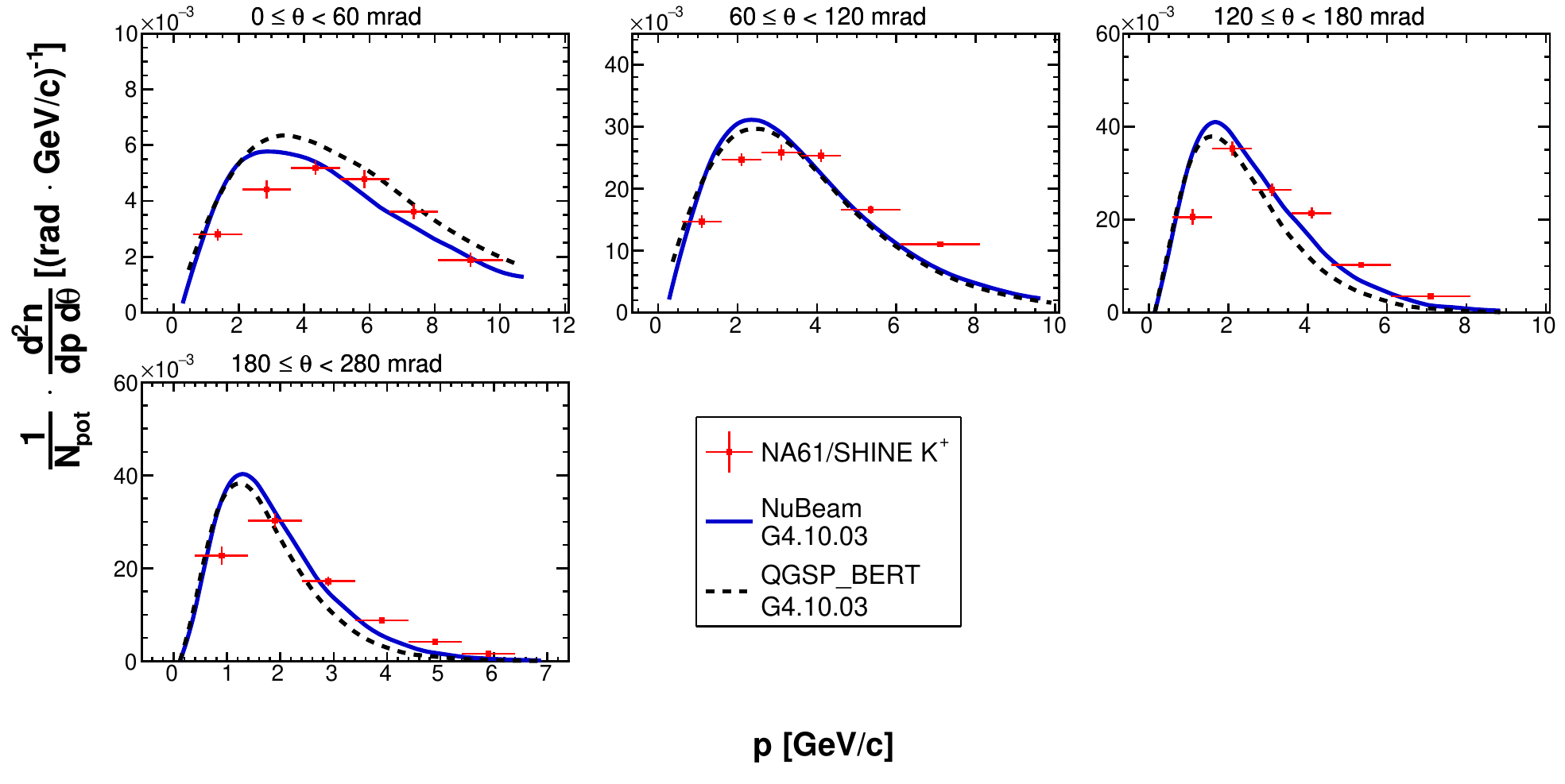}
    \caption{Double differential yields of positively charged kaons for the second upstream longitudinal bin ($18\leq z < 36\:$\cm). Vertical bars represent the total uncertanties. Predictions from the \NuBeam (solid blue line) and \QGSP (dashed black line) physics lists from \GeantFT.03~\cite{GEANT4, GEANT4bis} are overlaid on top of the data.}\label{fig:Kplus2}
    \end{center}
\end{figure*}

\begin{figure*}[ht]
    \begin{center}
    \includegraphics[width=1\textwidth]{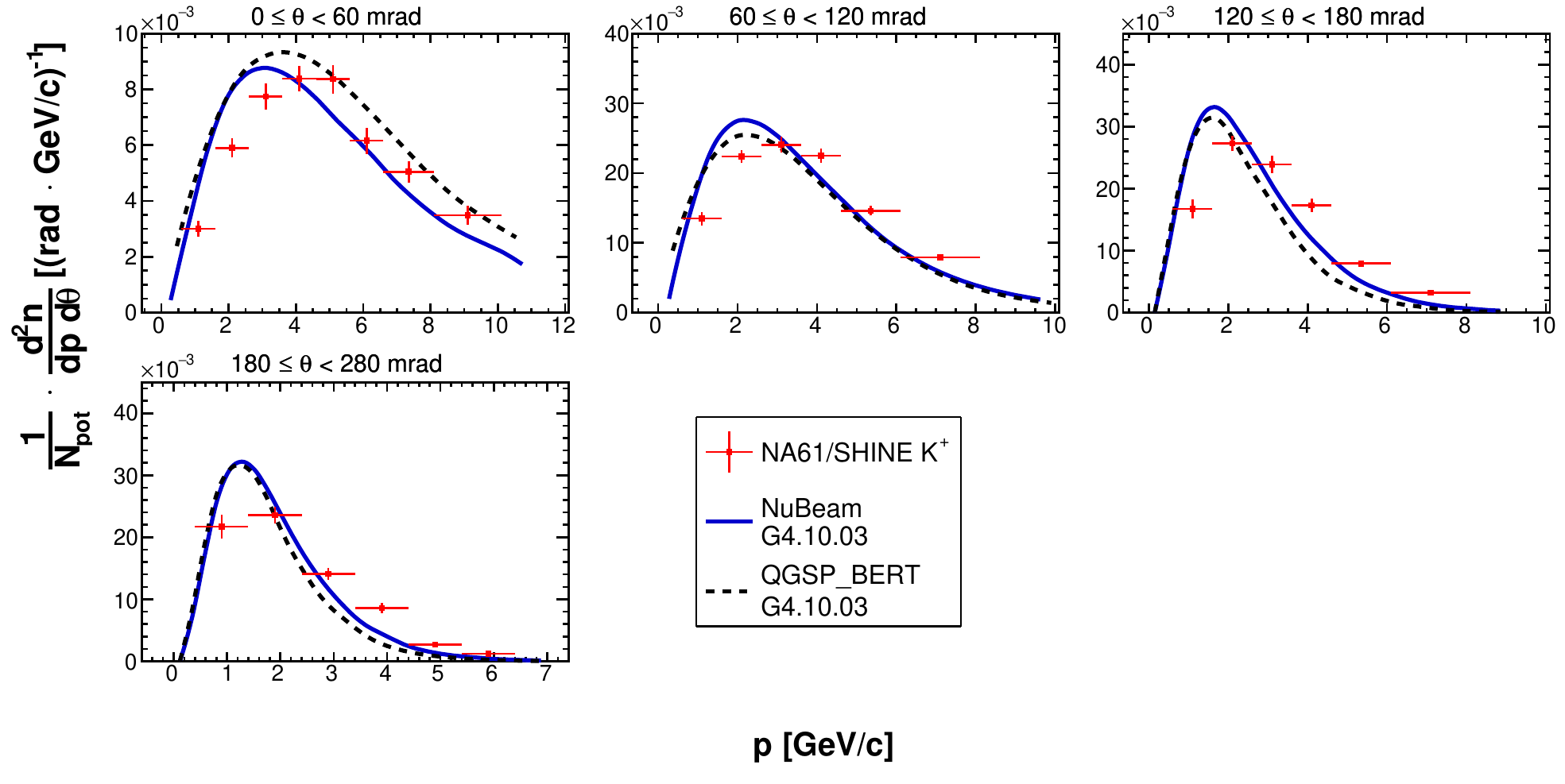}
    \caption{Double differential yields of positively charged kaons for the third upstream longitudinal bin ($36\leq z < 54\:$\cm). Vertical bars represent the total uncertanties. Predictions from the \NuBeam (solid blue line) and \QGSP (dashed black line) physics lists from \GeantFT.03~\cite{GEANT4, GEANT4bis} are overlaid on top of the data.}\label{fig:Kplus3}
    \end{center}
\end{figure*}

\begin{figure*}[ht]
    \begin{center}
    \includegraphics[width=1\textwidth]{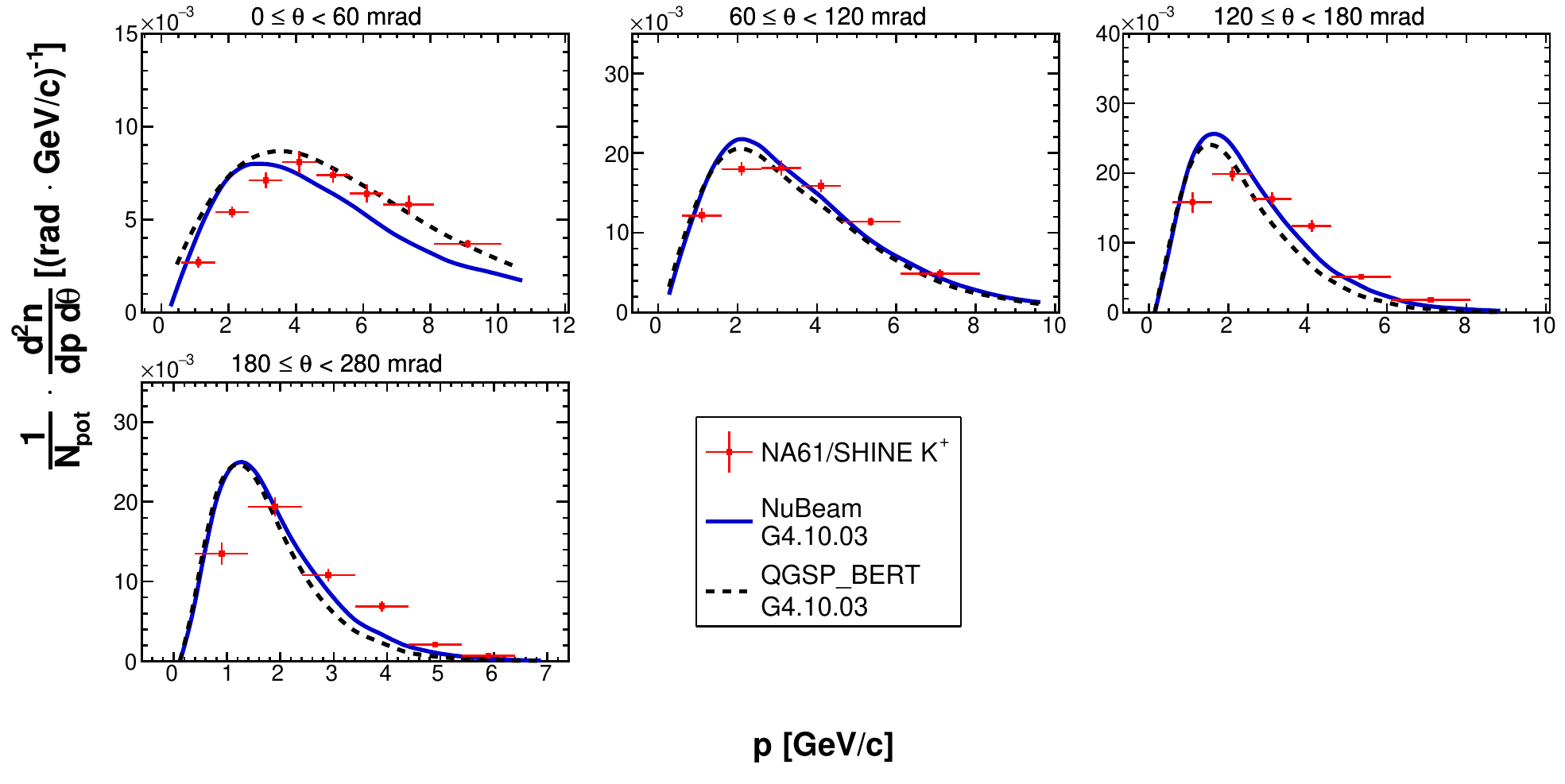}
    \caption{Double differential yields of positively charged kaons for the fourth upstream longitudinal bin ($54\leq z < 72\:$\cm). Vertical bars represent the total uncertanties. Predictions from the \NuBeam (solid blue line) and \QGSP (dashed black line) physics lists from \GeantFT.03~\cite{GEANT4, GEANT4bis} are overlaid on top of the data.}\label{fig:Kplus4}
    \end{center}
\end{figure*}

\begin{figure*}[ht]
    \begin{center}
    \includegraphics[width=1\textwidth]{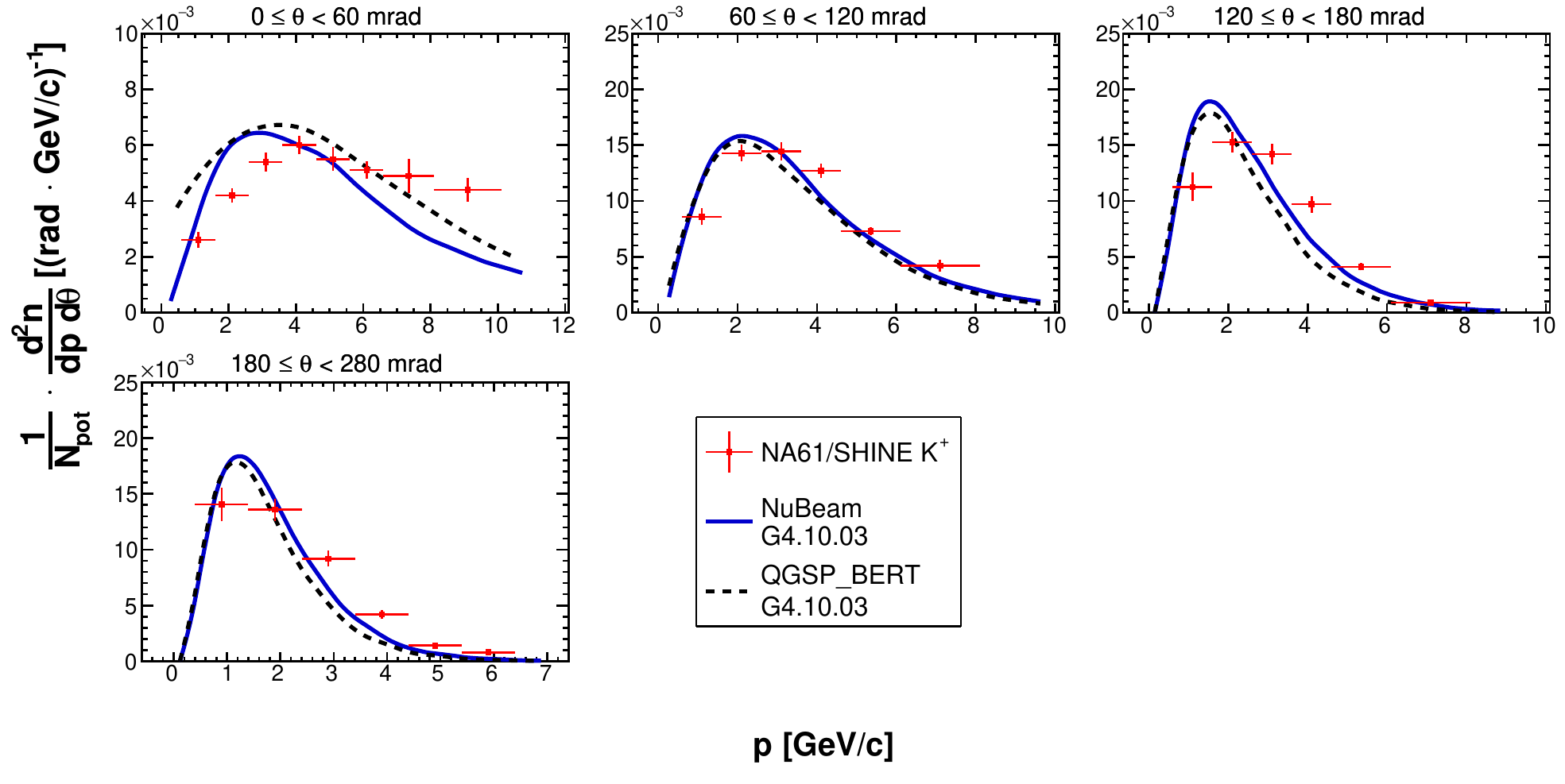}
    \caption{Double differential yields of positively charged kaons for the fifth upstream longitudinal bin ($72\leq z < 90\:$\cm). Vertical bars represent the total uncertanties. Predictions from the \NuBeam (solid blue line) and \QGSP (dashed black line) physics lists from \GeantFT.03~\cite{GEANT4, GEANT4bis} are overlaid on top of the data.}\label{fig:Kplus5}
    \end{center}
\end{figure*}

\begin{figure*}[ht]
    \begin{center}
    \includegraphics[width=1\textwidth]{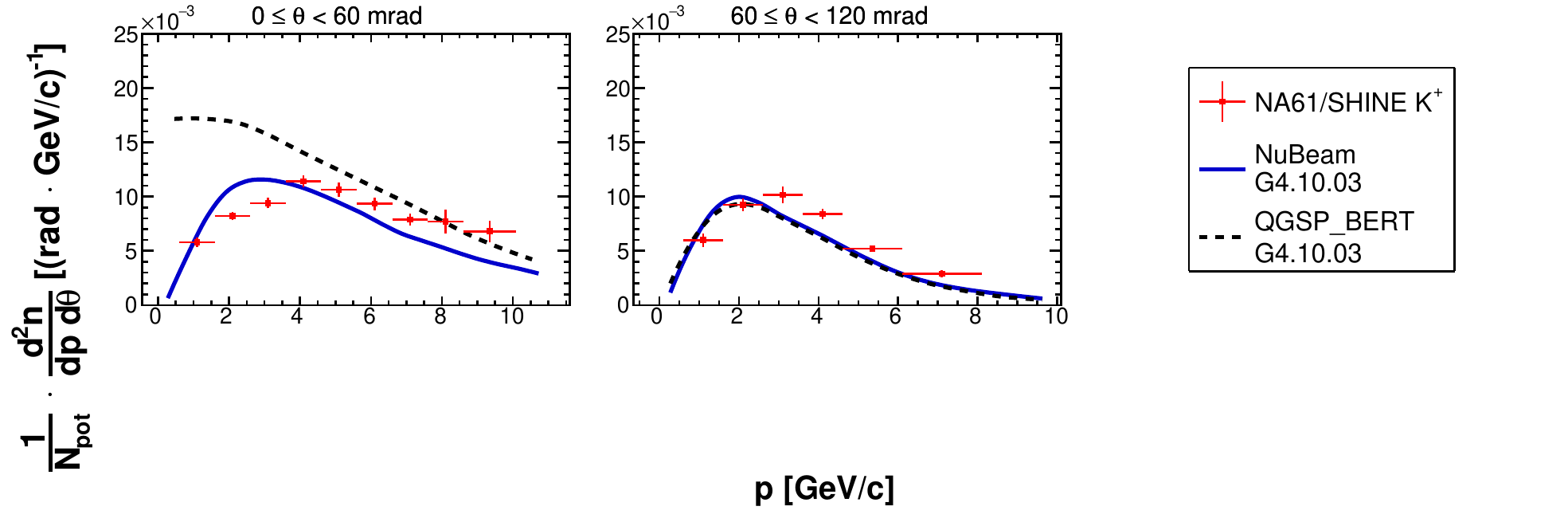}
    \caption{Double differential yields of positively charged kaons for the downstream target face ($z=90\:$\cm). Vertical bars represent the total uncertanties. Predictions from the \NuBeam (solid blue line) and \QGSP (dashed black line) physics lists from \GeantFT.03~\cite{GEANT4, GEANT4bis} are overlaid on top of the data.}\label{fig:Kplus6}
    \end{center}
\end{figure*}

\begin{figure*}[ht]
    \begin{center}
    \includegraphics[width=1\textwidth]{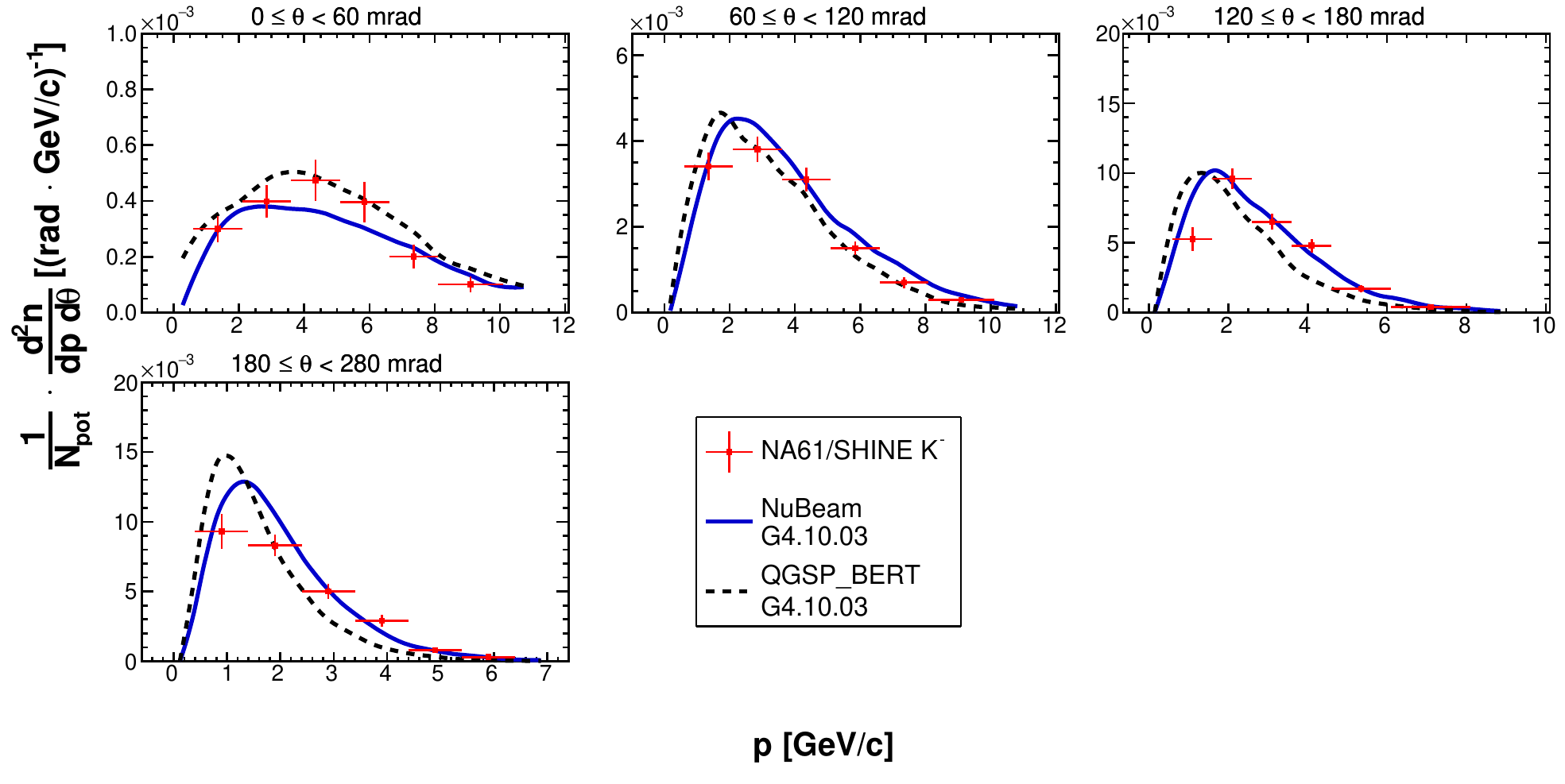}
    \caption{Double differential yields of negatively charged kaons for the most upstream longitudinal bin ($0\leq z < 18\:$\cm). Vertical bars represent the total uncertanties. Predictions from the \NuBeam (solid blue line) and \QGSP (dashed black line) physics lists from \GeantFT.03~\cite{GEANT4, GEANT4bis} are overlaid on top of the data.}\label{fig:KMinus1}
    \end{center}
\end{figure*}

\begin{figure*}[ht]
    \begin{center}
    \includegraphics[width=1\textwidth]{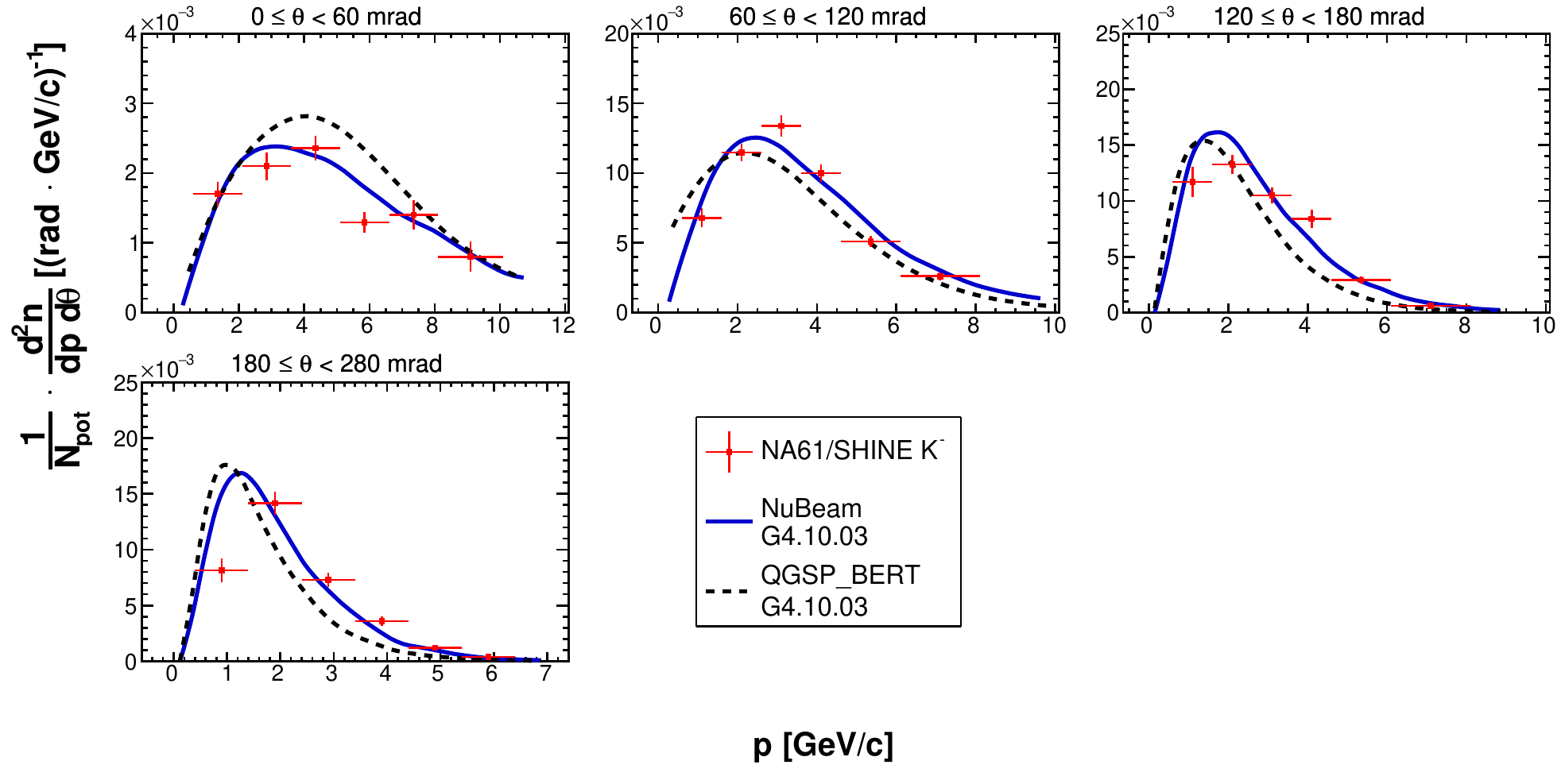}
    \caption{Double differential yields of negatively charged kaons for the second upstream longitudinal bin ($18\leq z < 36\:$\cm). Vertical bars represent the total uncertanties. Predictions from the \NuBeam (solid blue line) and \QGSP (dashed black line) physics lists from \GeantFT.03~\cite{GEANT4, GEANT4bis} are overlaid on top of the data.}\label{fig:KMinus2}
    \end{center}
\end{figure*}

\begin{figure*}[ht]
    \begin{center}
    \includegraphics[width=1\textwidth]{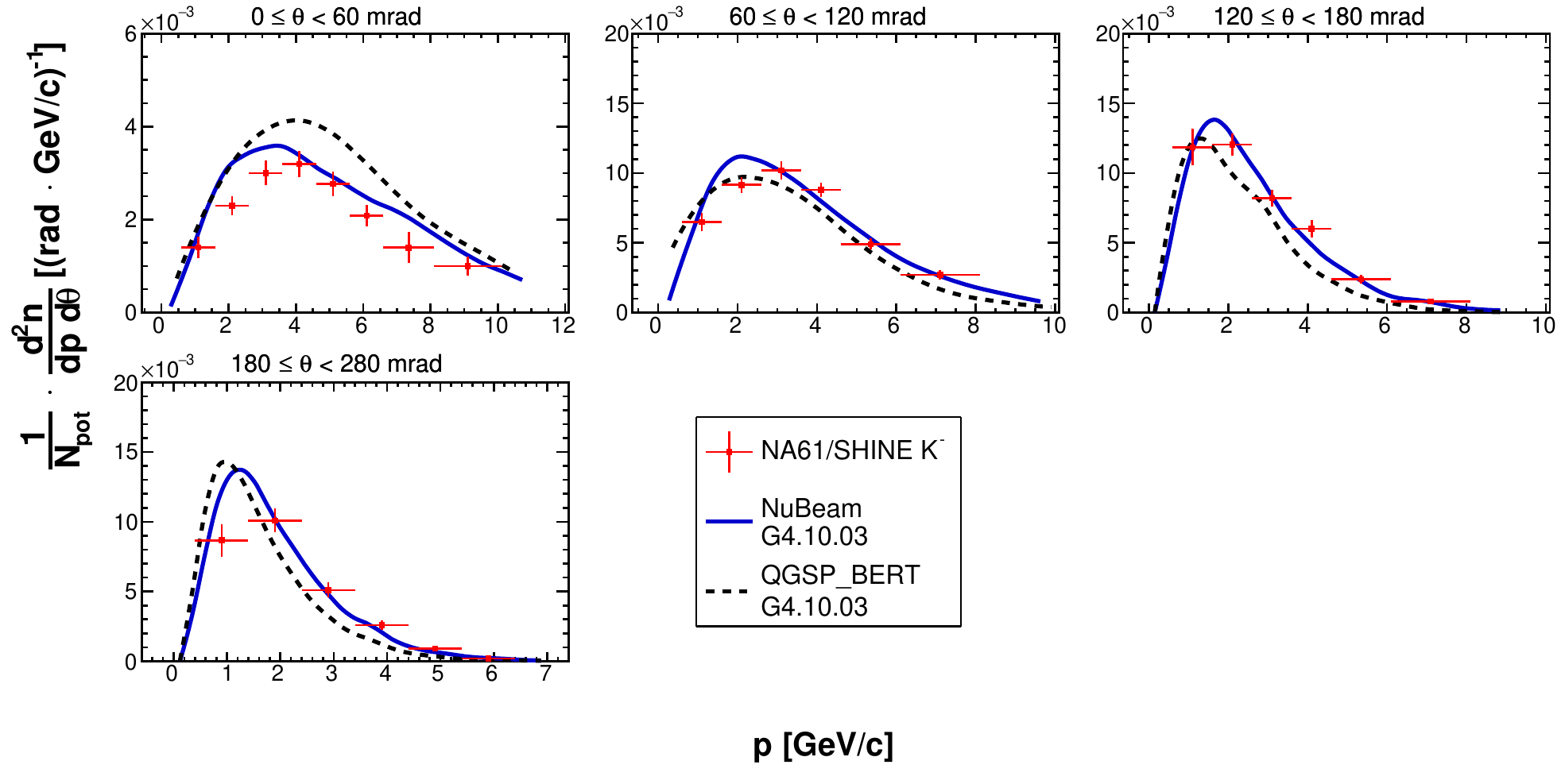}
    \caption{Double differential yields of negatively charged kaons for the third upstream longitudinal bin ($36\leq z < 54\:$\cm). Vertical bars represent the total uncertanties. Predictions from the \NuBeam (solid blue line) and \QGSP (dashed black line) physics lists from \GeantFT.03~\cite{GEANT4, GEANT4bis} are overlaid on top of the data.}\label{fig:KMinus3}
    \end{center}
\end{figure*}

\begin{figure*}[ht]
    \begin{center}
    \includegraphics[width=1\textwidth]{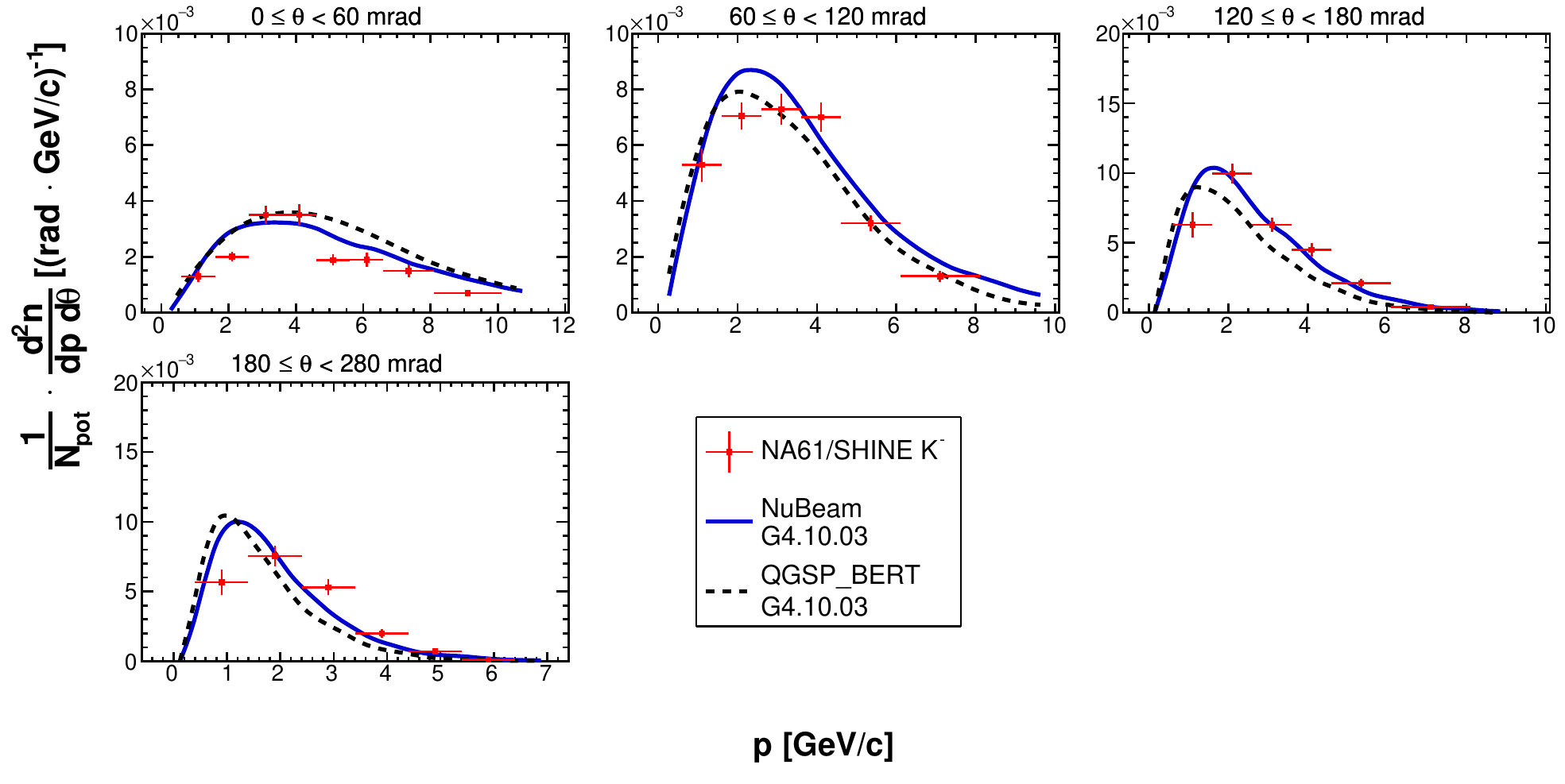}
    \caption{Double differential yields of negatively charged kaons for the fourth upstream longitudinal bin ($54\leq z < 72\:$\cm). Vertical bars represent the total uncertanties. Predictions from the \NuBeam (solid blue line) and \QGSP (dashed black line) physics lists from \GeantFT.03~\cite{GEANT4, GEANT4bis} are overlaid on top of the data.}\label{fig:KMinus4}
    \end{center}
\end{figure*}

\begin{figure*}[ht]
    \begin{center}
    \includegraphics[width=1\textwidth]{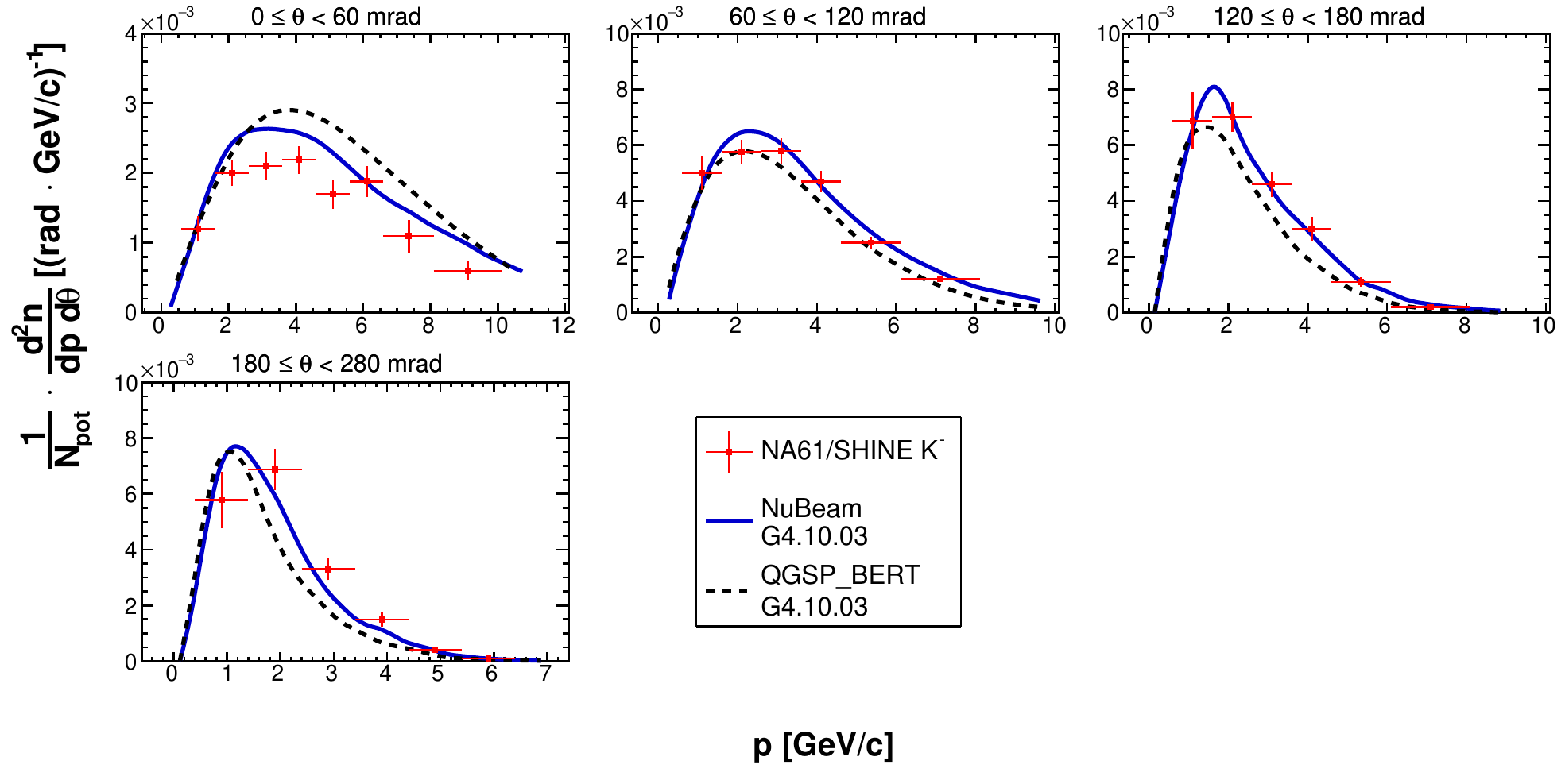}
    \caption{Double differential yields of negatively charged kaons for the fifth upstream longitudinal bin ($72\leq z < 90\:$\cm). Vertical bars represent the total uncertanties. Predictions from the \NuBeam (solid blue line) and \QGSP (dashed black line) physics lists from \GeantFT.03~\cite{GEANT4, GEANT4bis} are overlaid on top of the data.}\label{fig:KMinus5}
    \end{center}
\end{figure*}

\begin{figure*}[ht]
    \begin{center}
    \includegraphics[width=1\textwidth]{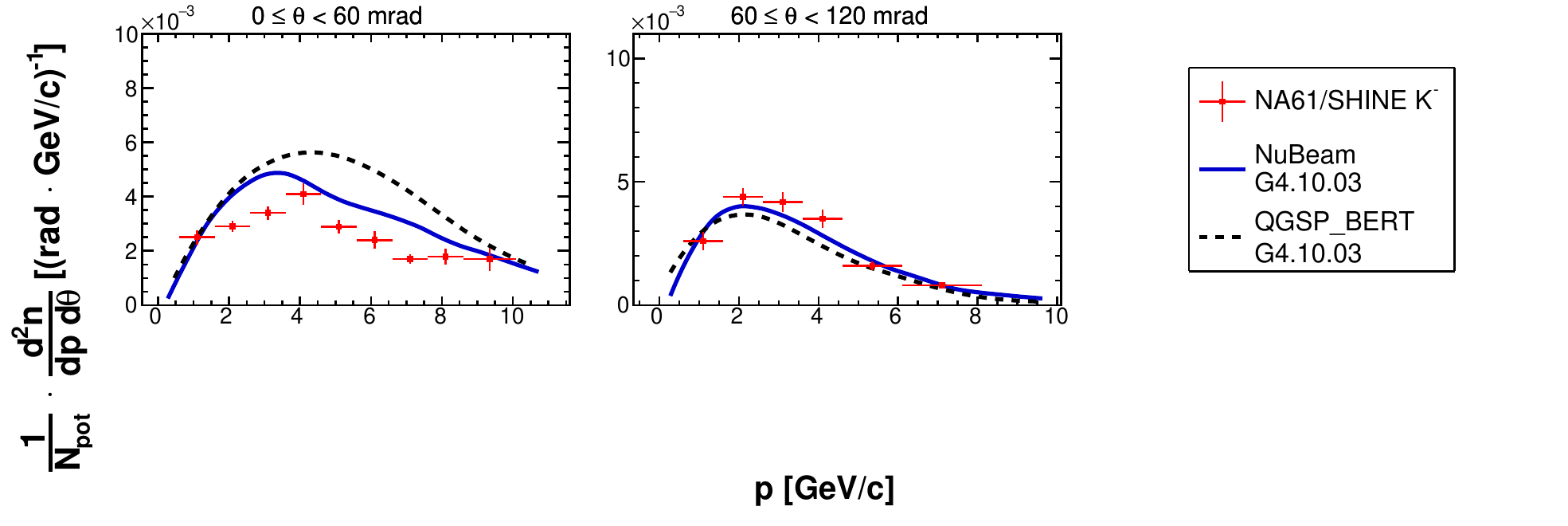}
    \caption{Double differential yields of negatively charged kaons for the downstream target face ($z=90\:$\cm). Vertical bars represent the total uncertanties. Predictions from the \NuBeam (solid blue line) and \QGSP (dashed black line) physics lists from \GeantFT.03~\cite{GEANT4, GEANT4bis} are overlaid on top of the data.}\label{fig:KMinus6}
    \end{center}
\end{figure*}

\begin{figure*}[ht]
    \begin{center}
    \includegraphics[width=1\textwidth]{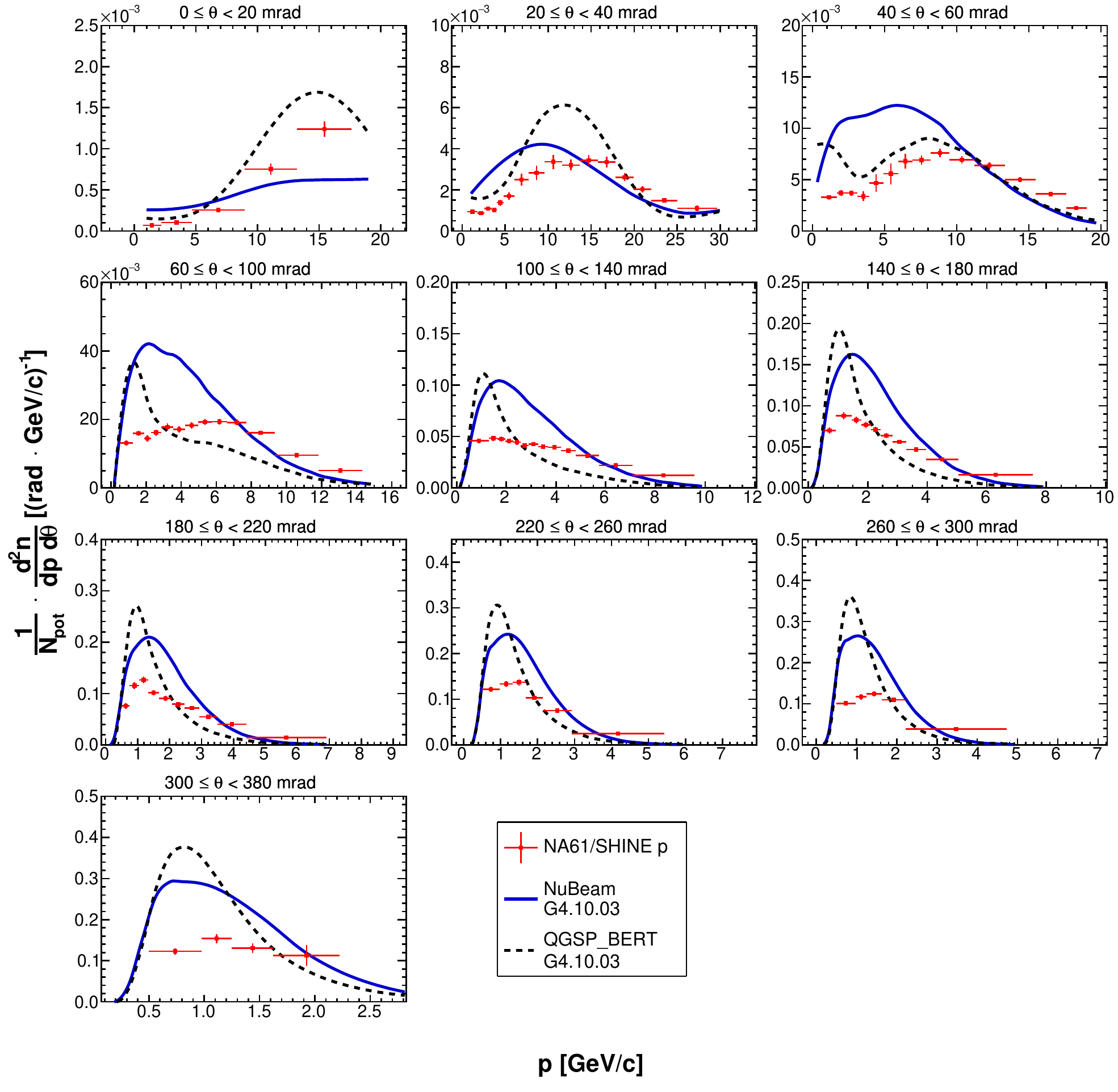}
    \caption{Double differential yields of protons for the most upstream longitudinal bin ($0\leq z < 18\:$\cm). Vertical bars represent the total uncertanties. Predictions from the \NuBeam (solid blue line) and \QGSP (dashed black line) physics lists from \GeantFT.03~\cite{GEANT4, GEANT4bis} are overlaid on top of the data.}\label{fig:proton1}
    \end{center}
\end{figure*}

\begin{figure*}[ht]
    \begin{center}
    \includegraphics[width=1\textwidth]{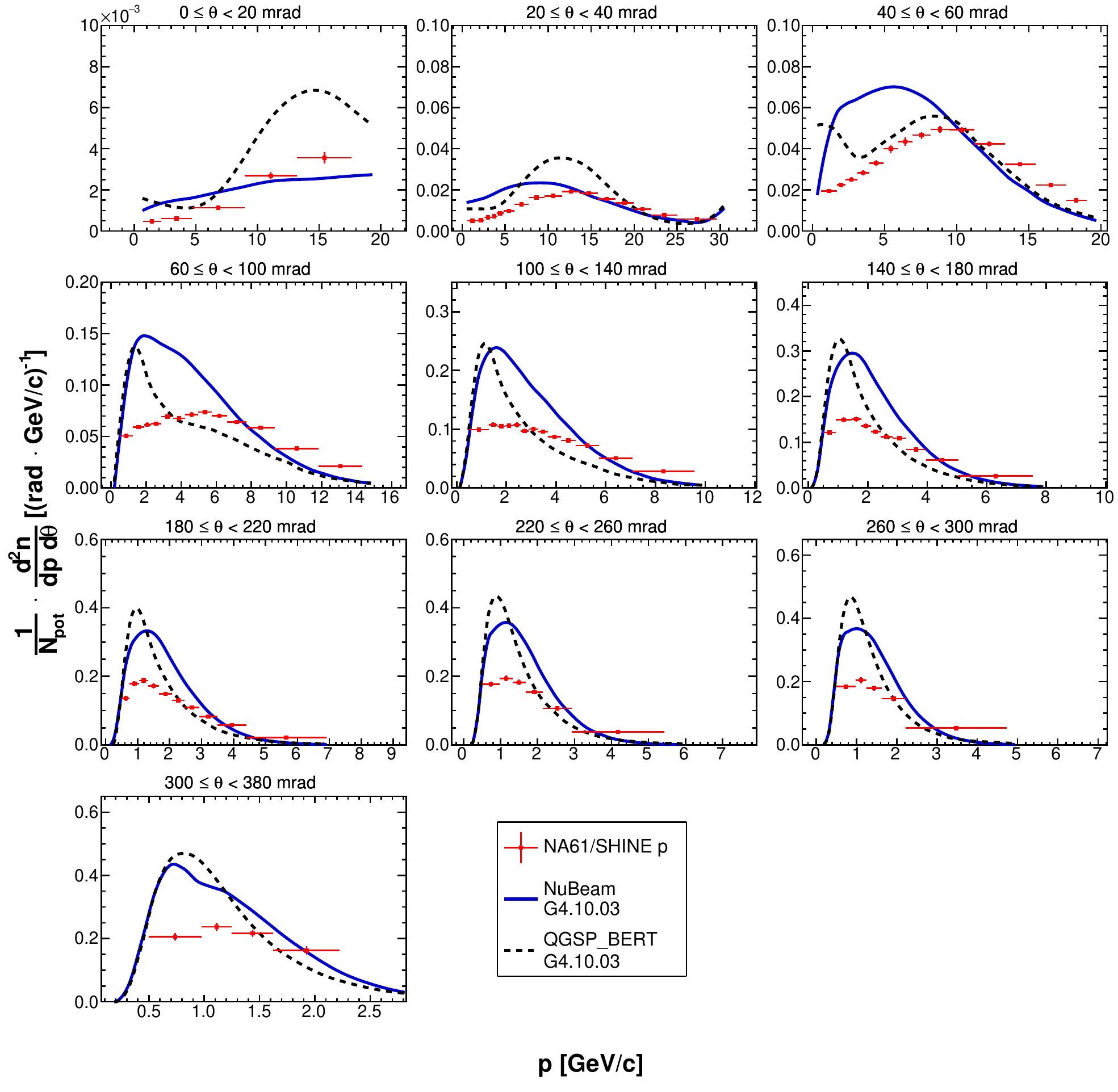}
    \caption{Double differential yields of protons for the second upstream longitudinal bin ($18\leq z < 36\:$\cm). Vertical bars represent the total uncertanties. Predictions from the \NuBeam (solid blue line) and \QGSP (dashed black line) physics lists from \GeantFT.03~\cite{GEANT4, GEANT4bis} are overlaid on top of the data.}\label{fig:proton2}
    \end{center}
\end{figure*}

\begin{figure*}[ht]
    \begin{center}
    \includegraphics[width=1\textwidth]{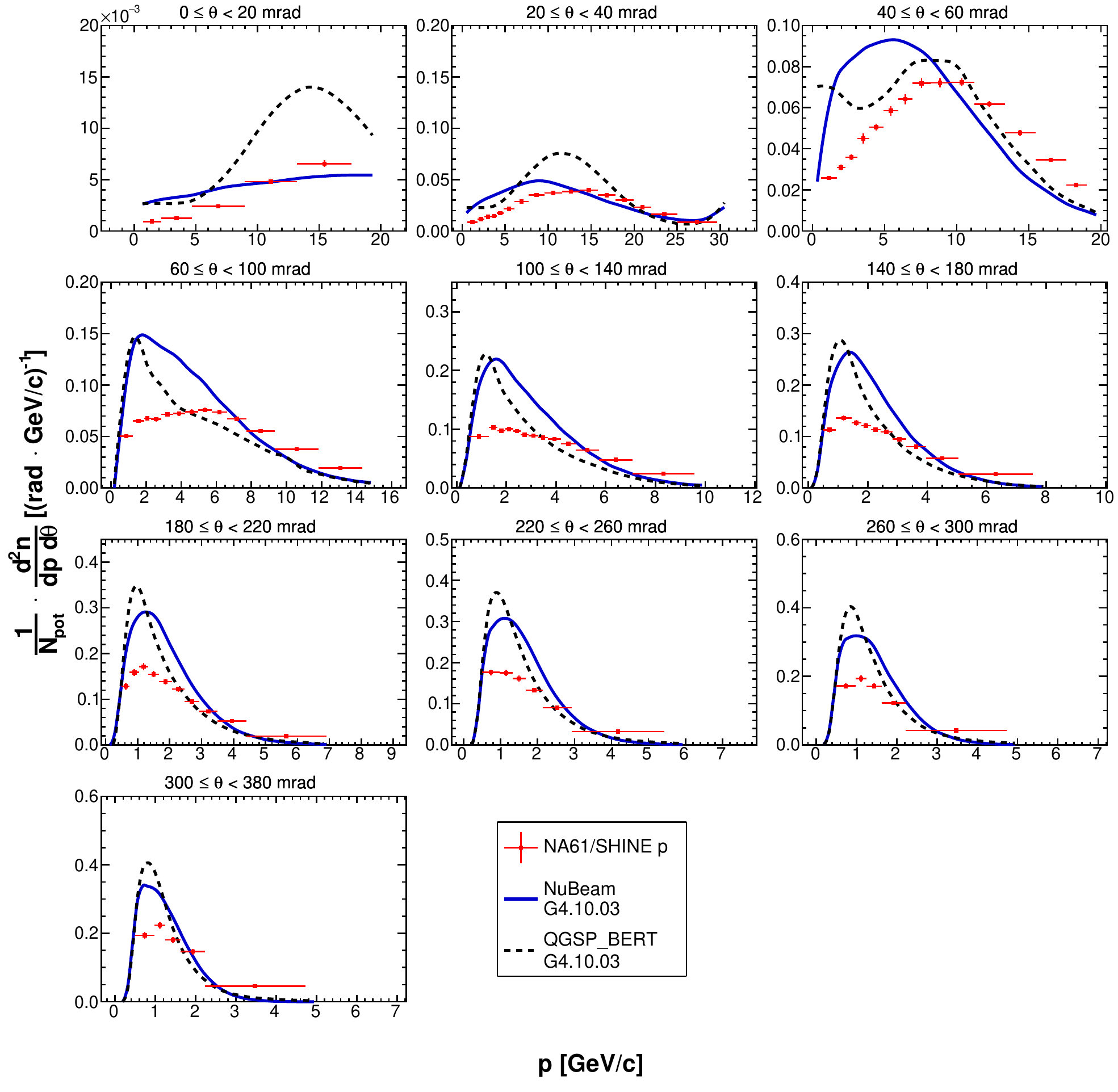}
    \caption{Double differential yields of protons kaons for the third upstream longitudinal bin ($36\leq z < 54\:$\cm). Vertical bars represent the total uncertanties. Predictions from the \NuBeam (solid blue line) and \QGSP (dashed black line) physics lists from \GeantFT.03~\cite{GEANT4, GEANT4bis} are overlaid on top of the data.}\label{fig:proton3}
    \end{center}
\end{figure*}

\begin{figure*}[ht]
    \begin{center}
    \includegraphics[width=1\textwidth]{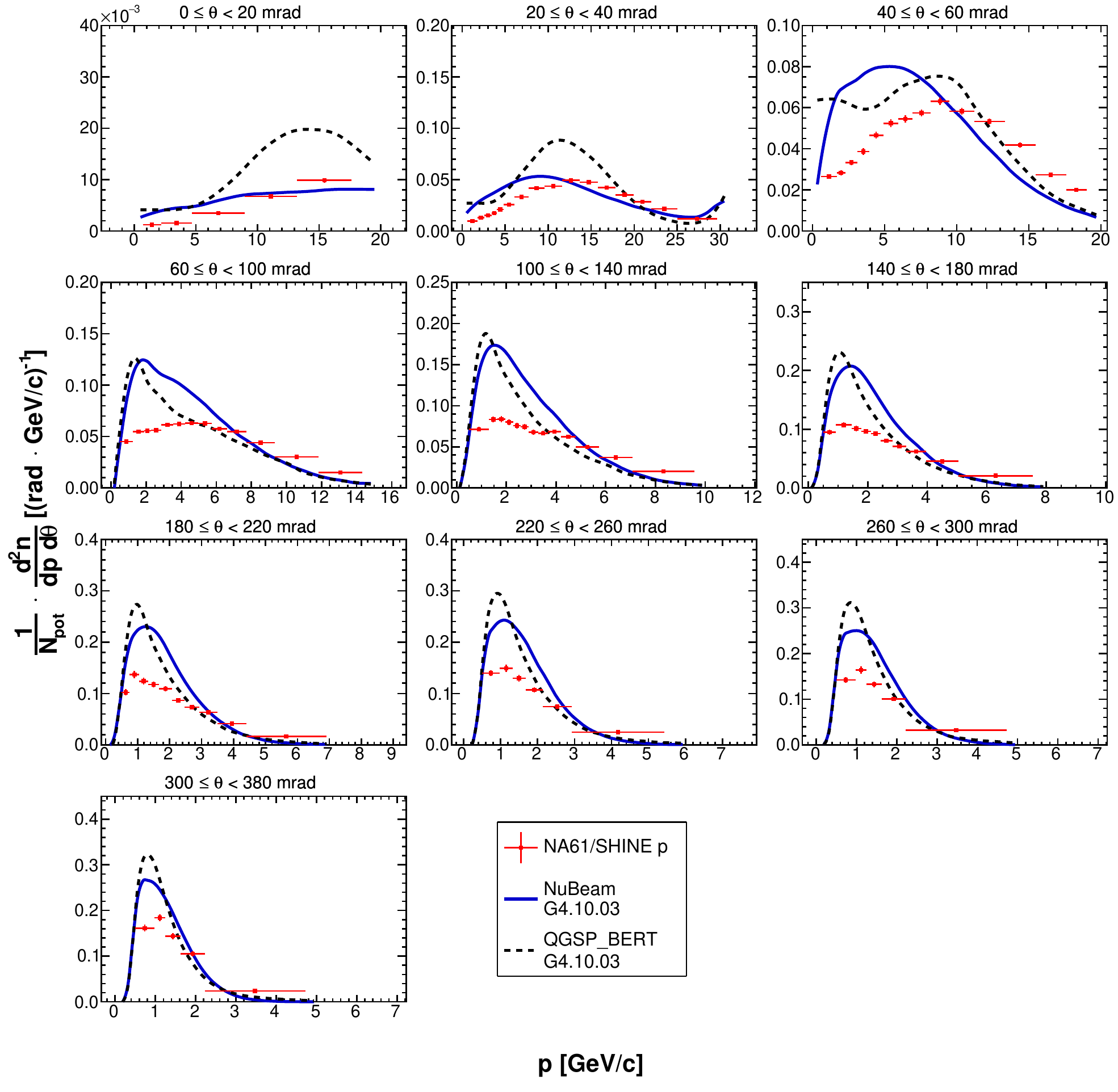}
    \caption{Double differential yields of protons for the fourth upstream longitudinal bin ($54\leq z < 72\:$\cm). Vertical bars represent the total uncertanties. Predictions from the \NuBeam (solid blue line) and \QGSP (dashed black line) physics lists from \GeantFT.03~\cite{GEANT4, GEANT4bis} are overlaid on top of the data.}\label{fig:proton4}
    \end{center}
\end{figure*}

\begin{figure*}[ht]
    \begin{center}
    \includegraphics[width=1\textwidth]{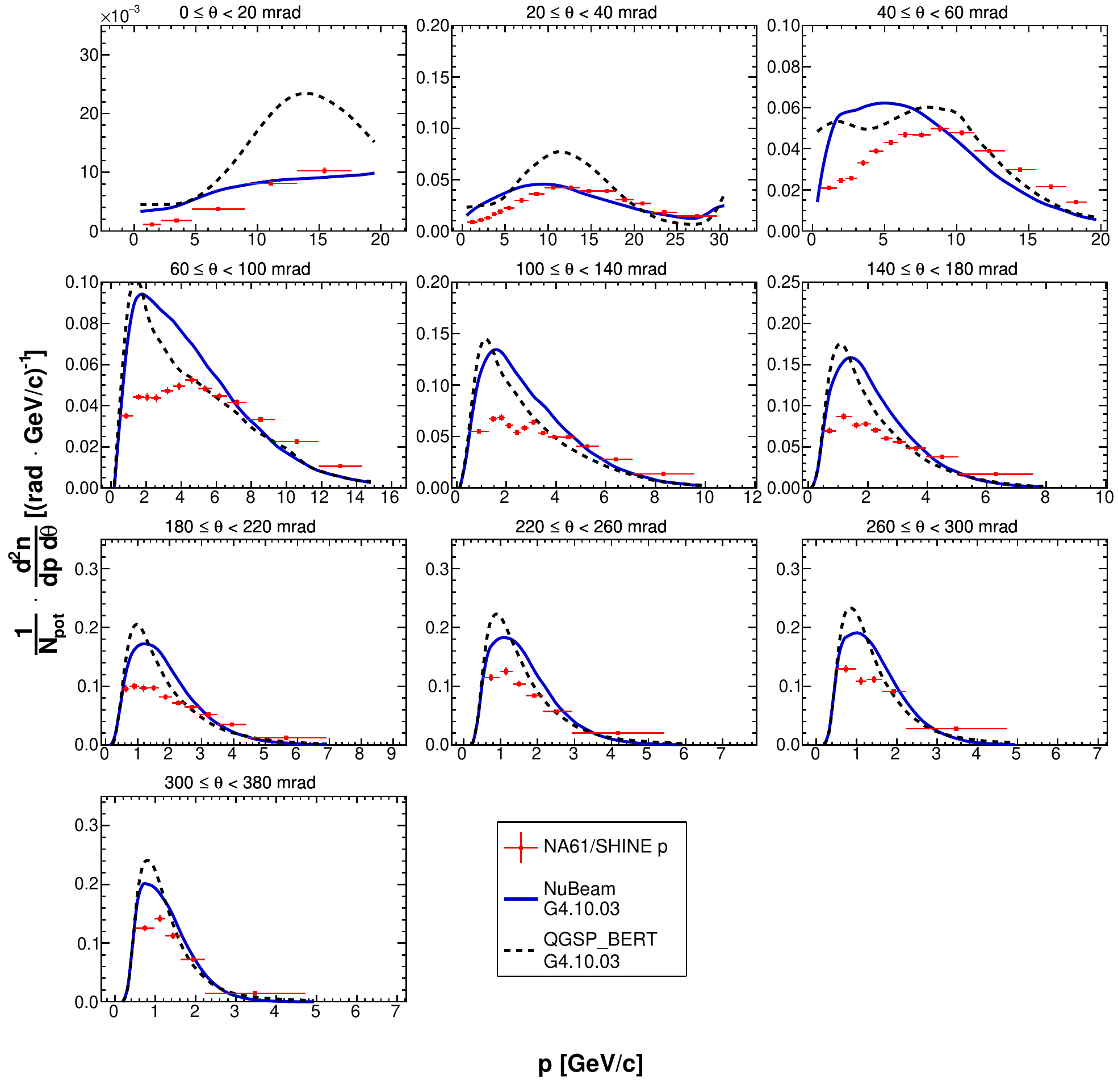}
    \caption{Double differential yields of protons for the fifth upstream longitudinal bin ($72\leq z < 90\:$\cm). Vertical bars represent the total uncertanties. Predictions from the \NuBeam (solid blue line) and \QGSP (dashed black line) physics lists from \GeantFT.03~\cite{GEANT4, GEANT4bis} are overlaid on top of the data.}\label{fig:proton5}
    \end{center}
\end{figure*}

\begin{figure*}[ht]
    \begin{center}
    \includegraphics[width=1\textwidth]{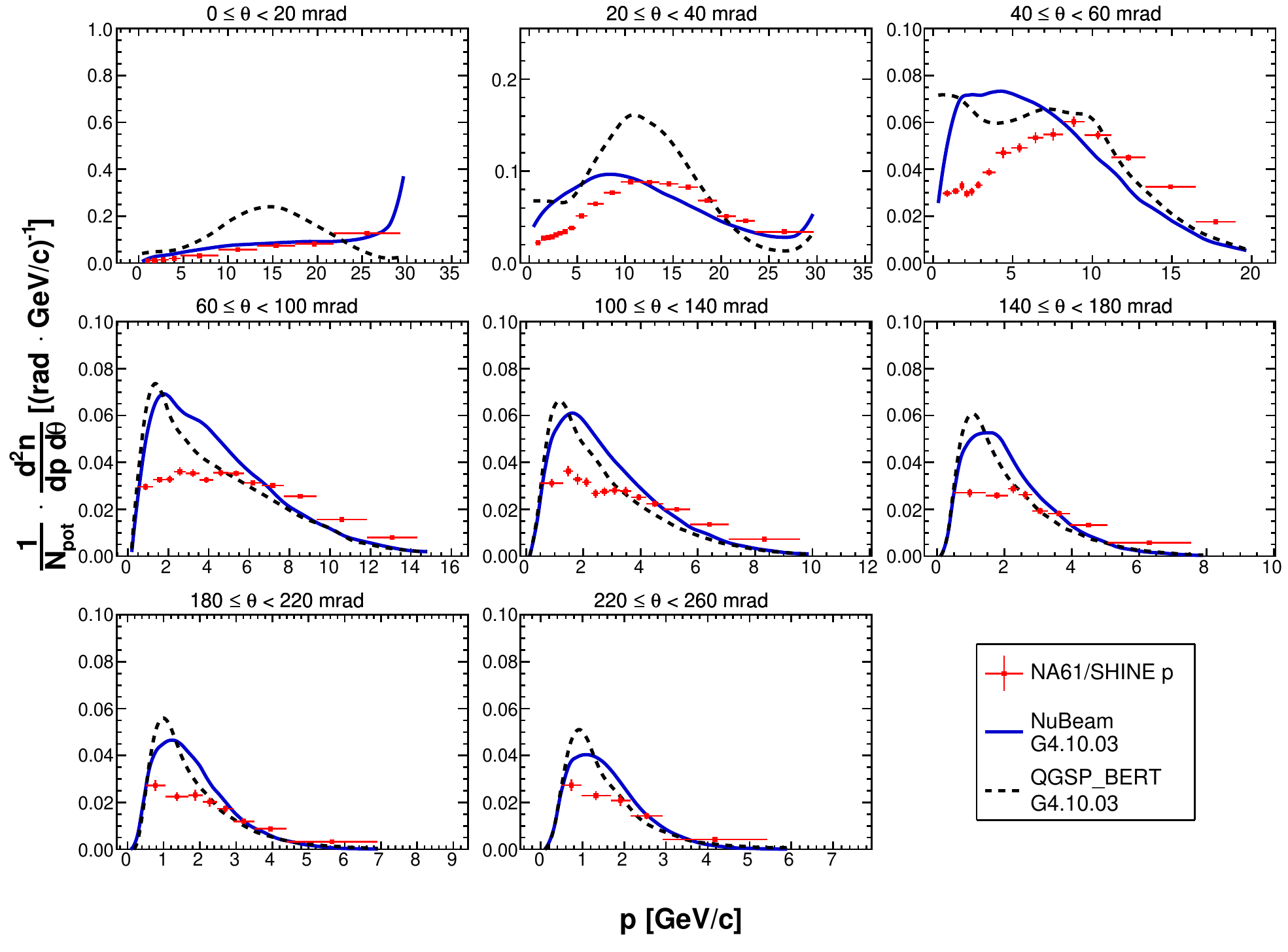}
    \caption{Double differential yields of protons for the downstream target face ($z=90\:$\cm). Vertical bars represent the total uncertanties. Predictions from the \NuBeam (solid blue line) and \QGSP (dashed black line) physics lists from \GeantFT.03~\cite{GEANT4, GEANT4bis} are overlaid on top of the data.}\label{fig:proton6}
    \end{center}
\end{figure*}

The results were compared with \NuBeam and \QGSP physics lists from \GeantFT.03~\cite{GEANT4, GEANT4bis}. Detailed comparisons are presented in Figs.~\ref{fig:piPlus1} -~\ref{fig:piPlus6} for \pip yields, in Figs.~\ref{fig:piMinus1} -~\ref{fig:piMinus6} for \pim yields, in Figs.~\ref{fig:Kplus1} -~\ref{fig:Kplus6} for \kp yields, in Figs.~\ref{fig:KMinus1} -~\ref{fig:KMinus6} for \km yields, and in Figs.~\ref{fig:proton1} -~\ref{fig:proton6} for \prot yields. The \NuBeam physics list provides better predictions of \pip, \pim, \kp, and \km yields. However, both physics lists fail to accurately predict \prot yields. Similar behaviour was noticed in the previous \NASixtyOne measurements of primary proton-carbon interactions at $31\:$\GeVc~\cite{thin2009paper}, where it was shown that most of the models poorly reproduce proton yields. 
The earlier observation~\cite{LTpaper2009} that a better agreement
with \FlukaEleven predictions can be obtained by lowering the production
cross section in \FlukaEleven is confirmed.

\subsection{Comparison with previous \NASixtyOne measurements}
\begin{figure*}[ht]
	\begin{center}
  \begin{subfigure}[t]{1\textwidth}
        \includegraphics[trim= 0 0 0 0, clip, width=1\textwidth]{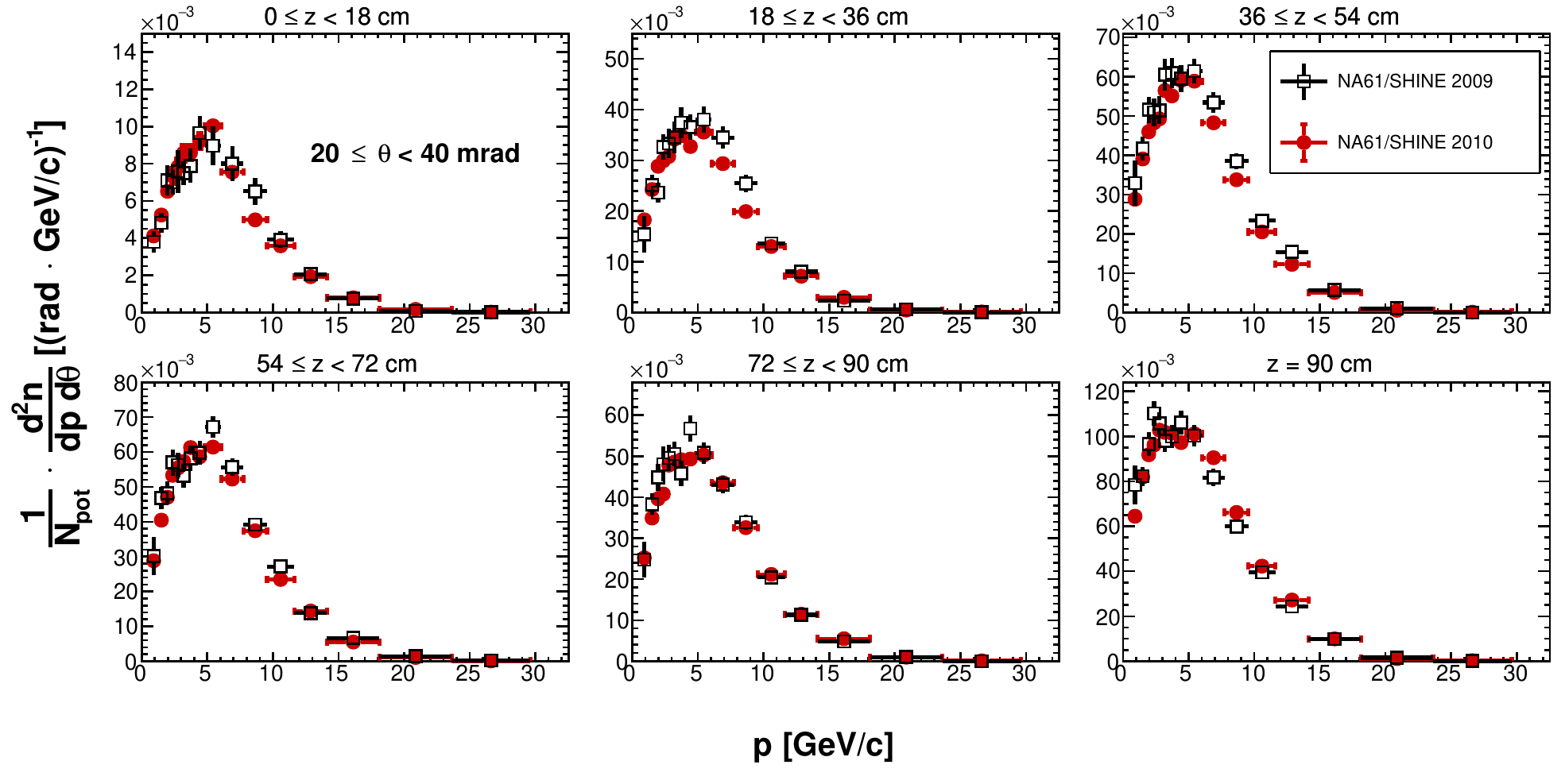}
    \caption{}
  \end{subfigure}	
  \begin{subfigure}[t]{1\textwidth}
        \includegraphics[trim= 0 0 0 0, clip, width=1\textwidth]{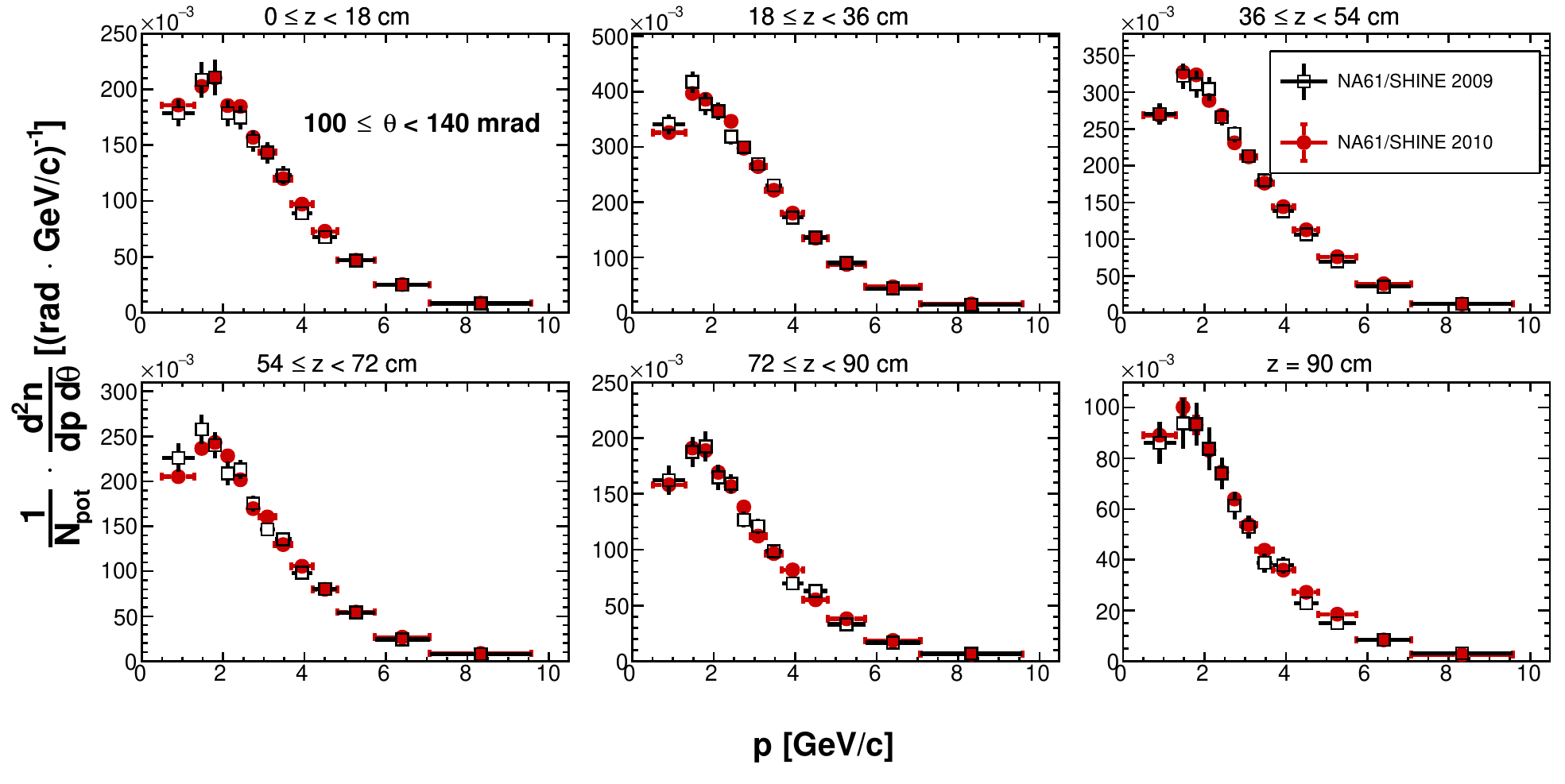}
    \caption{}
  \end{subfigure}	

  \end{center}
	\caption{Double differential yields of positively charged pions as a function of momentum for the polar angle between $20\:$\mrad and $40\:$\mrad (a) and between $100\:$\mrad and $140\:$\mrad (b). Each panel shows a different longitudinal bin. The 2009 measurements are shown in black, while the 2010 data are shown in red.}\label{fig:2009comppip}
\end{figure*}

\begin{figure*}[ht]
	\begin{center}
  \begin{subfigure}[t]{1\textwidth}
        \includegraphics[trim= 0 0 0 0, clip, width=1\textwidth]{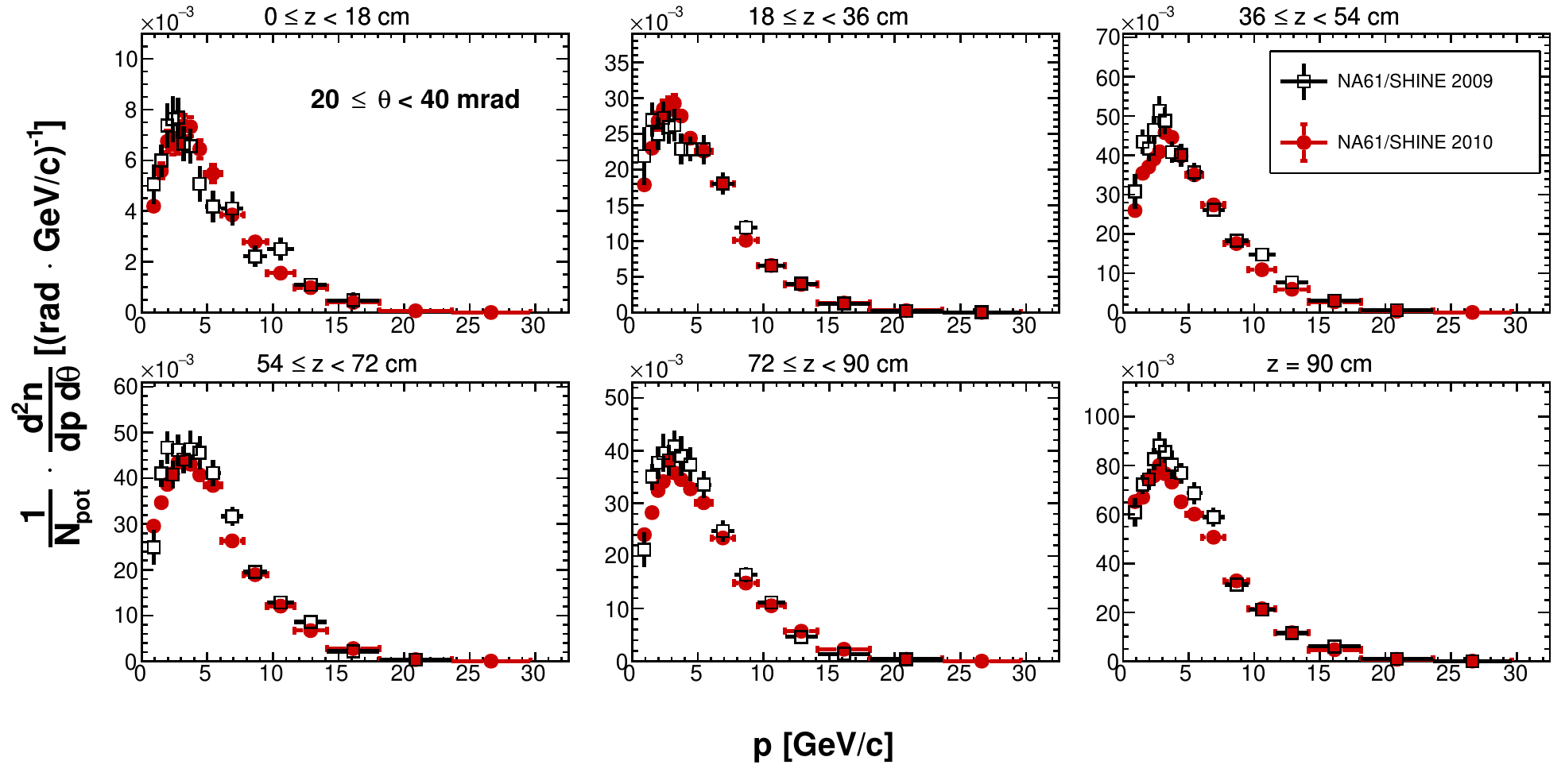}
    \caption{}
  \end{subfigure}	
  \begin{subfigure}[t]{1\textwidth}
        \includegraphics[trim= 0 0 0 0, clip, width=1\textwidth]{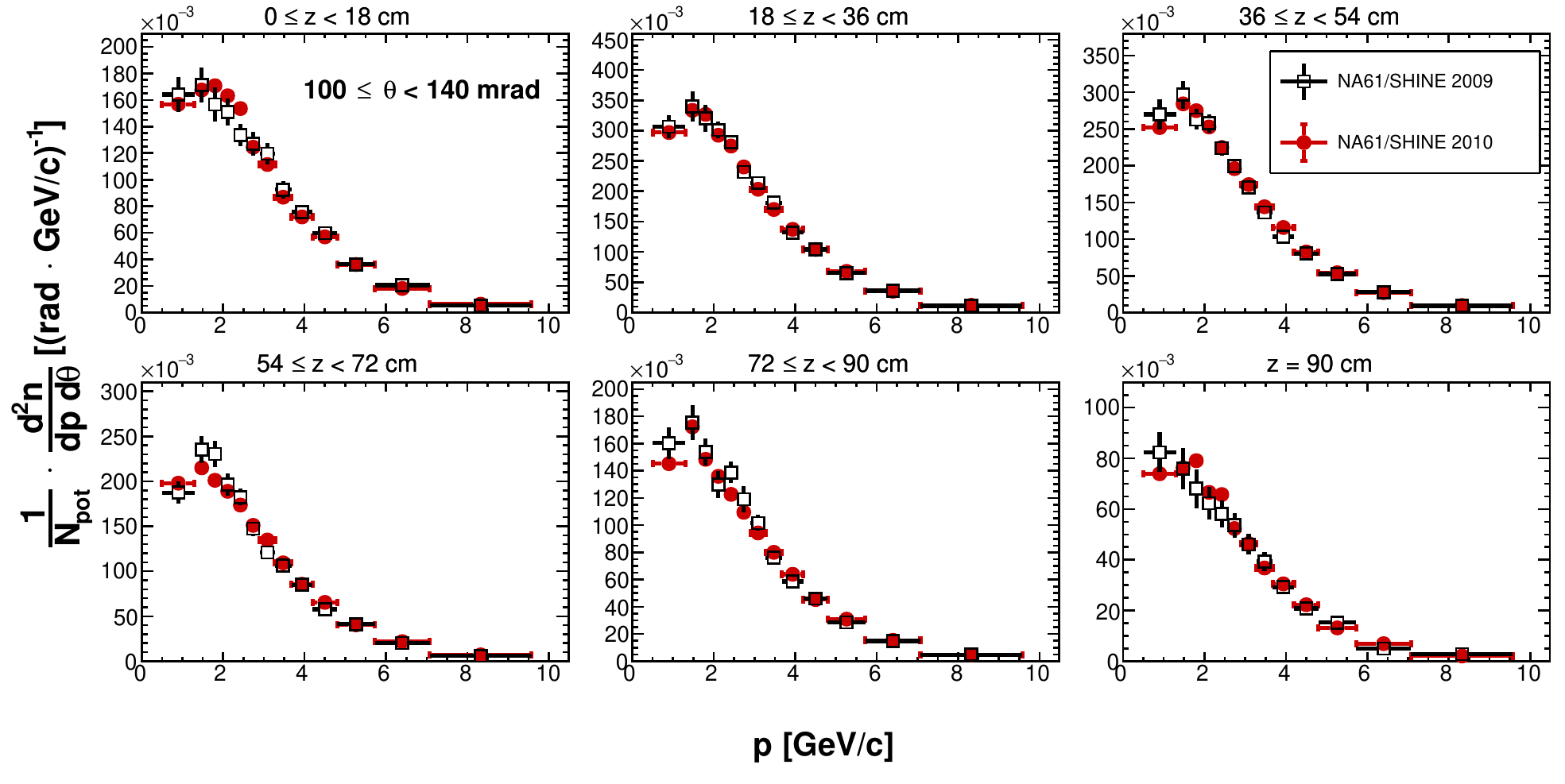}
    \caption{}
  \end{subfigure}	

  \end{center}
	\caption{Double differential yields of negatively charged pions as a function of momentum for the polar angle between $20\:$\mrad and $40\:$\mrad (a) and between $100\:$\mrad and $140\:$\mrad (b). Each panel shows a different longitudinal bin. The 2009 measurements are shown in black, while the 2010 data are shown in red.}\label{fig:2009comppim}
\end{figure*}

As previously mentioned, \NASixtyOne provided measurements
of the $\pi^\pm$ yields emitted from the T2K replica target using data
collected in 2009~\cite{LTpaper2009}.
Since the beam profiles for the data taken in 2009 and 2010 are different, a direct comparison between the 2009 measurements and the measurements presented in this paper is not possible. The beam profile for the 2009 measurements is wider, and the 2009 pion yields are expected to be larger for the low angle upstream longitudinal bins. The differences decrease for more downstream longitudinal bins and higher angles. But even for the downstream face of the target and high angles ($>100\:$\mrad) perfect agreement is not expected since Monte Carlo simulations show small differences in re-interactions for different beam profiles. Detailed Monte Carlo studies can be found in Ref.~\cite{PavinThesis}. However, under the assumption that major differences are only due to geometry - i.e. low angle secondaries cannot exit the target from the first longitudinal bin if the beam is very narrow, one can apply scaling to the radial beam distribution and achieve a similar radial distribution as in the previous measurement. The radial beam distribution was divided  into  $26$ bins, $0.5\:$\mm in size. A simple weight was applied to all selected tracks:

\begin{equation}
	w = \left(\frac{N_{POT, i}^{2009}}{N_{POT}^{2009}}\right)/\left(\frac{N_{POT, i}^{2010}}{N_{POT}^{2010}}\right),
\end{equation}
where $N_{POT, i}^{2009(2010)}$ is the number of beam protons hitting the target in the $i-th$ radial bin in data from $2009(2010)$ and $N_{POT}^{2009(2010)}$ is the total number of beam protons hitting the target. After scaling, the particle identification procedure was applied in the phase space binning as defined in Ref.~\cite{LTpaper2009}. Also, Monte Carlo and $tof$ correction factors were calculated, but the ad hoc correction was not applied. It is important to note that systematic uncertainties were not reevaluated for the new phase space bins. Qualitative comparisons of \pip and \pim yields can be seen in Figs.~\ref{fig:2009comppip} and~\ref{fig:2009comppim}, respectively. Only two polar angle bins are selected and presented for all longitudinal bins. A quantitative agreement between the two sets of measurements is not expected due to the simplified assumptions described above, as well as differences in the beam divergence which were not accounted for. Indeed, some differences in the \pip yields can be observed for small angles. However, in general, these two sets of measurements are consistent within about $20\%$. Note that the systematic uncertainties were not reevaluated for these special 2010 measurements and they are not shown on the plots. Additionally, only differences in the radial beam distribution were taken into account, while differences in the beam divergence were ignored.

In order to demonstrate the reliability of the computed $\pi^-$ yields, 
an independent full chain of data analysis was performed. 
By means of the software tools and procedures used in the analysis of data 
collected in 2009~\cite{LTpaper2009} the differential yields of negatively 
charged hadrons ($h^-$) were extracted. The analysis was done with the standard 
selection and reconstruction cuts. The quantities computed with 
the measured data were corrected with a corresponding Monte Carlo correction. 
The $h^-$ yields were computed with the most appropriate $T_2$ and $T_3$ trigger 
definitions. The resulting $h^-$ yields were compared with the $\pi^-$ yields 
obtained within the analysis framework described above. 
Such a comparison is reasonable since at the beam momentum of 31~\GeVc
the negatively-charged hadrons are strongly dominated 
by negatively-charged pions. It was concluded that both approaches 
provide yields in acceptable agreement. 
Some minor differences were observed only for the first longitudinal target bin, 
for small momentum (below 6~GeV/$c$) and polar angles ($\theta$ below 40~mrad), 
and in the $\theta$ interval [300; 400]~mrad for the first three longitudinal 
target bins.


\section{Dependence of the T2K re-weighting factors on the proton beam profile} \label{sec:ReWeightFactors}

The results presented in Section~\ref{sec:Results} will be used in the T2K neutrino beam simulation for re-weighting of the hadron yields on the target surface.
Details of the currently used re-weighting procedure based on \NASixtyOne thin-target measurements are presented in Ref.~\cite{T2Kflux}.

 The weights are simply calculated as the ratio of the data yields and the \FlukaEleven yields for each $(p,\:\theta,\: z)$  bin as follows:
\begin{equation}
	W_{ijk} = \left(\frac{1}{N_{POT}}\frac{n_{ijk}}{\Delta p_{ijk} \Delta \theta_{ij}}\right)_{data} / \left(\frac{1}{N_{POT}}\frac{n_{ijk}}{\Delta p_{ijk} \Delta \theta_{ij}}\right)_{MC}\label{eq_weightRT}
\end{equation}

\begin{figure*}[!htb]
	\begin{center}
  \begin{subfigure}[t]{0.8\textwidth}
        \includegraphics[trim= 0 0 0 0, clip, width=1\textwidth]{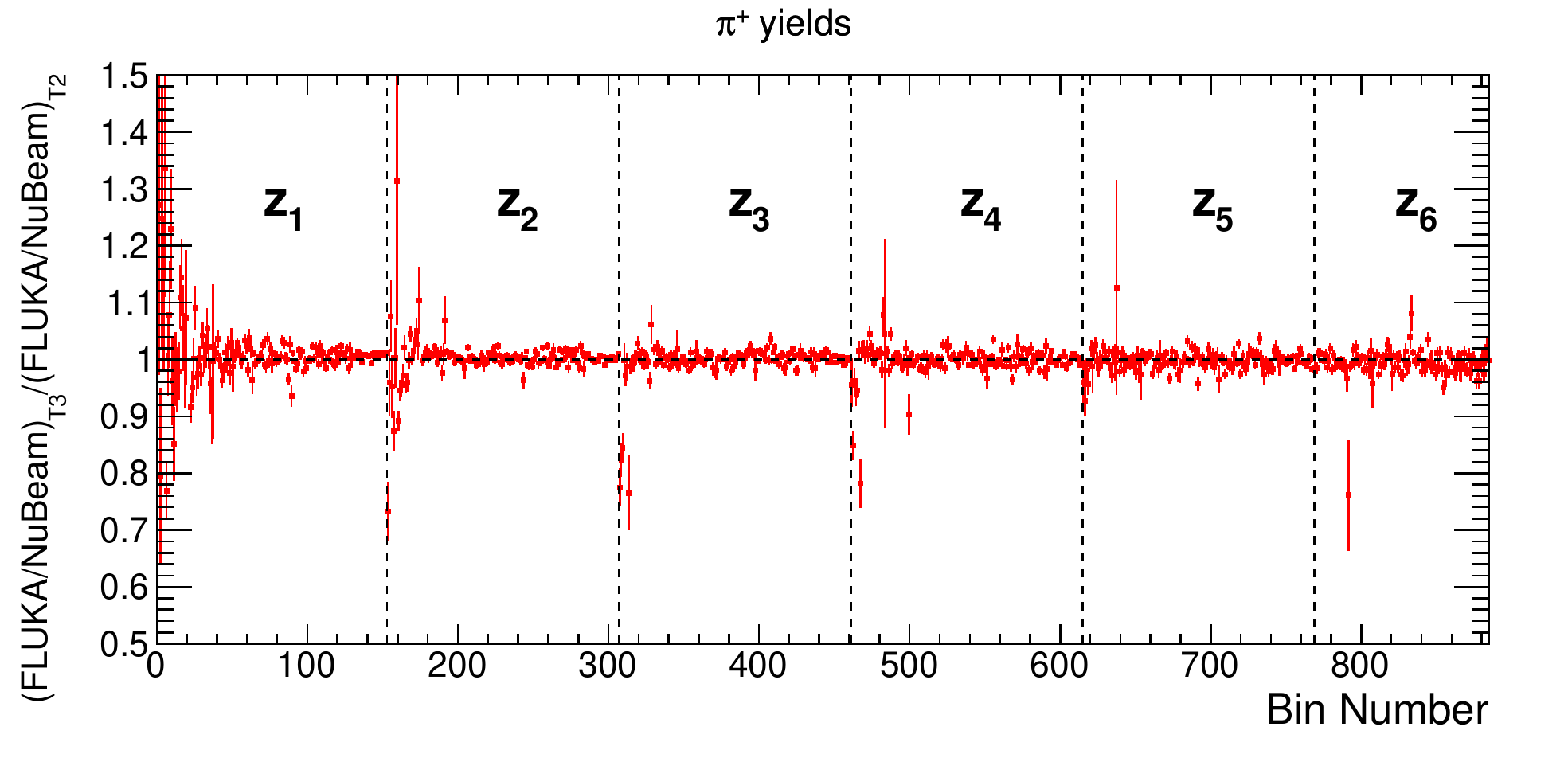}
    \caption{}
  \end{subfigure}	
  \begin{subfigure}[t]{0.8\textwidth}
        \includegraphics[trim= 0 0 0 0, clip, width=1\textwidth]{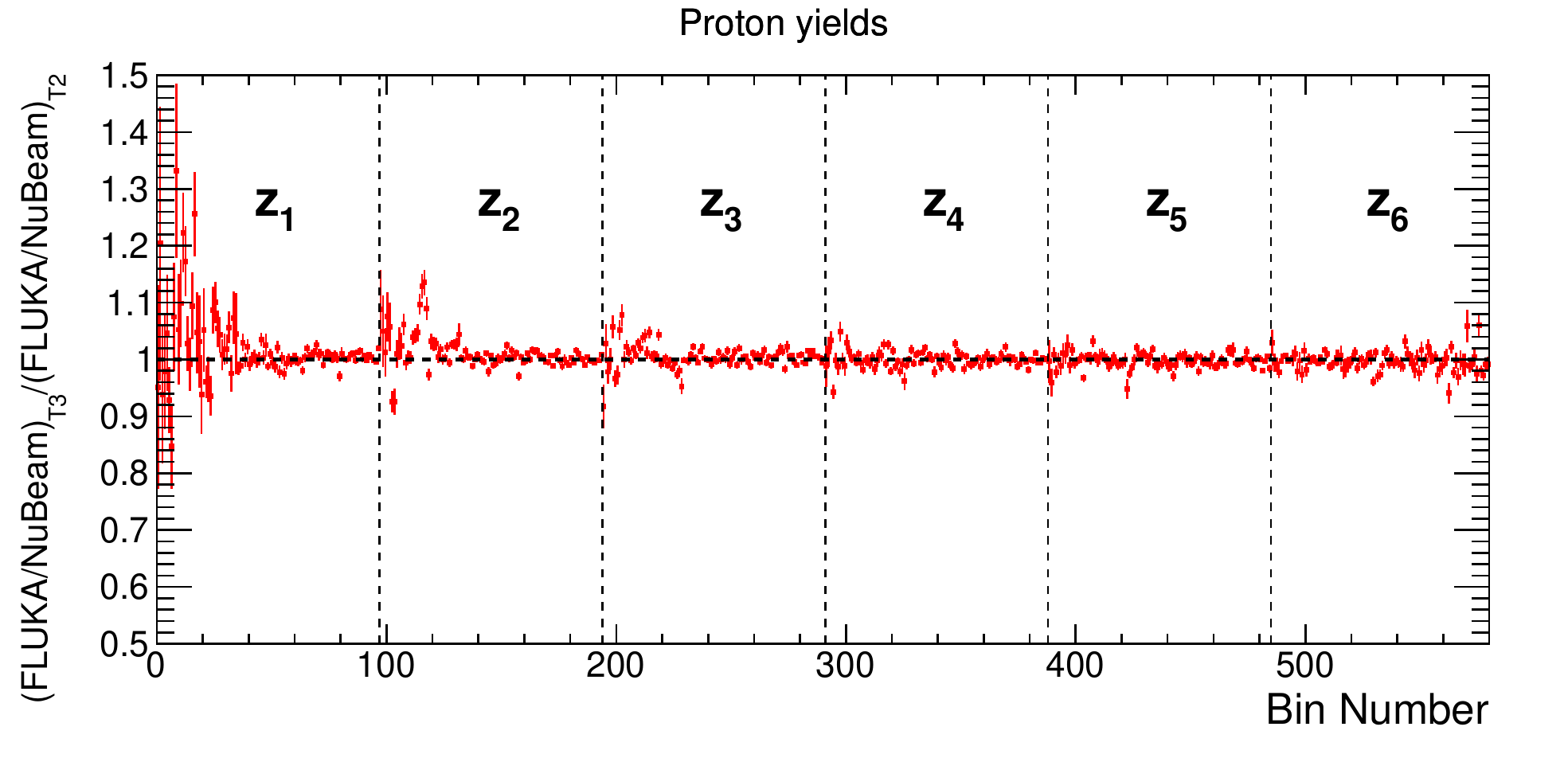}
    \caption{}
  \end{subfigure}	
 
 	\end{center}
\caption{Ratio of the simulated re-weighting factors produced with the $T_2$ and $T_3$ beam profiles for positively charged pions (a) and protons (b). }\label{fig:weightratio}	
\end{figure*}

These weights are applied to all neutrinos that have a hadron ancestor exiting the target surface in the phase space covered by \NASixtyOne. As previously mentioned, hadron yields coming from the surface of the long target depend on the width and position of the incoming proton beam hitting the upstream face of the target. Since it was shown that this effect is mostly geometrical, there is the possibility that the re-weighting factors are invariant under the beam profile change. To test this hypothesis, \FlukaEleven.2c.5 was used as input for data and the \NuBeam physics list from \GeantFT.03 as input for the Monte Carlo. Around $40 \times 10^6$ protons on target were simulated for \NASixtyOne $T_2$ and $T_3$ beam profiles using both models. Re-weighting factors were calculated for both beam profiles and compared. Examples of the ratios for \pip and \prot are shown in Fig.~\ref{fig:weightratio}. Some differences can be noticed for the upstream $z$ bins, while for other bins any differences are mostly below $2\%$. It is important to note that the differences between the $T_2$ and $T_3$ beam profiles are very large in comparison to differences between the $T_2$ and T2K beam profiles. 

Thus, one can
use the measured \NASixtyOne double differential hadron yields in the T2K flux simulation in the following way:
\begin{enumerate}[(i)]
\item Run the simulation of interactions inside the target for the \NASixtyOne $T_2$ beam profile. 
\item Calculate the re-weighting factors.
\item Run the simulation with the T2K beam profile.
\item Re-weight the new simulation using the previously calculated re-weighting factors.
\end{enumerate}

\section{Summary and conclusions} \label{sec:Conclusions}

Measurements of the double differential yields of \pip, \pim, \kp, \km and \prot emitted from the surface of a $90$-\cm-long carbon target (T2K replica) with incoming $31\:$\GeVc protons were performed by the \NASixtyOne experiment at the CERN SPS. Yields of \pip and \pim were measured with improved precision compared to the previously published \NASixtyOne results, while \kp, \km and \prot yields were obtained for the first time. These measurements are crucial for reducing the hadron production component of the T2K (anti)neutrino flux error,  which is the dominant component in the flux uncertainty.
Any reduction of the flux uncertainties will directly improve measurements of the (anti)neutrino-nucleus cross sections as well as of the (anti)neutrino oscillation parameters in T2K. A simple method of using these results in T2K which avoids problems with the dependence on the incident proton beam profile is proposed. 

The results were compared with the \NuBeam and \QGSP physics lists from \GeantFT.03 and show that none of the models predicts accurately the measured proton yields. This information can be used by model builders to improve hadronic Monte Carlo generators.

\begin{acknowledgements}

We would like to thank the CERN EP, BE and EN Departments for the
strong support of NA61/SHINE.

This work was supported by the Hungarian Scientific Research Fund (Grants
NKFIH 123842--123959), the J\'anos Bolyai Research Scholarship
of the Hungarian Academy of Sciences, the Polish Ministry of Science
and Higher Education (grants 667\slash N-CERN\slash2010\slash0,
NN\,202\,48\,4339 and NN\,202\,23\,1837), the Polish National Center
for Science (grants~2011\slash03\slash N\slash ST2\slash03691,
2013\slash11\slash N\slash ST2\slash03879, 2014\slash13\slash N\slash
ST2\slash02565, 2014\slash14\slash E\slash ST2\slash00018,
2014\slash15\slash B\slash ST2\slash02537 and
2015\slash18\slash M\slash ST2\slash00125, 2015\slash 19\slash N\slash ST2 \slash01689, 2016\slash23\slash B\slash ST2\slash00692, 
UMO-2014\slash 15\slash B\slash ST2\slash 03752),
the Russian Science Foundation, grant 16-12-10176, 
the Russian Academy of Science and the
Russian Foundation for Basic Research (grants 08-02-00018, 09-02-00664
and 12-02-91503-CERN), the Ministry of Science and
Education of the Russian Federation, grant No.\ 3.3380.2017\slash4.6,
 the National Research Nuclear
University MEPhI in the framework of the Russian Academic Excellence
Project (contract No.\ 02.a03.21.0005, 27.08.2013),
the Ministry of Education, Culture, Sports,
Science and Tech\-no\-lo\-gy, Japan, Grant-in-Aid for Sci\-en\-ti\-fic
Research (grants 18071005, 19034011, 19740162, 20740160 and 20039012),
the German Research Foundation (grant GA\,1480/2-2), the
Bulgarian Nuclear Regulatory Agency and the Joint Institute for
Nuclear Research, Dubna (bilateral contract No. 4418-1-15\slash 17),
Bulgarian National Science Fund (grant DN08/11), Ministry of Education
and Science of the Republic of Serbia (grant OI171002), Swiss
Nationalfonds Foundation (grant 200020\-117913/1), ETH Research Grant
TH-01\,07-3 and the U.S.\ Department of Energy.

\end{acknowledgements}

\bibliographystyle{spphys} 

\bibliography{submitEPJC}

\begin{thebibliography}{10}
\providecommand{\url}[1]{{#1}}
\providecommand{\urlprefix}{URL }
\expandafter\ifx\csname urlstyle\endcsname\relax
  \providecommand{\doi}[1]{DOI \discretionary{}{}{}#1}\else
  \providecommand{\doi}{DOI \discretionary{}{}{}\begingroup
  \urlstyle{rm}\Url}\fi

\bibitem{NA61detector_paper}
N.~Abgrall, et~al., JINST \textbf{9}, P06005 (2014).
\newblock \doi{10.1088/1748-0221/9/06/P06005}

\bibitem{T2K}
K.~Abe, et~al., Nucl. Instrum. Meth. \textbf{A659}, 106 (2011).
\newblock \doi{10.1016/j.nima.2011.06.067}

\bibitem{jparc:2003}
J-parc tdr, kek-report 2002-13 and jaeri-tech 2003-044 (2003).
\newblock Http://hadron.kek.jp/~accelerator/TDA/tdr2003/index2.html

\bibitem{T2Kflux}
K.~Abe, et~al., Phys. Rev. \textbf{D87}, 012001 (2013).
\newblock \doi{10.1103/PhysRevD.87.012001, 10.1103/PhysRevD.87.019902}

\bibitem{V0_2007}
N.~Abgrall, et~al., Phys. Rev. \textbf{C89}, 025205 (2014).
\newblock \doi{10.1103/PhysRevC.89.025205}

\bibitem{pion_paper}
N.~Abgrall, et~al., Phys. Rev. \textbf{C84}, 034604 (2011).
\newblock \doi{10.1103/PhysRevC.84.034604}

\bibitem{kaon_paper}
N.~Abgrall, et~al., Phys. Rev. \textbf{C85}, 035210 (2012).
\newblock \doi{10.1103/PhysRevC.85.035210}

\bibitem{thin2009paper}
N.~Abgrall, et~al., Eur. Phys. J. \textbf{C76}(2), 84 (2016).
\newblock \doi{10.1140/epjc/s10052-016-3898-y}

\bibitem{LTpaper}
N.~Abgrall, et~al., Nucl. Instrum. Meth. \textbf{A701}, 99 (2013).
\newblock \doi{10.1016/j.nima.2012.10.079}

\bibitem{LTpaper2009}
N.~Abgrall, et~al., Eur. Phys. J. \textbf{C76}(11), 617 (2016).
\newblock \doi{10.1140/epjc/s10052-016-4440-y}

\bibitem{Hasler:2039148}
A.~Hasler, Ph.D. Thesis, University of Geneva  (2015).
\newblock \urlprefix\url{https://cds.cern.ch/record/2039148}.
\newblock Presented 22 Jun 2015

\bibitem{Zambelli:2017pvq}
L.~Zambelli, A.~Fiorentini, T.~Vladisavljevic, J. Phys. Conf. Ser.
  \textbf{888}(1), 012067 (2017).
\newblock \doi{10.1088/1742-6596/888/1/012067}

\bibitem{PavinThesis}
M.~Pavin.
\newblock {Measurements of hadron yields from the T2K replica target in the
  NA61/SHINE experiment for neutrino flux prediction in T2K} (2017).
\newblock \urlprefix\url{https://cds.cern.ch/record/2292904}.
\newblock Ph.D. Thesis, University of Paris VI

\bibitem{CEDAR}
C.~Bovet, et~al.,   (1982).
\newblock CERN-YELLOW-82-13

\bibitem{toyotanso}
Toyotanso graphite.
\newblock
  \urlprefix\url{http://www.toyotanso.co.jp/Products/Special\_graphite/data\_en.html}

\bibitem{NA49-NIM}
S.~Afanasiev, et~al., Nucl.Instrum.Meth. \textbf{A430}, 210 (1999).
\newblock \doi{10.1016/S0168-9002(99)00239-9}

\bibitem{Gorbunov:2006pe}
S.~Gorbunov, I.~Kisel, Nucl. Instrum. Meth. \textbf{A559}, 148 (2006).
\newblock \doi{10.1016/j.nima.2005.11.133}

\bibitem{Wolin:1992ti}
E.J. Wolin, L.L. Ho, Nucl. Instrum. Meth. \textbf{A329}, 493 (1993).
\newblock \doi{10.1016/0168-9002(93)91285-U}

\bibitem{Fluka}
G.~Battistoni, et~al., AIP Conf. Proc. \textbf{896}, 31 (2007).
\newblock \doi{10.1063/1.2720455}

\bibitem{Fluka_CERN}
A.~Ferrari, P.R. Sala, A.~Fasso, J.~Ranft, CERN-2005-010, SLAC-R-773,
  INFN-TC-05-11

\bibitem{Fluka_new}
T.~B\"ohlen, et~al., Nuclear Data Sheets 120, 211-214  (2014)

\bibitem{Status_Report_2017}
A.~Aduszkiewicz, et~al., CERN-SPSC-2017-038 ; SPSC-SR-221  (2017)

\bibitem{Addendum2020+}
A.~Aduszkiewicz, et~al., CERN-SPSC-2018-008 ; SPSC-P-330-ADD-10  (2018)

\bibitem{Verkerke:2003ir}
W.~Verkerke, D.P. Kirkby, eConf \textbf{C0303241}, MOLT007 (2003).
\newblock [,186(2003)]

\bibitem{Brun:1987ma}
R.~Brun, F.~Bruyant, M.~Maire, A.~McPherson, P.~Zanarini, CERN-DD-EE-84-1
  (1987)

\bibitem{numtables}
Tables with numerical results for the t2k replica target 2010 paper.
\newblock \urlprefix\url{https://edms.cern.ch/document/1828979/1}

\bibitem{GEANT4}
S.~Agostinelli, et~al., Nucl. Instrum. Meth. \textbf{A506}, 250 (2003)

\bibitem{GEANT4bis}
J.~Allison, et~al., IEEE Transactions on Nuclear Science \textbf{53}(1), 270
  (2006)

\end{thebibliography}

\end{document}